\documentclass[aps,twocolumn,prd,10pt,showpacs,showkeys,preprintnumbers,superscriptaddress,nobibnotes,floatfix,longbibliography]{revtex4-1}

\pdfoutput=1
\usepackage{amsmath,amsfonts,amssymb,mathrsfs,graphicx,color,longtable}
\usepackage{hyperref}
\usepackage{graphicx}
\usepackage{amsfonts,amsmath,amssymb,bm}
\usepackage{array}
\usepackage{color}
\usepackage{enumitem}
\usepackage[explicit]{titlesec}
\usepackage[utf8]{inputenc}

\newcolumntype{C}[1]{>{\centering\let\newline\\\arraybackslash\hspace{4pt}}m{#1}}

\graphicspath{{figures/}}

\begin{document}

\title{DUNE as the Next-Generation Solar Neutrino Experiment}

\author{Francesco Capozzi}
\affiliation{Center for Cosmology and AstroParticle Physics (CCAPP), Ohio State University, Columbus, OH 43210}
\affiliation{Department of Physics, Ohio State University, Columbus, OH 43210}
\affiliation{Max-Planck-Institut f\"ur Physik (Werner-Heisenberg-Institut), 80805 M\"unchen, Germany}

\author{Shirley Weishi Li}
\affiliation{Center for Cosmology and AstroParticle Physics (CCAPP), Ohio State University, Columbus, OH 43210}
\affiliation{Department of Physics, Ohio State University, Columbus, OH 43210}
\affiliation{SLAC National Accelerator Laboratory, Menlo Park, CA, 94025}

\author{Guanying Zhu}
\affiliation{Center for Cosmology and AstroParticle Physics (CCAPP), Ohio State University, Columbus, OH 43210}
\affiliation{Department of Physics, Ohio State University, Columbus, OH 43210}

\author{John F. Beacom}
\affiliation{Center for Cosmology and AstroParticle Physics (CCAPP), Ohio State University, Columbus, OH 43210}
\affiliation{Department of Physics, Ohio State University, Columbus, OH 43210}
\affiliation{Department of Astronomy, Ohio State University, Columbus, OH 43210}

\date{4 September 2018; revised 11 June 2019}

\begin{abstract}
We show that the Deep Underground Neutrino Experiment (DUNE), with significant but feasible new efforts,  has the potential to deliver world-leading results in solar neutrinos. With a 100~kton-year exposure, DUNE could detect $\gtrsim 10^5$ signal events above 5~MeV electron energy.  Separate precision measurements of neutrino-mixing parameters and the $^8$B flux could be made using two detection channels ($\nu_e + \, ^{40}$Ar and $\nu_{e,\mu,\tau} + e^-$) and the day-night effect ($> 10 \sigma$).  New particle physics may be revealed through the comparison of solar neutrinos (with matter effects) and reactor neutrinos (without), which is discrepant by $\sim 2 \sigma$ (and could become $5.6 \sigma$).  New astrophysics may be revealed through the most precise measurement of the $^8$B flux (to 2.5\%) and the first detection of the {\it hep} flux (to 11\%).  {\it DUNE is required:} No other experiment, even proposed, has been shown capable of fully realizing these discovery opportunities.
\end{abstract}

\maketitle


\begin{figure}[t]
\begin{center}                  
\includegraphics[width=\columnwidth]{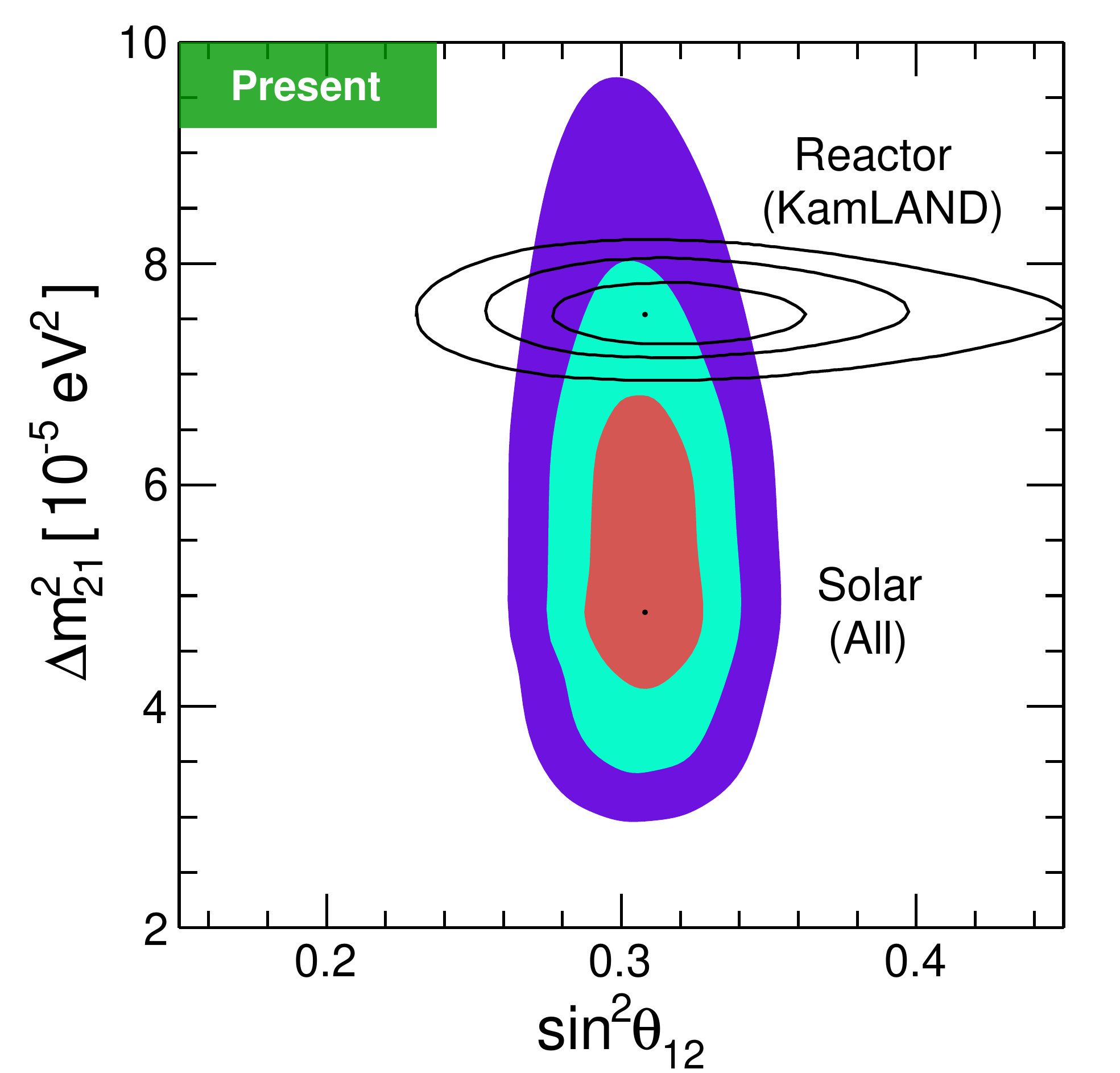}
\caption{Present measurements (1, 2, 3-$\sigma$) of neutrino mixing with solar~\cite{Cleveland:1998nv, Altmann:2005ix, Abdurashitov:2009tn, Bellini:2011rx, Aharmim:2011vm, Abe:2016nxk} and reactor~\cite{Gando:2013nba} neutrinos.}
\label{fig:contour_solar_kamland}
\end{center}
\end{figure}

\begin{figure}
\begin{center}                  
\includegraphics[width=\columnwidth]{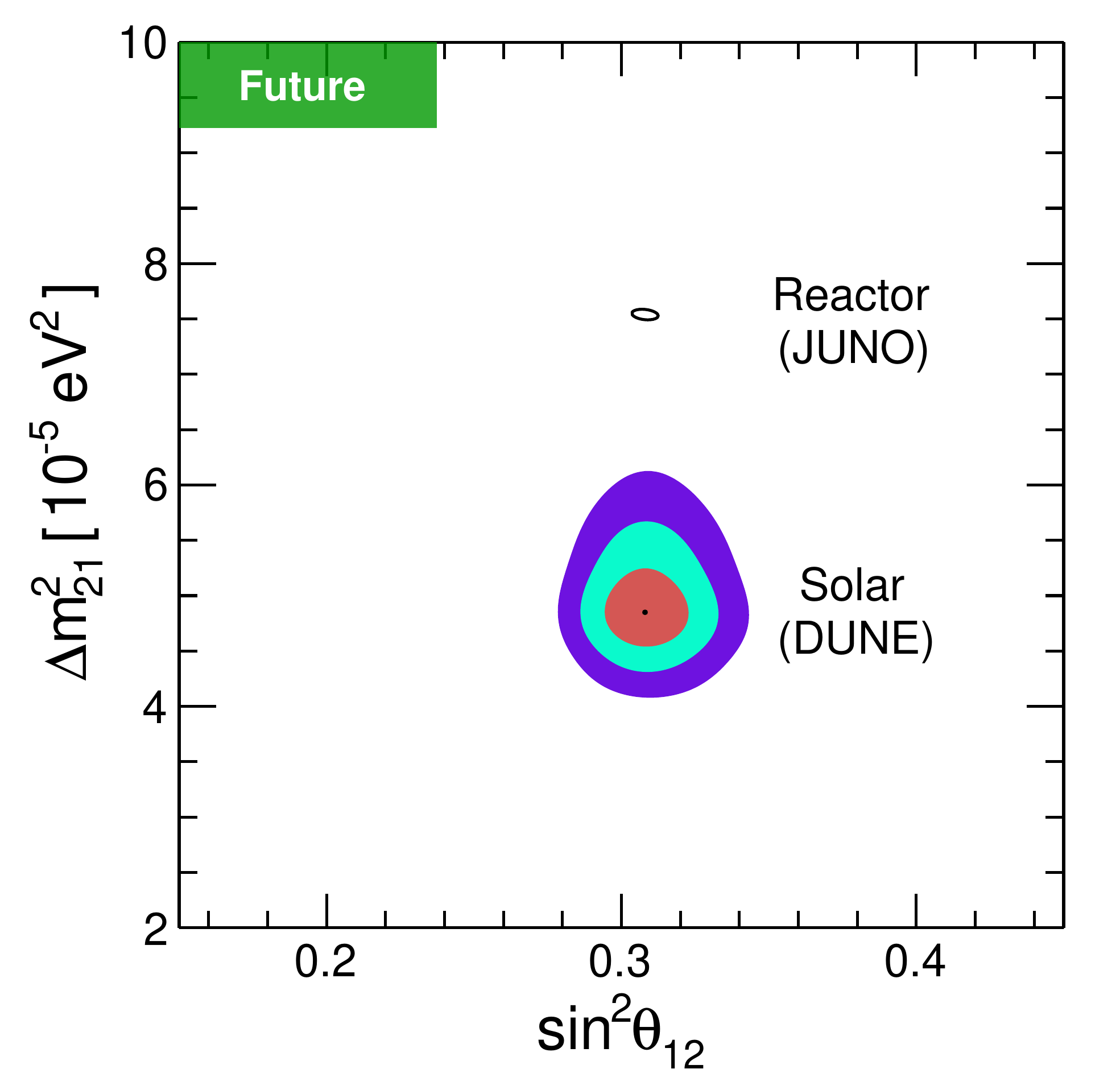}
\caption{Future precision of neutrino mixing with solar (DUNE alone; 1, 2, 3-$\sigma$) and reactor (JUNO alone; 3-$\sigma$~\cite{An:2015jdp, Capozzi:2015bpa}) neutrinos, using present best-fit points and 100 kton-year for each.}
\label{fig:contour_DUNE_JUNO}
\end{center}
\vspace{-0.4cm}
\end{figure}

{\bf Introduction.---}
Tremendous scientific opportunities remain in solar neutrinos.  What are the particle properties of neutrinos?  What are the nuclear processes that power our Sun and other stars?  Although the basics are known~\cite{Cleveland:1998nv, Altmann:2005ix, Abdurashitov:2009tn, Bellini:2011rx, Aharmim:2011vm, Abe:2016nxk}, there are multiple unknowns and discrepancies.  To progress, we need precise measurements of all neutrino-producing processes, plus ways to isolate new physics from new astrophysics.  Here we focus on high energies ($> 5$~MeV electron energy).

For particle physics, the primary opportunity is to test for new physics through a precision comparison of neutrino-mixing parameters~\cite{Barenboim:2001ac, deGouvea:2004va, Friedland:2004pp, Cirelli:2005sg, Palazzo:2011rj, Maltoni:2015kca, Capozzi:2017auw,Seo:2018rrb} measured in solar versus reactor experiments. Figures~\ref{fig:contour_solar_kamland} and~\ref{fig:contour_DUNE_JUNO} preview this.  There is a $\sim 2$-$\sigma$ discrepancy for $\Delta m^2_{21}$~\cite{Gando:2013nba, Esteban:2016qun, Abe:2016nxk, Capozzi:2018ubv}.  The reactor measurement will soon be greatly improved by the JUNO experiment~\cite{An:2015jdp}, but testing new physics depends on improving the solar measurement too.  The contrast in physical conditions is striking: neutrinos versus antineutrinos, matter versus vacuum mixing, plus a much larger distance, giving sensitivity to CPT violation~\cite{Barenboim:2001ac, deGouvea:2004va}, non-standard neutrino interactions~\cite{Friedland:2004pp, Liao:2017awz}, neutrino decay~\cite{Beacom:2002cb, Berryman:2014qha}, and more.

For astrophysics, the primary opportunity is to make an independent precise measurement of the $^8$B flux, which is extremely sensitive to the solar core temperature ($\sim T_c^{25}$~\cite{Bahcall:1996vj}), and which is an important ingredient for resolving the solar-metallicity discrepancy (requiring also progress on the $^7$Be and {\it CNO} fluxes)~\cite{Grevesse1998, Bahcall:2004pz, Bahcall_website, Asplund2009, Vinyoles:2016djt}. Discovery of the {\it hep} flux \cite{Grevesse1998, Bahcall:2004pz, Bahcall_website, Asplund2009, Vinyoles:2016djt, Marcucci:2000xy, Park:2002yp}, the highest-energy neutrino process, would probe physical conditions far from the solar center while still having large matter effects.

How can these opportunities be realized, especially simultaneously? This requires a new multi-10-kton scale experiment, plus breakthroughs in detection strategy.

We propose that DUNE --- intended to make transformative studies of GeV long-baseline neutrino mixing, proton decay, and supernova neutrino bursts~\cite{Acciarri:2015uup, Strait:2016mof, Acciarri:2016ooe} --- has the potential to do the same for solar neutrinos.  The budgeted plans for DUNE provide a large active volume, a huge overburden, and excellent technical capabilities, including at MeV energies (for supernovae)~\cite{Acciarri:2015uup, Strait:2016mof, Acciarri:2016ooe}.  For solar neutrinos, DUNE would need new investments, detailed below, that would also enhance its planned programs.  Building on prior work on solar-neutrino detection in liquid argon~\cite{Bahcall:1986ry, Arneodo:2000fa, Franco:2015pha, Ioannisian:2017dkx}, our paper goes much further.

We review the challenges in solar neutrinos, outline our proposed strategy for DUNE, calculate the signals and backgrounds, calculate the physics reach, define technical requirements, and conclude.  To show what DUNE could achieve, we calculate our main results under optimistic but feasible assumptions; we also discuss the impact of varying these assumptions.  In Supplemental Material (S.M.) and a separate paper on backgrounds~\cite{Zhu:2018rwc}, we provide further details.  Further technical studies will be needed.


{\bf Solar neutrinos: status and obstacles.---}
The fundamental challenge in solar neutrinos is disentangling neutrino-mixing effects and source properties.  Super-Kamiokande (Super-K) and Sudbury Neutrino Observatory (SNO) measurements of $^8$B neutrinos dominate the precision of solar determinations of $\sin^2\theta_{12}$ and $\Delta m^2_{21}$, as well as $\phi$($^8$B), the total $^8$B flux~\cite{Fogli:2005cq, Aharmim:2005gt,Aharmim:2011vm, Maltoni:2015kca, Abe:2016nxk}.  The {\it hep} flux, $\phi(hep) \sim 10^{-3} \, \phi$($^8$B), has not been detected~\cite{Hosaka:2005um, Mastbaum2016}.

Super-K and SNO measurements are consistent with an energy independent $\nu_e$ survival probability, $P_{ee} \simeq \sin^2\theta_{12}$; the lack of an observed upturn in $P_{ee}$ at low energies sets a weak upper limit on $\Delta m^2_{21}$~\cite{Aharmim:2011vm, Abe:2016nxk}.  Within the theoretical framework of matter-affected neutrino mixing~\cite{Wolfenstein:1977ue, Mikheev:1986gs, Mikheev:1986wj, Haxton:1986dm, Parke:1986jy, Giunti:2007ry}, these results are consistent with lower-energy solar neutrino data~\cite{Cleveland:1998nv, Altmann:2005ix, Abdurashitov:2009tn, Bellini:2011rx}.  Two other results were key:

$\bullet$
SNO separately measured $\phi$($^8$B) and $\sin^2\theta_{12}$ using two channels: $\nu_{e,\mu,\tau} + d \rightarrow \nu_{e,\mu,\tau} + p + n$, which is equally sensitive to all active flavors, and hence measures the total flux, and $\nu_e + d \rightarrow e^- + p + p$, from which they can then extract the mixing angle.  Progress on $\sin^2\theta_{12}$ is limited primarily by SNO's final precision for $\phi$($^8$B) of $\simeq 4\%$ ($\simeq 3\%$ statistical) and partially by the 1.7\% systematic uncertainty on the elastic-scattering channel in Super-K~\cite{Aharmim:2011vm, Fogli:2005cq, Maltoni:2015kca, Abe:2016nxk}.

$\bullet$
Super-K best constrains $\Delta m^2_{21}$ by measuring the day-night flux asymmetry (at $\simeq 3\sigma$)~\cite{Abe:2016nxk} with the $\nu_{e,\mu,\tau} + e^- \rightarrow \nu_{e,\mu,\tau} + e^-$ channel, where $P_{ee}$ at night is increased by several percent due to the matter effect in Earth~\cite{Lisi:1997yc, Maris:1997nk, Giunti:2007ry, daytime_models}.  Progress is limited by the slow increase in statistics after 20 years of exposure.


{\bf Unique advantages of DUNE.---}
DUNE will be in the Homestake mine in South Dakota (4300 m.w.e).  Each of two liquid-argon (LAr) modules (eventually four) will have a fiducial mass of 10 kton, surrounded by $\sim 1$ m LAr shielding (details depend on single- or dual-phase).  Readout is by the time-projection technique --- here drifting charge deposited in the volume onto wire planes at the boundaries --- plus prompt detection of scintillation light~\cite{Marchionni:2013tfa, Acciarri:2015uup, Strait:2016mof, Acciarri:2016ooe}.

{\it DUNE can simultaneously measure neutrino-mixing parameters and solar-neutrino fluxes}. Here, we first state our underlying ideas and simple estimates.

\begin{figure}[t]
\begin{center}
\includegraphics[width=\columnwidth]{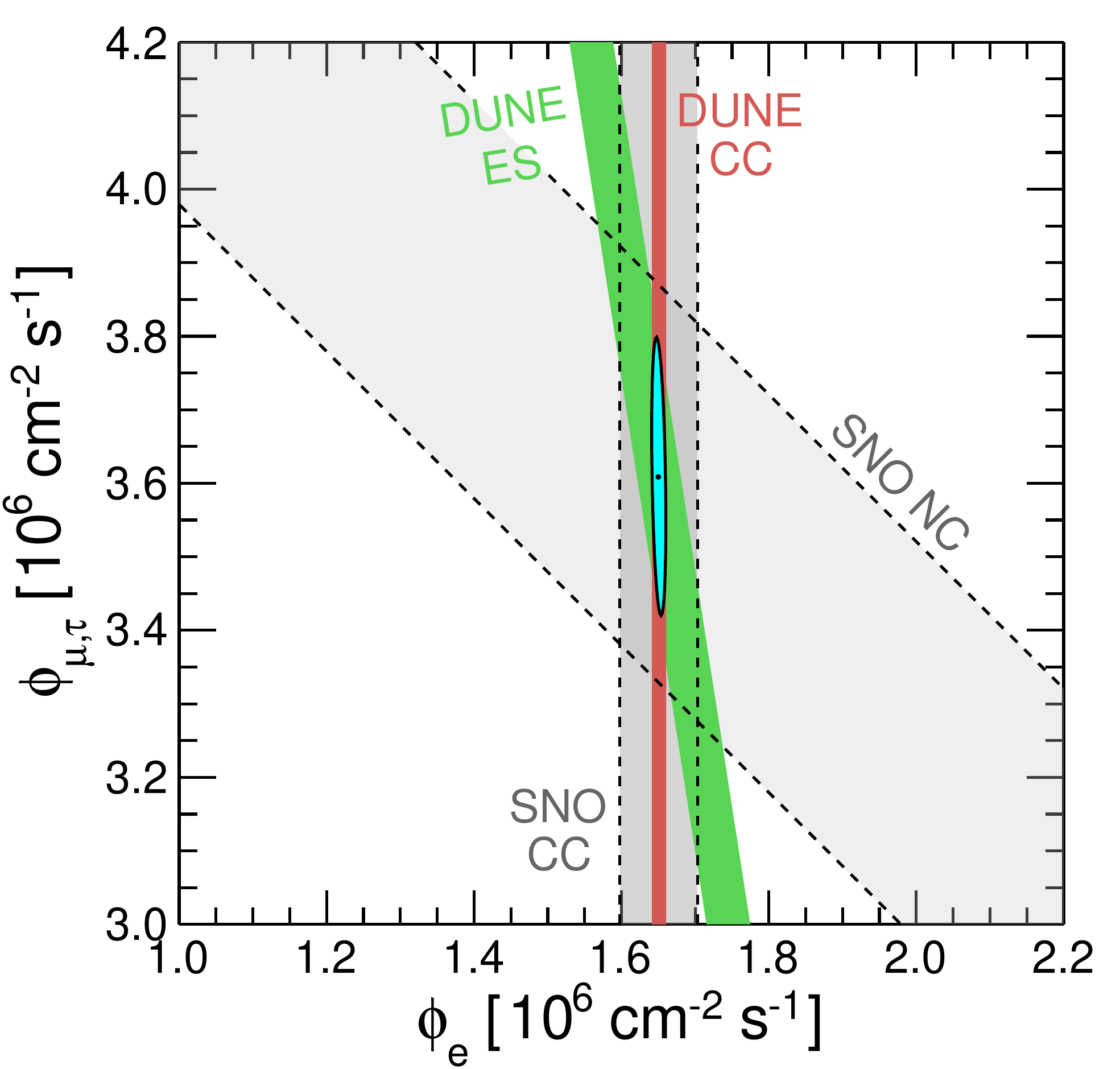}
\caption{Estimated precision of the $\nu_e$ and $\nu_{\mu,\tau}$ content of the $^8$B flux, present (SNO~\cite{Aharmim:2011vm, Aharmim:2011yq}) and future (DUNE), with the ellipse for DUNE alone.  Based on a simplified analysis, with only statistical uncertainties ($1 \sigma$) but assuming 2~d.o.f., and with SNO fluxes slightly rescaled to match their global-fit $^8$B flux.  Note small axis ranges.  Full analysis in text.}
\label{fig:flux_crossing}
\end{center}
\end{figure}

$\bullet$
The degeneracy between $\sin^2\theta_{12}$ and $\phi$($^8$B) can be broken using two detection channels:
\begin{equation}
	\nu_e + \, ^{40}{\rm Ar} \rightarrow e^- + \, ^{40}{\rm K}^*\,,
\end{equation}
where the rate $R_{\rm Ar} \propto \phi(^8{\rm B}) \times \sin^2\theta_{12}$, and
\begin{equation}
	\nu_{e,\mu,\tau} + e^- \rightarrow \nu_{e,\mu,\tau} + e^-\,,
\end{equation}
where $R_e \propto \phi(^8{\rm B}) \times \left(\sin^2\theta_{12} + \frac{1}{6}\cos^2\theta_{12}\right)$.  These channels can be adequately separated with a crude angular cut.  Though the dependence on the $\nu_{\mu,\tau}$ content is weak, DUNE can improve on SNO due to its huge statistics.  Figure~\ref{fig:flux_crossing} illustrates this.

$\bullet$
$\Delta m^2_{21}$ can be isolated through the day-night flux asymmetry, $A_{D/N} = (D - N)/\frac{1}{2}(D + N)$, which scales as $\propto E_\nu/\Delta m^2_{21}$.  For the solar $\Delta m^2_{21}$, an exposure of 100 kton-year, and using only events above 6 MeV electron energy (effective threshold; see below) and outside the forward cone, we expect $D = 3.04 \times 10^4$ and $N = 3.29 \times 10^4$ signal $\nu_e + \, ^{40}{\rm Ar} $ events, along with $0.83 \times 10^4$ background events in total (as detailed below, and conservatively including $\nu_{e,\mu,\tau} + e^-$ events).  Considering for now only statistical uncertainties, we expect $A_{D/N} \simeq -(7.9 \pm 0.8)$\% ($\sim 10 \sigma$). DUNE can improve on Super-K because the $\nu_e + \, ^{40}{\rm Ar}$ channel has a larger cross section, emphasizes larger neutrino energies, and a tighter relation between neutrino and electron energy.

{\bf Solar neutrinos in DUNE.---}
The MeV-range capabilities of DUNE~\cite{Acciarri:2015uup, Strait:2016mof, Acciarri:2016ooe} are designed for detecting supernova neutrinos.  Above 5~MeV, we assume electrons can be detected with high efficiency and 7\% energy-independent energy resolution~\cite{Acciarri:2016ooe}.  For solar signals, electrons lose energy dominantly by ionization, as the critical energy of LAr is 32~MeV~\cite{Patrignani:2016xqp, Amoruso:2003sw, Acciarri:2017sjy}.  The angular resolution of DUNE is uncertain; we adopt $25^\circ$, based on ICARUS simulations~\cite{Arneodo:2000fa}. Below, we discuss the impact of different assumptions.

We use neutrino spectra from Refs.~\cite{Winter:2004kf,Bahcall:1997eg} and radial distributions from the BS05(OP) model of Refs.~\cite{Bahcall:2004pz, Bahcall_website}.  As nominal fluxes, we use $\phi(^8{\rm B}) = 5.25 \times 10^6$~cm$^{-2}$~s$^{-1}$ (4\% uncertainty; from SNO~\cite{Aharmim:2011vm}) and $\phi(hep) = 8.25 \times 10^3$~cm$^{-2}$~s$^{-1}$ (30\% uncertainty; from theory~\cite{Vinyoles:2016djt, Park:2002yp}).  The endpoint energies are $\simeq 15$ and $\simeq 19$~MeV.

For the charged-current (CC) channel $\nu_e + \, ^{40}$Ar~\cite{Konopinski1950, Ormand:1994js, Bhattacharya:1998hc, Beacom:1998fj, Vogel:1999zy, Beacom:2001hr, Kurylov:2002vj, GilBotella:2003sz, Kolbe:2003ys, Bhattacharya:2009zz, Cheoun:2011zza, Suzuki2014, Hardy:2014qxa, Karakoc:2014awa, Akimov:2017ade}, the ground state threshold is $Q_\text{gs} = 1.5$~MeV~\cite{NNDC}, but this transition is forbidden. The cross section is dominated by transitions to nuclear excited states in $^{40}$K$^*$ (a super-allowed Fermi transition with $\Delta E_i = 4.4$~MeV, plus several Gamow-Teller transitions), which promptly produce gamma rays by nuclear de-excitation.  Due to these nuclear thresholds, DUNE is most sensitive to $E_\nu \gtrsim 9$~MeV.  We define the detectable energy of an event as the electron kinetic energy $T_e$, given by $T_e = E_\nu - Q$, where $Q = Q_\text{gs} + \Delta E_i$, conservatively neglecting the detectability of the $\Delta E_i$ in gamma rays (if these gamma rays were detectable, that would dramatically improve event identification, background rejection, energy reconstruction, and sensitivity).  The electrons are emitted near-isotropically.  Details, including cross section uncertainties, are discussed below and in S.M.

For the elastic-scattering (ES) channel $\nu_{e,\mu,\tau} + e^-$, there is no threshold and the cross section is known with sub-percent precision \cite{Bahcall:1995mm}.  All flavors participate, but the sensitivity to the $\nu_{\mu,\tau}$ content is reduced, as these have only neutral-current couplings. DUNE is sensitive to $E_\nu \gtrsim 5$~MeV, though the broad differential cross section effectively raises that.  The direction of the scattered electron is well correlated to the neutrino direction, with a maximum scattering angle of about 20$^\circ$.  We adequately separate $\nu_{e,\mu,\tau} + e^-$ and $\nu_e + \, ^{40}$Ar events by defining a forward cone of half-angle 40$^\circ$, maximizing the signal to background ratios for both event categories in the cone away from the Sun and its complement.  Inside the cone, which includes 81\% of $\nu_{e,\mu,\tau} + e^-$ events~\cite{Arneodo:2000fa}, they dominate; outside the cone, which includes 88\% of $\nu_e + \, ^{40}$Ar events, they dominate.

In principle, DUNE could use the neutral-current (NC) channel $\nu_{e,\mu,\tau} + \, ^{40}{\rm Ar} \rightarrow \nu_{e,\mu,\tau} + \, ^{40}{\rm Ar}^*$, where the final state is detected through nuclear gamma rays~\cite{Raghavan:1986fg, Kolbe:2003ys, Cheoun:2011zza}.  We treat this as a background because the cross section seems small.  If not, this could be an important new signal.

Backgrounds must be mitigated with standard MeV-detector techniques: defining a fiducial volume, removing U/Th from liquids and Rn from air, selecting low-background materials, applying deadtime after high-energy events, etc.~\cite{Cleveland:1998nv, Takeuchi:1999zq, Blevis:2003ih, Altmann:2005ix, Abdurashitov:2009tn, Bellini:2011rx, Aharmim:2011vm, Abe:2016nxk, Gando:2013nba, An:2015jdp}.  Three important backgrounds will remain (details in S.M. and Ref.~\cite{Zhu:2018rwc}).  First, neutron captures on $^{40}$Ar, releasing a total of 6.1~MeV in several gamma rays~\cite{Hardell1970, Nesaraja:2016ktw, CapGam}; these Compton scatter or pair-produce electrons.  These neutrons, most less than a few MeV, are dominantly produced by $(\alpha,n)$ interactions in the rock following U/Th-chain decays~\cite{Arneodo:2000fa, Wulandari:2003cr, deViveiros:2010fnb}; muon-induced neutrons are relatively negligible~\cite{Zhu:2018rwc}.  Once neutrons enter the detector, they fill the volume, due to their small cross section on argon.  We assume a hermetic, passive water (/oil/plastic) shield of 40-cm thickness, reducing this background by $\sim 4 \times 10^3$.  Below, we show that even no shielding is acceptable. Second, neutral-current $\nu_{e,\mu,\tau} + \, ^{40}{\rm Ar}$ events cause a peak near 9~MeV~\cite{Raghavan:1986fg}.  Third, at the highest energies, beta-decaying radioactivities induced by muons~\cite{Li:2014sea, Li:2015kpa, Li:2015lxa, Zhu:2018rwc}, for which we apply simple cuts.  The pileup rates from these and other backgrounds (e.g., $^{39}$Ar and $^{42}$Ar decays) are negligible~\cite{Zhu:2018rwc}.

\begin{figure*}[t]
\begin{center}
\includegraphics[width=1.9\columnwidth]{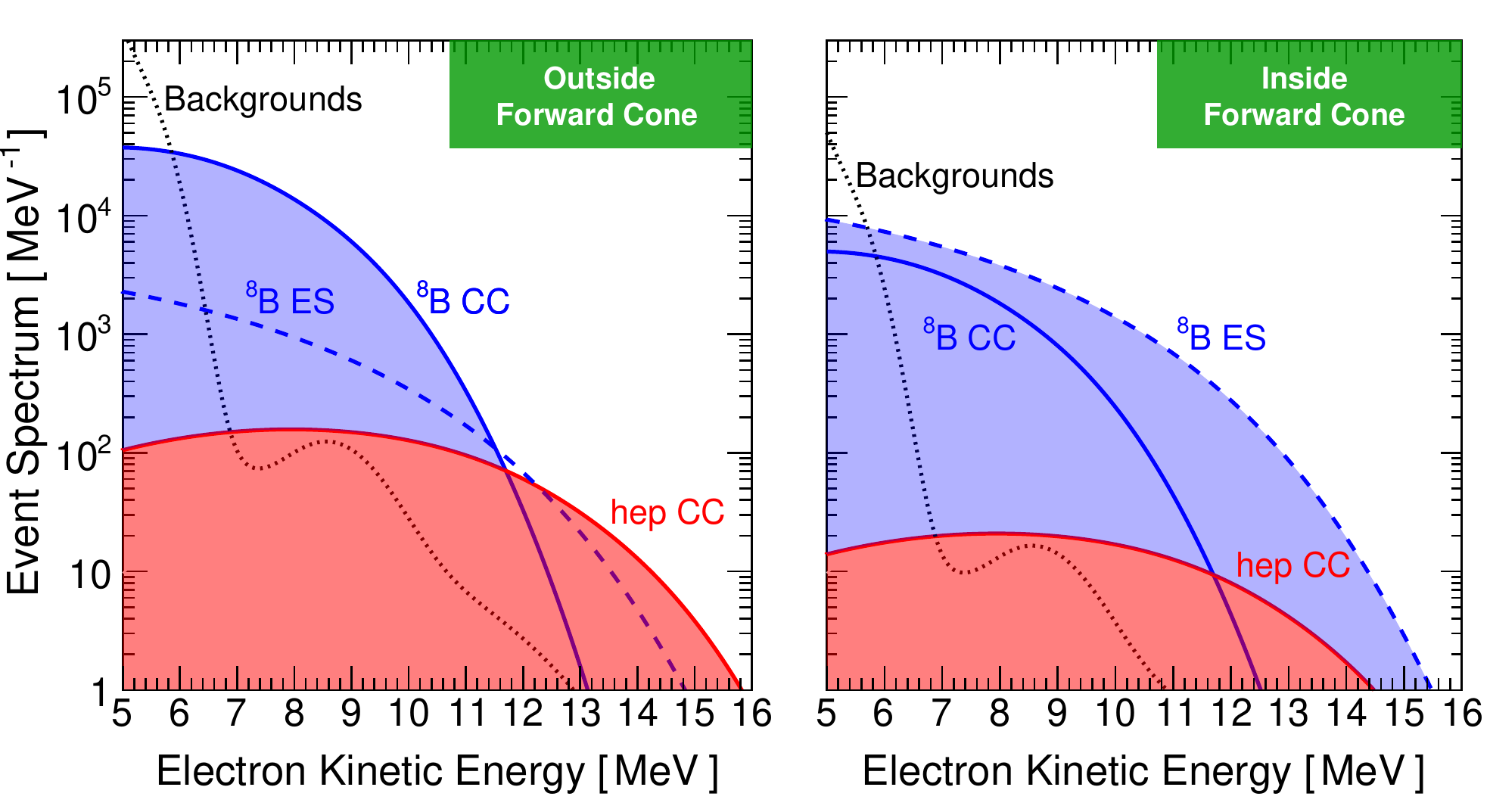}
\caption{Predicted solar-neutrino signals and backgrounds in DUNE for 100~kton-year, using a forward cone of half-angle 40$^{\circ}$ and (here only) combining day and night data.  We include all factors discussed in the text.}
\label{fig:event_spectrum}
\end{center}
\end{figure*}

Figure~\ref{fig:event_spectrum} shows the solar-neutrino signal and background spectra in DUNE.  Our calculations include three-flavor neutrino mixing~\cite{Wolfenstein:1977ue, Mikheev:1986gs, Mikheev:1986wj, Haxton:1986dm, Parke:1986jy, Lisi:1997yc, Maris:1997nk, Giunti:2007ry}, realistic detection effects (differential cross sections~\cite{Behrens1969, Schenter1983, Bhattacharya:2009zz, Hayes:2016qnu}, energy smearing, angular cuts~\cite{Arneodo:2000fa}, background reduction~\cite{rogers1990geology, shultis2002, Ferrari:2005zk, Battistoni:2007zzb, Lesko:2011qk, Chan:rock, Ricci:2014qpa, Heise:2015vza, Westerdale:2017kml, JinpingNeutrinoExperimentgroup:2016nol, Gaisser:2002jj, Abe:2017aap, Beacom:2010kk, Bays:2011si, NNDC}), and a 100~kton-year exposure.

For $^8$B events, the two channels are well separated and have superb yields above 5~MeV electron energy.  For $\nu_e + \, ^{40}$Ar, there are $9.9 \times 10^4$ events outside the forward cone.  For $\nu_{e,\mu,\tau} + e^-$, there are $2.6 \times 10^4$ events inside the forward cone.  This channel provides better sensitivity to lower-energy neutrinos and the only sensitivity to $\nu_{\mu, \tau}$.  For {\it hep} events, the $\nu_e + \, ^{40}$Ar channel allows a clear separation at high electron energies, with 150 events above 11~MeV (the small $\nu_{e,\mu,\tau} + e^-$ channel is not shown).


{\bf DUNE physics reach.---}
DUNE can significantly improve the precision of solar neutrino observables. We jointly fit (without priors) four parameters: $\sin^2\theta_{12}$, $\Delta m^2_{21}$, $\phi(^8\text{B})$, $\phi(hep)$.  When reporting projected uncertainties for $n$ parameters, we marginalize over the others, adopting $\Delta \chi^2$ confidence levels for $n$ d.o.f.  We assume that new physics affecting solar neutrinos is reflected in mixing-parameter values that differ from the reactor values. We use the theoretically expected counts for signals and backgrounds, with Poisson uncertainties (below, we discuss systematics).  We partition data into bins of energy, bins of Earth zenith angle (night only), and outside/inside the forward cone (as in Fig.~\ref{fig:event_spectrum}).  We use the total electron spectra, assuming only statistical separation of the components, and not requiring neutrino-energy reconstruction.

Figure~\ref{fig:contour_DUNE_JUNO} shows the projected precision for DUNE's measurement of neutrino mixing parameters, assuming the solar best-fit values ($\sin^2\theta_{12}$ = 0.308, $\Delta m^2_{21} = 4.85 \times 10^{-5}$ eV$^2$~\cite{Abe:2016nxk}).  The uncertainties are 3.0\% and 5.9\%, a factor of $\simeq 1.5$ and $\simeq 3$ better than from all solar experiments to date, respectively, as shown in Fig.~\ref{fig:contour_solar_kamland}.  The sensitivity to $\Delta m^2_{21}$ comes primarily from the day-night effect (10.4$\sigma$).  The $^8$B flux (marginalizing over other parameters) can be measured to $2.5\%$, a factor $\simeq 1.6$ better than from SNO.  DUNE can make a robust first detection of {\it  hep} neutrinos, with a precision of 11\%, a factor $\simeq 3$ better than the current theoretical uncertainty.


{\bf Going forward.---}
New investments are needed to enhance the MeV capabilities of DUNE.  At the trigger level, this includes enhancing data acquisition, storage, and processing for a steady rate of MeV events.  Calibration at MeV energies across the detector volume will be crucial to controlling systematics.  An enhanced light-detection system would enhance MeV detection.

Backgrounds must be controlled, and the biggest concern, after standard cuts~\cite{Cleveland:1998nv, Takeuchi:1999zq, Blevis:2003ih, Altmann:2005ix, Abdurashitov:2009tn, Bellini:2011rx, Aharmim:2011vm, Abe:2016nxk, Gando:2013nba, An:2015jdp}, is neutron captures~\cite{Zhu:2018rwc}.  We see three possible strategies.

1. 40~cm of shielding, as assumed above.  This allows a low threshold ($\simeq 5.8$~MeV) to test for shape distortions in the spectrum and to enhance particle-identification techniques.  

2. Additional runtime.  With less or no shielding, the effective analysis threshold would be higher.  This can be compensated by a larger exposure than 100 kton-year.  With 30, 20, 10, or 0~cm of shielding, the effective analysis threshold is $\simeq$ 6.2, 6.5, 6.9, or 7.2~MeV, and the exposure needed for comparable results increases at most of a factor of $\sim 2$.

3. Better particle-identification techniques.  We assume neutrino-interaction events and neutron-capture events with the same electron energy are indistinguishable.  This is too conservative because $\nu_{e,\mu,\tau} + e^-$, $\nu_e + \, ^{40}$Ar, and neutron-capture events would have one electron, an electron with gammas, and multiple gammas, respectively.  For gamma rays, the radiation length is 14~cm and Compton-scattering dominates~\cite{Patrignani:2016xqp}, so these event classes should be distinct.

The impact of varying the assumptions made above is detailed in S.M.  Here we summarize some key points. The sensitivity to $\Delta m^2_{21}$ is very robust because it is determined from the day-night effect, which cancels many inputs. The uncertainty on the $\nu_e +\,^{40}{\rm Ar}$ cross section (presently 10\%) and the detector systematics should be reduced to 1\%.  Without this, DUNE alone cannot break the degeneracy between $\sin^2\theta_{12}$ and $\phi(^8$B), though its measurement of $\Delta m^2_{21}$ would be unaffected and its measurement of $\phi(hep)$ only be modestly affected. Importantly, the intended precision of $\sin^2\theta_{12}$ and $\phi(^8$B) could be largely restored by combining with existing solar data. If the energy resolution is 20\%, the intended precision of $\phi(hep)$ would degrade to 18\%, though that of $\sin^2\theta_{12}$, $\phi(^8$B) and $\Delta m^2_{21}$ could be largely restored by increasing exposure by $\sim$2. DUNE's sensitivity would be lost if the energy resolution is poor {\it and} backgrounds are not reduced.


{\bf Concluding perspectives.---}
This is the first study to detail how DUNE, with a new, challenging but feasible solar-neutrino program, would open substantial discovery space in both particle physics and astrophysics. With DUNE's precision measurements of $\sin^2\theta_{12}$ and $\Delta m^2_{21}$, the comparison to JUNO's may reveal new particle physics of neutrinos.  Simultaneously, with DUNE's precision measurements of $\phi(^8\text{B})$ and $\phi(hep)$, it may reveal new astrophysics.  Further studies are needed to evaluate our proposal and to optimize its sensitivity.

No other planned experiment has been shown potentially capable of meeting all of these goals.  Of proposed experiments, Hyper-Kamiokande (Hyper-K)~\cite{Abe:2011ts, Abe:2018uyc} stands out, and it can nicely complement DUNE.  Hyper-K would have only one channel, $\nu_{e,\mu,\tau} + e^-$, but huge statistics.  DUNE and Hyper-K would measure $\Delta m^2_{21}$ (from the day-night asymmetry) comparably well.  Hyper-K would have a significant advantage on measuring the upturn in the $\nu_e$ survival probability.  DUNE would measure $\phi(hep)$ much better.  Their combined impact would be significantly enhanced by new experiments for low-energy solar neutrinos~\cite{Franco:2015pha, Andringa:2015tza, Aalbers:2016jon}.

Solar neutrino studies, begun long ago, are not done.  DUNE can lead the next generation of discoveries.


 \medskip
\begin{acknowledgments}
For helpful discussions, we are grateful to
Manojeet Bhattacharya,
Ed Blucher,
Steve Brice,
Mauricio Bustamante,
Gustavo Cancelo,
David Caratelli,
Flavio Cavanna,
Alex Friedland,
Dick Furnstahl,
Cristiano Galbiati,
Alejandro Garcia,
Mathew Graham,
Jeff Hartnell, 
Alex Himmel,
Brennan Jordan,
Josh Klein,
Pierre Lasorak,
Eligio Lisi,
Joe Lykken,
Kenny Ng,
Gabriel Orebi Gann,
Ornella Palamara,
Wendy Panero,
Stephen Parke,
Ryan Patterson,
Simon Peeters,
Georg Raffelt,
Juergen Reichenbacher,
Elizabeth Ricard-McCutchan,
David Schmitz,
Kate Scholberg,
Michael Smy,
Bob Svoboda,
Mark Vagins,
Francesco Vissani,
Petr Vogel,
and
Kyle Wendt.

The work of all authors was supported by NSF Grants PHY-1404311 and PHY-1714479 awarded to JFB.  FC was later supported by Deutsche Forschungsgemeinschaft Grants EXC 153 and SFB 1258, as well as by European Union Grant H2020-MSCA-ITN-2015/674896.  SWL was also supported by an Ohio State Presidential Fellowship, and later at SLAC by the Department of Energy under contract number DE-AC02-76SF00515.

\vspace{0.7cm}
\centerline{\bf Author Information}
\medskip
\phantom{foobar}

\noindent {\tt capozzi@mpp.mpg.de}\,, \href{https://orcid.org/0000-0001-6135-1531}{\tt 0000-0001-6135-1531}

\noindent {\tt shirleyl@slac.stanford.edu}\,, \href{https://orcid.org/0000-0002-2157-8982}{\tt 0000-0002-2157-8982}

\noindent {\tt zhu.1475@osu.edu}\,, \href{https://orcid.org/0000-0003-0031-634X}{\tt 0000-0003-0031-634X}

\noindent {\tt beacom.7@osu.edu}\,, \href{https://orcid.org/0000-0002-0005-2631}{\tt 0000-0002-0005-2631}

\medskip

We speak for ourselves as theorists, not on behalf of the DUNE Collaboration.  This work is based on our ideas, our calculations, and publicly available information.

\end{acknowledgments}



\onecolumngrid

\newpage

\twocolumngrid

\bibliographystyle{apsrev4-1}
\bibliography{refs}

\begin{thebibliography}{107}%
\makeatletter
\providecommand \@ifxundefined [1]{%
 \@ifx{#1\undefined}
}%
\providecommand \@ifnum [1]{%
 \ifnum #1\expandafter \@firstoftwo
 \else \expandafter \@secondoftwo
 \fi
}%
\providecommand \@ifx [1]{%
 \ifx #1\expandafter \@firstoftwo
 \else \expandafter \@secondoftwo
 \fi
}%
\providecommand \natexlab [1]{#1}%
\providecommand \enquote  [1]{``#1''}%
\providecommand \bibnamefont  [1]{#1}%
\providecommand \bibfnamefont [1]{#1}%
\providecommand \citenamefont [1]{#1}%
\providecommand \href@noop [0]{\@secondoftwo}%
\providecommand \href [0]{\begingroup \@sanitize@url \@href}%
\providecommand \@href[1]{\@@startlink{#1}\@@href}%
\providecommand \@@href[1]{\endgroup#1\@@endlink}%
\providecommand \@sanitize@url [0]{\catcode `\\12\catcode `\$12\catcode
  `\&12\catcode `\#12\catcode `\^12\catcode `\_12\catcode `\%12\relax}%
\providecommand \@@startlink[1]{}%
\providecommand \@@endlink[0]{}%
\providecommand \url  [0]{\begingroup\@sanitize@url \@url }%
\providecommand \@url [1]{\endgroup\@href {#1}{\urlprefix }}%
\providecommand \urlprefix  [0]{URL }%
\providecommand \Eprint [0]{\href }%
\providecommand \doibase [0]{http://dx.doi.org/}%
\providecommand \selectlanguage [0]{\@gobble}%
\providecommand \bibinfo  [0]{\@secondoftwo}%
\providecommand \bibfield  [0]{\@secondoftwo}%
\providecommand \translation [1]{[#1]}%
\providecommand \BibitemOpen [0]{}%
\providecommand \bibitemStop [0]{}%
\providecommand \bibitemNoStop [0]{.\EOS\space}%
\providecommand \EOS [0]{\spacefactor3000\relax}%
\providecommand \BibitemShut  [1]{\csname bibitem#1\endcsname}%
\let\auto@bib@innerbib\@empty
\bibitem [{\citenamefont {Cleveland}\ \emph {et~al.}(1998)\citenamefont
  {Cleveland}, \citenamefont {Daily}, \citenamefont {Davis}, \citenamefont
  {Distel}, \citenamefont {Lande}, \citenamefont {Lee}, \citenamefont
  {Wildenhain},\ and\ \citenamefont {Ullman}}]{Cleveland:1998nv}%
  \BibitemOpen
  \bibfield  {author} {\bibinfo {author} {\bibfnamefont {B.~T.}\ \bibnamefont
  {Cleveland}}, \bibinfo {author} {\bibfnamefont {T.}~\bibnamefont {Daily}},
  \bibinfo {author} {\bibfnamefont {R.}~\bibnamefont {Davis}, \bibfnamefont
  {Jr.}}, \bibinfo {author} {\bibfnamefont {J.~R.}\ \bibnamefont {Distel}},
  \bibinfo {author} {\bibfnamefont {K.}~\bibnamefont {Lande}}, \bibinfo
  {author} {\bibfnamefont {C.~K.}\ \bibnamefont {Lee}}, \bibinfo {author}
  {\bibfnamefont {P.~S.}\ \bibnamefont {Wildenhain}}, \ and\ \bibinfo {author}
  {\bibfnamefont {J.}~\bibnamefont {Ullman}},\ }\href {\doibase 10.1086/305343}
  {\bibfield  {journal} {\bibinfo  {journal} {Astrophys. J.}\ }\textbf
  {\bibinfo {volume} {496}},\ \bibinfo {pages} {505} (\bibinfo {year}
  {1998})}\BibitemShut {NoStop}%
\bibitem [{\citenamefont {Altmann}\ \emph {et~al.}(2005)\citenamefont {Altmann}
  \emph {et~al.}}]{Altmann:2005ix}%
  \BibitemOpen
  \bibfield  {author} {\bibinfo {author} {\bibfnamefont {M.}~\bibnamefont
  {Altmann}} \emph {et~al.} (\bibinfo {collaboration} {GNO}),\ }\href {\doibase
  10.1016/j.physletb.2005.04.068} {\bibfield  {journal} {\bibinfo  {journal}
  {Phys. Lett. B}\ }\textbf {\bibinfo {volume} {616}},\ \bibinfo {pages} {174}
  (\bibinfo {year} {2005})},\ \Eprint {http://arxiv.org/abs/hep-ex/0504037}
  {arXiv:hep-ex/0504037 [hep-ex]} \BibitemShut {NoStop}%
\bibitem [{\citenamefont {Abdurashitov}\ \emph {et~al.}(2009)\citenamefont
  {Abdurashitov} \emph {et~al.}}]{Abdurashitov:2009tn}%
  \BibitemOpen
  \bibfield  {author} {\bibinfo {author} {\bibfnamefont {J.~N.}\ \bibnamefont
  {Abdurashitov}} \emph {et~al.} (\bibinfo {collaboration} {SAGE}),\ }\href
  {\doibase 10.1103/PhysRevC.80.015807} {\bibfield  {journal} {\bibinfo
  {journal} {Phys. Rev. C}\ }\textbf {\bibinfo {volume} {80}},\ \bibinfo
  {pages} {015807} (\bibinfo {year} {2009})},\ \Eprint
  {http://arxiv.org/abs/0901.2200} {arXiv:0901.2200 [nucl-ex]} \BibitemShut
  {NoStop}%
\bibitem [{\citenamefont {Bellini}\ \emph {et~al.}(2011)\citenamefont {Bellini}
  \emph {et~al.}}]{Bellini:2011rx}%
  \BibitemOpen
  \bibfield  {author} {\bibinfo {author} {\bibfnamefont {G.}~\bibnamefont
  {Bellini}} \emph {et~al.} (\bibinfo {collaboration} {Borexino}),\ }\href
  {\doibase 10.1103/PhysRevLett.107.141302} {\bibfield  {journal} {\bibinfo
  {journal} {Phys. Rev. Lett.}\ }\textbf {\bibinfo {volume} {107}},\ \bibinfo
  {pages} {141302} (\bibinfo {year} {2011})},\ \Eprint
  {http://arxiv.org/abs/1104.1816} {arXiv:1104.1816 [hep-ex]} \BibitemShut
  {NoStop}%
\bibitem [{\citenamefont {Aharmim}\ \emph
  {et~al.}(2013{\natexlab{a}})\citenamefont {Aharmim} \emph
  {et~al.}}]{Aharmim:2011vm}%
  \BibitemOpen
  \bibfield  {author} {\bibinfo {author} {\bibfnamefont {B.}~\bibnamefont
  {Aharmim}} \emph {et~al.} (\bibinfo {collaboration} {SNO}),\ }\href {\doibase
  10.1103/PhysRevC.88.025501} {\bibfield  {journal} {\bibinfo  {journal} {Phys.
  Rev. C}\ }\textbf {\bibinfo {volume} {88}},\ \bibinfo {pages} {025501}
  (\bibinfo {year} {2013}{\natexlab{a}})},\ \Eprint
  {http://arxiv.org/abs/1109.0763} {arXiv:1109.0763 [nucl-ex]} \BibitemShut
  {NoStop}%
\bibitem [{\citenamefont {Abe}\ \emph {et~al.}(2016)\citenamefont {Abe} \emph
  {et~al.}}]{Abe:2016nxk}%
  \BibitemOpen
  \bibfield  {author} {\bibinfo {author} {\bibfnamefont {K.}~\bibnamefont
  {Abe}} \emph {et~al.} (\bibinfo {collaboration} {Super-Kamiokande}),\ }\href
  {\doibase 10.1103/PhysRevD.94.052010} {\bibfield  {journal} {\bibinfo
  {journal} {Phys. Rev. D}\ }\textbf {\bibinfo {volume} {94}},\ \bibinfo
  {pages} {052010} (\bibinfo {year} {2016})},\ \Eprint
  {http://arxiv.org/abs/1606.07538} {arXiv:1606.07538 [hep-ex]} \BibitemShut
  {NoStop}%
\bibitem [{\citenamefont {Gando}\ \emph {et~al.}(2013)\citenamefont {Gando}
  \emph {et~al.}}]{Gando:2013nba}%
  \BibitemOpen
  \bibfield  {author} {\bibinfo {author} {\bibfnamefont {A.}~\bibnamefont
  {Gando}} \emph {et~al.} (\bibinfo {collaboration} {KamLAND}),\ }\href
  {\doibase 10.1103/PhysRevD.88.033001} {\bibfield  {journal} {\bibinfo
  {journal} {Phys. Rev. D}\ }\textbf {\bibinfo {volume} {88}},\ \bibinfo
  {pages} {033001} (\bibinfo {year} {2013})},\ \Eprint
  {http://arxiv.org/abs/1303.4667} {arXiv:1303.4667 [hep-ex]} \BibitemShut
  {NoStop}%
\bibitem [{\citenamefont {An}\ \emph {et~al.}(2016)\citenamefont {An} \emph
  {et~al.}}]{An:2015jdp}%
  \BibitemOpen
  \bibfield  {author} {\bibinfo {author} {\bibfnamefont {F.}~\bibnamefont {An}}
  \emph {et~al.} (\bibinfo {collaboration} {JUNO}),\ }\href {\doibase
  10.1088/0954-3899/43/3/030401} {\bibfield  {journal} {\bibinfo  {journal} {J.
  Phys. G}\ }\textbf {\bibinfo {volume} {43}},\ \bibinfo {pages} {030401}
  (\bibinfo {year} {2016})},\ \Eprint {http://arxiv.org/abs/1507.05613}
  {arXiv:1507.05613 [physics.ins-det]} \BibitemShut {NoStop}%
\bibitem [{\citenamefont {Capozzi}\ \emph {et~al.}(2015)\citenamefont
  {Capozzi}, \citenamefont {Lisi},\ and\ \citenamefont
  {Marrone}}]{Capozzi:2015bpa}%
  \BibitemOpen
  \bibfield  {author} {\bibinfo {author} {\bibfnamefont {F.}~\bibnamefont
  {Capozzi}}, \bibinfo {author} {\bibfnamefont {E.}~\bibnamefont {Lisi}}, \
  and\ \bibinfo {author} {\bibfnamefont {A.}~\bibnamefont {Marrone}},\ }\href
  {\doibase 10.1103/PhysRevD.92.093011} {\bibfield  {journal} {\bibinfo
  {journal} {Phys. Rev. D}\ }\textbf {\bibinfo {volume} {92}},\ \bibinfo
  {pages} {093011} (\bibinfo {year} {2015})},\ \Eprint
  {http://arxiv.org/abs/1508.01392} {arXiv:1508.01392 [hep-ph]} \BibitemShut
  {NoStop}%
\bibitem [{\citenamefont {Barenboim}\ \emph {et~al.}(2002)\citenamefont
  {Barenboim}, \citenamefont {Borissov}, \citenamefont {Lykken},\ and\
  \citenamefont {Smirnov}}]{Barenboim:2001ac}%
  \BibitemOpen
  \bibfield  {author} {\bibinfo {author} {\bibfnamefont {G.}~\bibnamefont
  {Barenboim}}, \bibinfo {author} {\bibfnamefont {L.}~\bibnamefont {Borissov}},
  \bibinfo {author} {\bibfnamefont {J.~D.}\ \bibnamefont {Lykken}}, \ and\
  \bibinfo {author} {\bibfnamefont {A.~Y.}\ \bibnamefont {Smirnov}},\ }\href
  {\doibase 10.1088/1126-6708/2002/10/001} {\bibfield  {journal} {\bibinfo
  {journal} {JHEP}\ }\textbf {\bibinfo {volume} {10}},\ \bibinfo {pages} {001}
  (\bibinfo {year} {2002})},\ \Eprint {http://arxiv.org/abs/hep-ph/0108199}
  {arXiv:hep-ph/0108199 [hep-ph]} \BibitemShut {NoStop}%
\bibitem [{\citenamefont {de~Gouvea}\ and\ \citenamefont
  {Pe{\~n}a-Garay}(2005)}]{deGouvea:2004va}%
  \BibitemOpen
  \bibfield  {author} {\bibinfo {author} {\bibfnamefont {A.}~\bibnamefont
  {de~Gouvea}}\ and\ \bibinfo {author} {\bibfnamefont {C.}~\bibnamefont
  {Pe{\~n}a-Garay}},\ }\href {\doibase 10.1103/PhysRevD.71.093002} {\bibfield
  {journal} {\bibinfo  {journal} {Phys. Rev. D}\ }\textbf {\bibinfo {volume}
  {71}},\ \bibinfo {pages} {093002} (\bibinfo {year} {2005})},\ \Eprint
  {http://arxiv.org/abs/hep-ph/0406301} {arXiv:hep-ph/0406301 [hep-ph]}
  \BibitemShut {NoStop}%
\bibitem [{\citenamefont {Friedland}\ \emph {et~al.}(2004)\citenamefont
  {Friedland}, \citenamefont {Lunardini},\ and\ \citenamefont
  {Pe{\~n}a-Garay}}]{Friedland:2004pp}%
  \BibitemOpen
  \bibfield  {author} {\bibinfo {author} {\bibfnamefont {A.}~\bibnamefont
  {Friedland}}, \bibinfo {author} {\bibfnamefont {C.}~\bibnamefont
  {Lunardini}}, \ and\ \bibinfo {author} {\bibfnamefont {C.}~\bibnamefont
  {Pe{\~n}a-Garay}},\ }\href {\doibase 10.1016/j.physletb.2004.05.047}
  {\bibfield  {journal} {\bibinfo  {journal} {Phys. Lett. B}\ }\textbf
  {\bibinfo {volume} {594}},\ \bibinfo {pages} {347} (\bibinfo {year}
  {2004})},\ \Eprint {http://arxiv.org/abs/hep-ph/0402266}
  {arXiv:hep-ph/0402266 [hep-ph]} \BibitemShut {NoStop}%
\bibitem [{\citenamefont {Cirelli}\ \emph {et~al.}(2005)\citenamefont
  {Cirelli}, \citenamefont {Gonz{\'a}lez-Garc{\'i}a},\ and\ \citenamefont
  {Pe{\~n}a-Garay}}]{Cirelli:2005sg}%
  \BibitemOpen
  \bibfield  {author} {\bibinfo {author} {\bibfnamefont {M.}~\bibnamefont
  {Cirelli}}, \bibinfo {author} {\bibfnamefont {M.~C.}\ \bibnamefont
  {Gonz{\'a}lez-Garc{\'i}a}}, \ and\ \bibinfo {author} {\bibfnamefont
  {C.}~\bibnamefont {Pe{\~n}a-Garay}},\ }\href {\doibase
  10.1016/j.nuclphysb.2005.04.034} {\bibfield  {journal} {\bibinfo  {journal}
  {Nucl. Phys. B}\ }\textbf {\bibinfo {volume} {719}},\ \bibinfo {pages} {219}
  (\bibinfo {year} {2005})},\ \Eprint {http://arxiv.org/abs/hep-ph/0503028}
  {arXiv:hep-ph/0503028 [hep-ph]} \BibitemShut {NoStop}%
\bibitem [{\citenamefont {Palazzo}(2011)}]{Palazzo:2011rj}%
  \BibitemOpen
  \bibfield  {author} {\bibinfo {author} {\bibfnamefont {A.}~\bibnamefont
  {Palazzo}},\ }\href {\doibase 10.1103/PhysRevD.83.113013} {\bibfield
  {journal} {\bibinfo  {journal} {Phys. Rev. D}\ }\textbf {\bibinfo {volume}
  {83}},\ \bibinfo {pages} {113013} (\bibinfo {year} {2011})},\ \Eprint
  {http://arxiv.org/abs/1105.1705} {arXiv:1105.1705 [hep-ph]} \BibitemShut
  {NoStop}%
\bibitem [{\citenamefont {Maltoni}\ and\ \citenamefont
  {Smirnov}(2016)}]{Maltoni:2015kca}%
  \BibitemOpen
  \bibfield  {author} {\bibinfo {author} {\bibfnamefont {M.}~\bibnamefont
  {Maltoni}}\ and\ \bibinfo {author} {\bibfnamefont {A.~{\relax Yu}.}\
  \bibnamefont {Smirnov}},\ }\href {\doibase 10.1140/epja/i2016-16087-0}
  {\bibfield  {journal} {\bibinfo  {journal} {Eur. Phys. J. A}\ }\textbf
  {\bibinfo {volume} {52}},\ \bibinfo {pages} {87} (\bibinfo {year} {2016})},\
  \Eprint {http://arxiv.org/abs/1507.05287} {arXiv:1507.05287 [hep-ph]}
  \BibitemShut {NoStop}%
\bibitem [{\citenamefont {Capozzi}\ \emph {et~al.}(2017)\citenamefont
  {Capozzi}, \citenamefont {Shoemaker},\ and\ \citenamefont
  {Vecchi}}]{Capozzi:2017auw}%
  \BibitemOpen
  \bibfield  {author} {\bibinfo {author} {\bibfnamefont {F.}~\bibnamefont
  {Capozzi}}, \bibinfo {author} {\bibfnamefont {I.~M.}\ \bibnamefont
  {Shoemaker}}, \ and\ \bibinfo {author} {\bibfnamefont {L.}~\bibnamefont
  {Vecchi}},\ }\href {\doibase 10.1088/1475-7516/2017/07/021} {\bibfield
  {journal} {\bibinfo  {journal} {JCAP}\ }\textbf {\bibinfo {volume} {1707}},\
  \bibinfo {pages} {021} (\bibinfo {year} {2017})},\ \Eprint
  {http://arxiv.org/abs/1702.08464} {arXiv:1702.08464 [hep-ph]} \BibitemShut
  {NoStop}%
\bibitem [{\citenamefont {Seo}\ and\ \citenamefont
  {Parke}(2019)}]{Seo:2018rrb}%
  \BibitemOpen
  \bibfield  {author} {\bibinfo {author} {\bibfnamefont {S.-H.}\ \bibnamefont
  {Seo}}\ and\ \bibinfo {author} {\bibfnamefont {S.~J.}\ \bibnamefont
  {Parke}},\ }\href {\doibase 10.1103/PhysRevD.99.033012} {\bibfield  {journal}
  {\bibinfo  {journal} {Phys. Rev. D}\ }\textbf {\bibinfo {volume} {99}},\
  \bibinfo {pages} {033012} (\bibinfo {year} {2019})},\ \Eprint
  {http://arxiv.org/abs/1808.09150} {arXiv:1808.09150 [hep-ex]} \BibitemShut
  {NoStop}%
\bibitem [{\citenamefont {Esteban}\ \emph {et~al.}(2017)\citenamefont
  {Esteban}, \citenamefont {Gonz{\'a}lez-Garc{\'i}a}, \citenamefont {Maltoni},
  \citenamefont {Mart{\'i}nez-Soler},\ and\ \citenamefont
  {Schwetz}}]{Esteban:2016qun}%
  \BibitemOpen
  \bibfield  {author} {\bibinfo {author} {\bibfnamefont {I.}~\bibnamefont
  {Esteban}}, \bibinfo {author} {\bibfnamefont {M.~C.}\ \bibnamefont
  {Gonz{\'a}lez-Garc{\'i}a}}, \bibinfo {author} {\bibfnamefont
  {M.}~\bibnamefont {Maltoni}}, \bibinfo {author} {\bibfnamefont
  {I.}~\bibnamefont {Mart{\'i}nez-Soler}}, \ and\ \bibinfo {author}
  {\bibfnamefont {T.}~\bibnamefont {Schwetz}},\ }\href {\doibase
  10.1007/JHEP01(2017)087} {\bibfield  {journal} {\bibinfo  {journal} {JHEP}\
  }\textbf {\bibinfo {volume} {01}},\ \bibinfo {pages} {087} (\bibinfo {year}
  {2017})},\ \Eprint {http://arxiv.org/abs/1611.01514} {arXiv:1611.01514
  [hep-ph]} \BibitemShut {NoStop}%
\bibitem [{\citenamefont {Capozzi}\ \emph {et~al.}(2018)\citenamefont
  {Capozzi}, \citenamefont {Lisi}, \citenamefont {Marrone},\ and\ \citenamefont
  {Palazzo}}]{Capozzi:2018ubv}%
  \BibitemOpen
  \bibfield  {author} {\bibinfo {author} {\bibfnamefont {F.}~\bibnamefont
  {Capozzi}}, \bibinfo {author} {\bibfnamefont {E.}~\bibnamefont {Lisi}},
  \bibinfo {author} {\bibfnamefont {A.}~\bibnamefont {Marrone}}, \ and\
  \bibinfo {author} {\bibfnamefont {A.}~\bibnamefont {Palazzo}},\ }\href
  {\doibase 10.1016/j.ppnp.2018.05.005} {\bibfield  {journal} {\bibinfo
  {journal} {Prog. Part. Nucl. Phys.}\ }\textbf {\bibinfo {volume} {102}},\
  \bibinfo {pages} {48} (\bibinfo {year} {2018})},\ \Eprint
  {http://arxiv.org/abs/1804.09678} {arXiv:1804.09678 [hep-ph]} \BibitemShut
  {NoStop}%
\bibitem [{\citenamefont {Liao}\ \emph {et~al.}(2017)\citenamefont {Liao},
  \citenamefont {Marfatia},\ and\ \citenamefont {Whisnant}}]{Liao:2017awz}%
  \BibitemOpen
  \bibfield  {author} {\bibinfo {author} {\bibfnamefont {J.}~\bibnamefont
  {Liao}}, \bibinfo {author} {\bibfnamefont {D.}~\bibnamefont {Marfatia}}, \
  and\ \bibinfo {author} {\bibfnamefont {K.}~\bibnamefont {Whisnant}},\ }\href
  {\doibase 10.1016/j.physletb.2017.05.054} {\bibfield  {journal} {\bibinfo
  {journal} {Phys. Lett.}\ }\textbf {\bibinfo {volume} {B771}},\ \bibinfo
  {pages} {247} (\bibinfo {year} {2017})},\ \Eprint
  {http://arxiv.org/abs/1704.04711} {arXiv:1704.04711 [hep-ph]} \BibitemShut
  {NoStop}%
\bibitem [{\citenamefont {Beacom}\ and\ \citenamefont
  {Bell}(2002)}]{Beacom:2002cb}%
  \BibitemOpen
  \bibfield  {author} {\bibinfo {author} {\bibfnamefont {J.~F.}\ \bibnamefont
  {Beacom}}\ and\ \bibinfo {author} {\bibfnamefont {N.~F.}\ \bibnamefont
  {Bell}},\ }\href {\doibase 10.1103/PhysRevD.65.113009} {\bibfield  {journal}
  {\bibinfo  {journal} {Phys. Rev. D}\ }\textbf {\bibinfo {volume} {65}},\
  \bibinfo {pages} {113009} (\bibinfo {year} {2002})},\ \Eprint
  {http://arxiv.org/abs/hep-ph/0204111} {arXiv:hep-ph/0204111 [hep-ph]}
  \BibitemShut {NoStop}%
\bibitem [{\citenamefont {Berryman}\ \emph {et~al.}(2015)\citenamefont
  {Berryman}, \citenamefont {de~Gouvea},\ and\ \citenamefont
  {Hernandez}}]{Berryman:2014qha}%
  \BibitemOpen
  \bibfield  {author} {\bibinfo {author} {\bibfnamefont {J.~M.}\ \bibnamefont
  {Berryman}}, \bibinfo {author} {\bibfnamefont {A.}~\bibnamefont {de~Gouvea}},
  \ and\ \bibinfo {author} {\bibfnamefont {D.}~\bibnamefont {Hernandez}},\
  }\href {\doibase 10.1103/PhysRevD.92.073003} {\bibfield  {journal} {\bibinfo
  {journal} {Phys. Rev. D}\ }\textbf {\bibinfo {volume} {92}},\ \bibinfo
  {pages} {073003} (\bibinfo {year} {2015})},\ \Eprint
  {http://arxiv.org/abs/1411.0308} {arXiv:1411.0308 [hep-ph]} \BibitemShut
  {NoStop}%
\bibitem [{\citenamefont {Bahcall}\ and\ \citenamefont
  {Ulmer}(1996)}]{Bahcall:1996vj}%
  \BibitemOpen
  \bibfield  {author} {\bibinfo {author} {\bibfnamefont {J.~N.}\ \bibnamefont
  {Bahcall}}\ and\ \bibinfo {author} {\bibfnamefont {A.}~\bibnamefont
  {Ulmer}},\ }\href {\doibase 10.1103/PhysRevD.53.4202} {\bibfield  {journal}
  {\bibinfo  {journal} {Phys. Rev. D}\ }\textbf {\bibinfo {volume} {53}},\
  \bibinfo {pages} {4202} (\bibinfo {year} {1996})},\ \Eprint
  {http://arxiv.org/abs/astro-ph/9602012} {arXiv:astro-ph/9602012 [astro-ph]}
  \BibitemShut {NoStop}%
\bibitem [{\citenamefont {{Grevesse}}\ and\ \citenamefont
  {{Sauval}}(1998)}]{Grevesse1998}%
  \BibitemOpen
  \bibfield  {author} {\bibinfo {author} {\bibfnamefont {N.}~\bibnamefont
  {{Grevesse}}}\ and\ \bibinfo {author} {\bibfnamefont {A.~J.}\ \bibnamefont
  {{Sauval}}},\ }\href {\doibase 10.1023/A:1005161325181} {\bibfield  {journal}
  {\bibinfo  {journal} {Space Sci. Rev.}\ }\textbf {\bibinfo {volume} {85}},\
  \bibinfo {pages} {161} (\bibinfo {year} {1998})}\BibitemShut {NoStop}%
\bibitem [{\citenamefont {Bahcall}\ \emph {et~al.}(2005)\citenamefont
  {Bahcall}, \citenamefont {Serenelli},\ and\ \citenamefont
  {Basu}}]{Bahcall:2004pz}%
  \BibitemOpen
  \bibfield  {author} {\bibinfo {author} {\bibfnamefont {J.~N.}\ \bibnamefont
  {Bahcall}}, \bibinfo {author} {\bibfnamefont {A.~M.}\ \bibnamefont
  {Serenelli}}, \ and\ \bibinfo {author} {\bibfnamefont {S.}~\bibnamefont
  {Basu}},\ }\href {\doibase 10.1086/428929} {\bibfield  {journal} {\bibinfo
  {journal} {Astrophys. J.}\ }\textbf {\bibinfo {volume} {621}},\ \bibinfo
  {pages} {L85} (\bibinfo {year} {2005})},\ \Eprint
  {http://arxiv.org/abs/astro-ph/0412440} {arXiv:astro-ph/0412440 [astro-ph]}
  \BibitemShut {NoStop}%
\bibitem [{\citenamefont {Bahcall}(2005)}]{Bahcall_website}%
  \BibitemOpen
  \bibfield  {author} {\bibinfo {author} {\bibfnamefont {J.~N.}\ \bibnamefont
  {Bahcall}},\ }\href {http://www.sns.ias.edu/~jnb/SNdata/sndata.html}
  {\enquote {\bibinfo {title} {{Software and data for solar neutrino
  research}},}\ } (\bibinfo {year} {2005}),\ \bibinfo {note} {[Online; accessed
  2017-9-30]}\BibitemShut {NoStop}%
\bibitem [{\citenamefont {{Asplund}}\ \emph {et~al.}(2009)\citenamefont
  {{Asplund}}, \citenamefont {{Grevesse}}, \citenamefont {{Sauval}},\ and\
  \citenamefont {{Scott}}}]{Asplund2009}%
  \BibitemOpen
  \bibfield  {author} {\bibinfo {author} {\bibfnamefont {M.}~\bibnamefont
  {{Asplund}}}, \bibinfo {author} {\bibfnamefont {N.}~\bibnamefont
  {{Grevesse}}}, \bibinfo {author} {\bibfnamefont {A.~J.}\ \bibnamefont
  {{Sauval}}}, \ and\ \bibinfo {author} {\bibfnamefont {P.}~\bibnamefont
  {{Scott}}},\ }\href {\doibase 10.1146/annurev.astro.46.060407.145222}
  {\bibfield  {journal} {\bibinfo  {journal} {Annu. Rev. Astron. Astrophys.}\
  }\textbf {\bibinfo {volume} {47}},\ \bibinfo {pages} {481} (\bibinfo {year}
  {2009})},\ \Eprint {http://arxiv.org/abs/0909.0948} {arXiv:0909.0948
  [astro-ph.SR]} \BibitemShut {NoStop}%
\bibitem [{\citenamefont {Vinyoles}\ \emph {et~al.}(2017)\citenamefont
  {Vinyoles}, \citenamefont {Serenelli}, \citenamefont {Villante},
  \citenamefont {Basu}, \citenamefont {Bergström}, \citenamefont
  {Gonz{\'a}lez-Garc{\'i}a}, \citenamefont {Maltoni}, \citenamefont
  {Peña-Garay},\ and\ \citenamefont {Song}}]{Vinyoles:2016djt}%
  \BibitemOpen
  \bibfield  {author} {\bibinfo {author} {\bibfnamefont {N.}~\bibnamefont
  {Vinyoles}}, \bibinfo {author} {\bibfnamefont {A.~M.}\ \bibnamefont
  {Serenelli}}, \bibinfo {author} {\bibfnamefont {F.~L.}\ \bibnamefont
  {Villante}}, \bibinfo {author} {\bibfnamefont {S.}~\bibnamefont {Basu}},
  \bibinfo {author} {\bibfnamefont {J.}~\bibnamefont {Bergström}}, \bibinfo
  {author} {\bibfnamefont {M.~C.}\ \bibnamefont {Gonz{\'a}lez-Garc{\'i}a}},
  \bibinfo {author} {\bibfnamefont {M.}~\bibnamefont {Maltoni}}, \bibinfo
  {author} {\bibfnamefont {C.}~\bibnamefont {Peña-Garay}}, \ and\ \bibinfo
  {author} {\bibfnamefont {N.}~\bibnamefont {Song}},\ }\href {\doibase
  10.3847/1538-4357/835/2/202} {\bibfield  {journal} {\bibinfo  {journal}
  {Astrophys. J.}\ }\textbf {\bibinfo {volume} {835}},\ \bibinfo {pages} {202}
  (\bibinfo {year} {2017})},\ \Eprint {http://arxiv.org/abs/1611.09867}
  {arXiv:1611.09867 [astro-ph.SR]} \BibitemShut {NoStop}%
\bibitem [{\citenamefont {Marcucci}\ \emph {et~al.}(2000)\citenamefont
  {Marcucci}, \citenamefont {Schiavilla}, \citenamefont {Viviani},
  \citenamefont {Kievsky}, \citenamefont {Rosati},\ and\ \citenamefont
  {Beacom}}]{Marcucci:2000xy}%
  \BibitemOpen
  \bibfield  {author} {\bibinfo {author} {\bibfnamefont {L.~E.}\ \bibnamefont
  {Marcucci}}, \bibinfo {author} {\bibfnamefont {R.}~\bibnamefont
  {Schiavilla}}, \bibinfo {author} {\bibfnamefont {M.}~\bibnamefont {Viviani}},
  \bibinfo {author} {\bibfnamefont {A.}~\bibnamefont {Kievsky}}, \bibinfo
  {author} {\bibfnamefont {S.}~\bibnamefont {Rosati}}, \ and\ \bibinfo {author}
  {\bibfnamefont {J.~F.}\ \bibnamefont {Beacom}},\ }\href {\doibase
  10.1103/PhysRevC.63.015801} {\bibfield  {journal} {\bibinfo  {journal} {Phys.
  Rev.}\ }\textbf {\bibinfo {volume} {C63}},\ \bibinfo {pages} {015801}
  (\bibinfo {year} {2000})},\ \Eprint {http://arxiv.org/abs/nucl-th/0006005}
  {arXiv:nucl-th/0006005 [nucl-th]} \BibitemShut {NoStop}%
\bibitem [{\citenamefont {Park}\ \emph {et~al.}(2003)\citenamefont {Park},
  \citenamefont {Marcucci}, \citenamefont {Schiavilla}, \citenamefont
  {Viviani}, \citenamefont {Kievsky}, \citenamefont {Rosati}, \citenamefont
  {Kubodera}, \citenamefont {Min},\ and\ \citenamefont {Rho}}]{Park:2002yp}%
  \BibitemOpen
  \bibfield  {author} {\bibinfo {author} {\bibfnamefont {T.~S.}\ \bibnamefont
  {Park}}, \bibinfo {author} {\bibfnamefont {L.~E.}\ \bibnamefont {Marcucci}},
  \bibinfo {author} {\bibfnamefont {R.}~\bibnamefont {Schiavilla}}, \bibinfo
  {author} {\bibfnamefont {M.}~\bibnamefont {Viviani}}, \bibinfo {author}
  {\bibfnamefont {A.}~\bibnamefont {Kievsky}}, \bibinfo {author} {\bibfnamefont
  {S.}~\bibnamefont {Rosati}}, \bibinfo {author} {\bibfnamefont
  {K.}~\bibnamefont {Kubodera}}, \bibinfo {author} {\bibfnamefont {D.~P.}\
  \bibnamefont {Min}}, \ and\ \bibinfo {author} {\bibfnamefont
  {M.}~\bibnamefont {Rho}},\ }\href {\doibase 10.1103/PhysRevC.67.055206}
  {\bibfield  {journal} {\bibinfo  {journal} {Phys. Rev.}\ }\textbf {\bibinfo
  {volume} {C67}},\ \bibinfo {pages} {055206} (\bibinfo {year} {2003})},\
  \Eprint {http://arxiv.org/abs/nucl-th/0208055} {arXiv:nucl-th/0208055
  [nucl-th]} \BibitemShut {NoStop}%
\bibitem [{\citenamefont {Acciarri}\ \emph {et~al.}(2015)\citenamefont
  {Acciarri} \emph {et~al.}}]{Acciarri:2015uup}%
  \BibitemOpen
  \bibfield  {author} {\bibinfo {author} {\bibfnamefont {R.}~\bibnamefont
  {Acciarri}} \emph {et~al.} (\bibinfo {collaboration} {DUNE}),\ }\href@noop {}
  {\  (\bibinfo {year} {2015})},\ \Eprint {http://arxiv.org/abs/1512.06148}
  {arXiv:1512.06148 [physics.ins-det]} \BibitemShut {NoStop}%
\bibitem [{\citenamefont {Strait}\ \emph {et~al.}(2016)\citenamefont {Strait}
  \emph {et~al.}}]{Strait:2016mof}%
  \BibitemOpen
  \bibfield  {author} {\bibinfo {author} {\bibfnamefont {J.}~\bibnamefont
  {Strait}} \emph {et~al.} (\bibinfo {collaboration} {DUNE}),\ }\href@noop {}
  {\  (\bibinfo {year} {2016})},\ \Eprint {http://arxiv.org/abs/1601.05823}
  {arXiv:1601.05823 [physics.ins-det]} \BibitemShut {NoStop}%
\bibitem [{\citenamefont {Acciarri}\ \emph {et~al.}(2016)\citenamefont
  {Acciarri} \emph {et~al.}}]{Acciarri:2016ooe}%
  \BibitemOpen
  \bibfield  {author} {\bibinfo {author} {\bibfnamefont {R.}~\bibnamefont
  {Acciarri}} \emph {et~al.} (\bibinfo {collaboration} {DUNE}),\ }\href@noop {}
  {\  (\bibinfo {year} {2016})},\ \Eprint {http://arxiv.org/abs/1601.02984}
  {arXiv:1601.02984 [physics.ins-det]} \BibitemShut {NoStop}%
\bibitem [{\citenamefont {Bahcall}\ \emph {et~al.}(1986)\citenamefont
  {Bahcall}, \citenamefont {Baldo-Ceolin}, \citenamefont {Cline},\ and\
  \citenamefont {Rubbia}}]{Bahcall:1986ry}%
  \BibitemOpen
  \bibfield  {author} {\bibinfo {author} {\bibfnamefont {J.~N.}\ \bibnamefont
  {Bahcall}}, \bibinfo {author} {\bibfnamefont {M.}~\bibnamefont
  {Baldo-Ceolin}}, \bibinfo {author} {\bibfnamefont {D.~B.}\ \bibnamefont
  {Cline}}, \ and\ \bibinfo {author} {\bibfnamefont {C.}~\bibnamefont
  {Rubbia}},\ }\href {\doibase 10.1016/0370-2693(86)91519-4} {\bibfield
  {journal} {\bibinfo  {journal} {Phys. Lett. B}\ }\textbf {\bibinfo {volume}
  {178}},\ \bibinfo {pages} {324} (\bibinfo {year} {1986})}\BibitemShut
  {NoStop}%
\bibitem [{\citenamefont {Arneodo}\ \emph {et~al.}(2000)\citenamefont {Arneodo}
  \emph {et~al.}}]{Arneodo:2000fa}%
  \BibitemOpen
  \bibfield  {author} {\bibinfo {author} {\bibfnamefont {F.}~\bibnamefont
  {Arneodo}} \emph {et~al.},\ }\href {\doibase 10.1016/S0168-9002(00)00520-9}
  {\bibfield  {journal} {\bibinfo  {journal} {Nucl. Instrum. Meth. A}\ }\textbf
  {\bibinfo {volume} {455}},\ \bibinfo {pages} {376} (\bibinfo {year}
  {2000})}\BibitemShut {NoStop}%
\bibitem [{\citenamefont {Franco}\ \emph {et~al.}(2016)\citenamefont {Franco}
  \emph {et~al.}}]{Franco:2015pha}%
  \BibitemOpen
  \bibfield  {author} {\bibinfo {author} {\bibfnamefont {D.}~\bibnamefont
  {Franco}} \emph {et~al.},\ }\href {\doibase 10.1088/1475-7516/2016/08/017}
  {\bibfield  {journal} {\bibinfo  {journal} {JCAP}\ }\textbf {\bibinfo
  {volume} {1608}},\ \bibinfo {pages} {017} (\bibinfo {year} {2016})},\ \Eprint
  {http://arxiv.org/abs/1510.04196} {arXiv:1510.04196 [physics.ins-det]}
  \BibitemShut {NoStop}%
\bibitem [{\citenamefont {Ioannisian}\ \emph {et~al.}(2017)\citenamefont
  {Ioannisian}, \citenamefont {Smirnov},\ and\ \citenamefont
  {Wyler}}]{Ioannisian:2017dkx}%
  \BibitemOpen
  \bibfield  {author} {\bibinfo {author} {\bibfnamefont {A.}~\bibnamefont
  {Ioannisian}}, \bibinfo {author} {\bibfnamefont {A.}~\bibnamefont {Smirnov}},
  \ and\ \bibinfo {author} {\bibfnamefont {D.}~\bibnamefont {Wyler}},\ }\href
  {\doibase 10.1103/PhysRevD.96.036005} {\bibfield  {journal} {\bibinfo
  {journal} {Phys. Rev. D}\ }\textbf {\bibinfo {volume} {96}},\ \bibinfo
  {pages} {036005} (\bibinfo {year} {2017})},\ \Eprint
  {http://arxiv.org/abs/1702.06097} {arXiv:1702.06097 [hep-ph]} \BibitemShut
  {NoStop}%
\bibitem [{\citenamefont {Zhu}\ \emph {et~al.}(2019)\citenamefont {Zhu},
  \citenamefont {Li},\ and\ \citenamefont {Beacom}}]{Zhu:2018rwc}%
  \BibitemOpen
  \bibfield  {author} {\bibinfo {author} {\bibfnamefont {G.}~\bibnamefont
  {Zhu}}, \bibinfo {author} {\bibfnamefont {S.~W.}\ \bibnamefont {Li}}, \ and\
  \bibinfo {author} {\bibfnamefont {J.~F.}\ \bibnamefont {Beacom}},\ }\href
  {\doibase 10.1103/PhysRevC.99.055810} {\bibfield  {journal} {\bibinfo
  {journal} {Phys. Rev. C}\ }\textbf {\bibinfo {volume} {99}},\ \bibinfo
  {pages} {055810} (\bibinfo {year} {2019})},\ \Eprint
  {http://arxiv.org/abs/1811.07912} {arXiv:1811.07912 [hep-ph]} \BibitemShut
  {NoStop}%
\bibitem [{\citenamefont {Fogli}\ \emph {et~al.}(2006)\citenamefont {Fogli},
  \citenamefont {Lisi}, \citenamefont {Marrone},\ and\ \citenamefont
  {Palazzo}}]{Fogli:2005cq}%
  \BibitemOpen
  \bibfield  {author} {\bibinfo {author} {\bibfnamefont {G.~L.}\ \bibnamefont
  {Fogli}}, \bibinfo {author} {\bibfnamefont {E.}~\bibnamefont {Lisi}},
  \bibinfo {author} {\bibfnamefont {A.}~\bibnamefont {Marrone}}, \ and\
  \bibinfo {author} {\bibfnamefont {A.}~\bibnamefont {Palazzo}},\ }\href
  {\doibase 10.1016/j.ppnp.2005.08.002} {\bibfield  {journal} {\bibinfo
  {journal} {Prog. Part. Nucl. Phys.}\ }\textbf {\bibinfo {volume} {57}},\
  \bibinfo {pages} {742} (\bibinfo {year} {2006})},\ \Eprint
  {http://arxiv.org/abs/hep-ph/0506083} {arXiv:hep-ph/0506083 [hep-ph]}
  \BibitemShut {NoStop}%
\bibitem [{\citenamefont {Aharmim}\ \emph {et~al.}(2005)\citenamefont {Aharmim}
  \emph {et~al.}}]{Aharmim:2005gt}%
  \BibitemOpen
  \bibfield  {author} {\bibinfo {author} {\bibfnamefont {B.}~\bibnamefont
  {Aharmim}} \emph {et~al.} (\bibinfo {collaboration} {SNO}),\ }\href {\doibase
  10.1103/PhysRevC.72.055502} {\bibfield  {journal} {\bibinfo  {journal} {Phys.
  Rev.}\ }\textbf {\bibinfo {volume} {C72}},\ \bibinfo {pages} {055502}
  (\bibinfo {year} {2005})},\ \Eprint {http://arxiv.org/abs/nucl-ex/0502021}
  {arXiv:nucl-ex/0502021 [nucl-ex]} \BibitemShut {NoStop}%
\bibitem [{\citenamefont {Hosaka}\ \emph {et~al.}(2006)\citenamefont {Hosaka}
  \emph {et~al.}}]{Hosaka:2005um}%
  \BibitemOpen
  \bibfield  {author} {\bibinfo {author} {\bibfnamefont {J.}~\bibnamefont
  {Hosaka}} \emph {et~al.} (\bibinfo {collaboration} {Super-Kamiokande}),\
  }\href {\doibase 10.1103/PhysRevD.73.112001} {\bibfield  {journal} {\bibinfo
  {journal} {Phys. Rev. D}\ }\textbf {\bibinfo {volume} {73}},\ \bibinfo
  {pages} {112001} (\bibinfo {year} {2006})},\ \Eprint
  {http://arxiv.org/abs/hep-ex/0508053} {arXiv:hep-ex/0508053 [hep-ex]}
  \BibitemShut {NoStop}%
\bibitem [{\citenamefont {Mastbaum}(2016)}]{Mastbaum2016}%
  \BibitemOpen
  \bibfield  {author} {\bibinfo {author} {\bibfnamefont {A.~T.}\ \bibnamefont
  {Mastbaum}},\ }\emph {\bibinfo {title} {{Constraining the Hep Solar Neutrino
  and Diffuse Supernova Neutrino Background Fluxes With the Sudbury Neutrino
  Observatory}}},\ \href {https://repository.upenn.edu/edissertations/1884/}
  {Ph.D. thesis},\ \bibinfo  {school} {U. of Pennsylvania} (\bibinfo {year}
  {2016})\BibitemShut {NoStop}%
\bibitem [{\citenamefont {Wolfenstein}(1978)}]{Wolfenstein:1977ue}%
  \BibitemOpen
  \bibfield  {author} {\bibinfo {author} {\bibfnamefont {L.}~\bibnamefont
  {Wolfenstein}},\ }\href {\doibase 10.1103/PhysRevD.17.2369} {\bibfield
  {journal} {\bibinfo  {journal} {Phys. Rev. D}\ }\textbf {\bibinfo {volume}
  {17}},\ \bibinfo {pages} {2369} (\bibinfo {year} {1978})}\BibitemShut
  {NoStop}%
\bibitem [{\citenamefont {Mikheev}\ and\ \citenamefont
  {Smirnov}(1985)}]{Mikheev:1986gs}%
  \BibitemOpen
  \bibfield  {author} {\bibinfo {author} {\bibfnamefont {S.~P.}\ \bibnamefont
  {Mikheev}}\ and\ \bibinfo {author} {\bibfnamefont {A.~{\relax Yu}.}\
  \bibnamefont {Smirnov}},\ }\href@noop {} {\bibfield  {journal} {\bibinfo
  {journal} {Sov. J. Nucl. Phys.}\ }\textbf {\bibinfo {volume} {42}},\ \bibinfo
  {pages} {913} (\bibinfo {year} {1985})},\ \bibinfo {note} {[Yad. Fiz.
  42,1441(1985)]}\BibitemShut {NoStop}%
\bibitem [{\citenamefont {Mikheev}\ and\ \citenamefont
  {Smirnov}(1986)}]{Mikheev:1986wj}%
  \BibitemOpen
  \bibfield  {author} {\bibinfo {author} {\bibfnamefont {S.~P.}\ \bibnamefont
  {Mikheev}}\ and\ \bibinfo {author} {\bibfnamefont {A.~{\relax Yu}.}\
  \bibnamefont {Smirnov}},\ }\href {\doibase 10.1007/BF02508049} {\bibfield
  {journal} {\bibinfo  {journal} {Nuovo Cim. C}\ }\textbf {\bibinfo {volume}
  {9}},\ \bibinfo {pages} {17} (\bibinfo {year} {1986})}\BibitemShut {NoStop}%
\bibitem [{\citenamefont {Haxton}(1986)}]{Haxton:1986dm}%
  \BibitemOpen
  \bibfield  {author} {\bibinfo {author} {\bibfnamefont {W.~C.}\ \bibnamefont
  {Haxton}},\ }\href {\doibase 10.1103/PhysRevLett.57.1271} {\bibfield
  {journal} {\bibinfo  {journal} {Phys. Rev. Lett.}\ }\textbf {\bibinfo
  {volume} {57}},\ \bibinfo {pages} {1271} (\bibinfo {year}
  {1986})}\BibitemShut {NoStop}%
\bibitem [{\citenamefont {Parke}(1986)}]{Parke:1986jy}%
  \BibitemOpen
  \bibfield  {author} {\bibinfo {author} {\bibfnamefont {S.~J.}\ \bibnamefont
  {Parke}},\ }\href {\doibase 10.1103/PhysRevLett.57.1275} {\bibfield
  {journal} {\bibinfo  {journal} {Phys. Rev. Lett.}\ }\textbf {\bibinfo
  {volume} {57}},\ \bibinfo {pages} {1275} (\bibinfo {year}
  {1986})}\BibitemShut {NoStop}%
\bibitem [{\citenamefont {Giunti}\ and\ \citenamefont
  {Kim}(2007)}]{Giunti:2007ry}%
  \BibitemOpen
  \bibfield  {author} {\bibinfo {author} {\bibfnamefont {C.}~\bibnamefont
  {Giunti}}\ and\ \bibinfo {author} {\bibfnamefont {C.~W.}\ \bibnamefont
  {Kim}},\ }\href@noop {} {\emph {\bibinfo {title} {{Fundamentals of Neutrino
  Physics and Astrophysics}}}}\ (\bibinfo  {publisher} {Oxford University
  Press},\ \bibinfo {year} {2007})\BibitemShut {NoStop}%
\bibitem [{\citenamefont {Lisi}\ and\ \citenamefont
  {Montanino}(1997)}]{Lisi:1997yc}%
  \BibitemOpen
  \bibfield  {author} {\bibinfo {author} {\bibfnamefont {E.}~\bibnamefont
  {Lisi}}\ and\ \bibinfo {author} {\bibfnamefont {D.}~\bibnamefont
  {Montanino}},\ }\href {\doibase 10.1103/PhysRevD.56.1792} {\bibfield
  {journal} {\bibinfo  {journal} {Phys. Rev. D}\ }\textbf {\bibinfo {volume}
  {56}},\ \bibinfo {pages} {1792} (\bibinfo {year} {1997})},\ \Eprint
  {http://arxiv.org/abs/hep-ph/9702343} {arXiv:hep-ph/9702343 [hep-ph]}
  \BibitemShut {NoStop}%
\bibitem [{\citenamefont {Maris}\ and\ \citenamefont
  {Petcov}(1997)}]{Maris:1997nk}%
  \BibitemOpen
  \bibfield  {author} {\bibinfo {author} {\bibfnamefont {M.}~\bibnamefont
  {Maris}}\ and\ \bibinfo {author} {\bibfnamefont {S.~T.}\ \bibnamefont
  {Petcov}},\ }\href {\doibase 10.1103/PhysRevD.56.7444} {\bibfield  {journal}
  {\bibinfo  {journal} {Phys. Rev. D}\ }\textbf {\bibinfo {volume} {56}},\
  \bibinfo {pages} {7444} (\bibinfo {year} {1997})},\ \Eprint
  {http://arxiv.org/abs/hep-ph/9705392} {arXiv:hep-ph/9705392 [hep-ph]}
  \BibitemShut {NoStop}%
\bibitem [{\citenamefont {Forsythe}\ \emph {et~al.}(1995)\citenamefont
  {Forsythe}, \citenamefont {Rykiel}, \citenamefont {Stahl}, \citenamefont
  {H.},\ and\ \citenamefont {M.}}]{daytime_models}%
  \BibitemOpen
  \bibfield  {author} {\bibinfo {author} {\bibfnamefont {W.~C.}\ \bibnamefont
  {Forsythe}}, \bibinfo {author} {\bibfnamefont {E.~J.~J.}\ \bibnamefont
  {Rykiel}}, \bibinfo {author} {\bibfnamefont {R.~S.}\ \bibnamefont {Stahl}},
  \bibinfo {author} {\bibfnamefont {W.}~\bibnamefont {H.}}, \ and\ \bibinfo
  {author} {\bibfnamefont {S.~R.}\ \bibnamefont {M.}},\ }\href {\doibase
  10.1016/0304-3800(94)00034-F} {\bibfield  {journal} {\bibinfo  {journal}
  {Ecological Modelling}\ }\textbf {\bibinfo {volume} {80}},\ \bibinfo {pages}
  {87} (\bibinfo {year} {1995})}\BibitemShut {NoStop}%
\bibitem [{\citenamefont {Marchionni}(2013)}]{Marchionni:2013tfa}%
  \BibitemOpen
  \bibfield  {author} {\bibinfo {author} {\bibfnamefont {A.}~\bibnamefont
  {Marchionni}},\ }\href {\doibase 10.1146/annurev.nucl.012809.104445}
  {\bibfield  {journal} {\bibinfo  {journal} {Ann. Rev. Nucl. Part. Sci.}\
  }\textbf {\bibinfo {volume} {63}},\ \bibinfo {pages} {269} (\bibinfo {year}
  {2013})},\ \Eprint {http://arxiv.org/abs/1307.6918} {arXiv:1307.6918
  [physics.ins-det]} \BibitemShut {NoStop}%
\bibitem [{\citenamefont {Aharmim}\ \emph
  {et~al.}(2013{\natexlab{b}})\citenamefont {Aharmim} \emph
  {et~al.}}]{Aharmim:2011yq}%
  \BibitemOpen
  \bibfield  {author} {\bibinfo {author} {\bibfnamefont {B.}~\bibnamefont
  {Aharmim}} \emph {et~al.} (\bibinfo {collaboration} {SNO}),\ }\href {\doibase
  10.1103/PhysRevC.87.015502} {\bibfield  {journal} {\bibinfo  {journal} {Phys.
  Rev. C}\ }\textbf {\bibinfo {volume} {87}},\ \bibinfo {pages} {015502}
  (\bibinfo {year} {2013}{\natexlab{b}})},\ \Eprint
  {http://arxiv.org/abs/1107.2901} {arXiv:1107.2901 [nucl-ex]} \BibitemShut
  {NoStop}%
\bibitem [{\citenamefont {Patrignani}\ \emph {et~al.}(2016)\citenamefont
  {Patrignani} \emph {et~al.}}]{Patrignani:2016xqp}%
  \BibitemOpen
  \bibfield  {author} {\bibinfo {author} {\bibfnamefont {C.}~\bibnamefont
  {Patrignani}} \emph {et~al.} (\bibinfo {collaboration} {Particle Data
  Group}),\ }\href {\doibase 10.1088/1674-1137/40/10/100001} {\bibfield
  {journal} {\bibinfo  {journal} {Chin. Phys. C}\ }\textbf {\bibinfo {volume}
  {40}},\ \bibinfo {pages} {100001} (\bibinfo {year} {2016})}\BibitemShut
  {NoStop}%
\bibitem [{\citenamefont {Amoruso}\ \emph {et~al.}(2004)\citenamefont {Amoruso}
  \emph {et~al.}}]{Amoruso:2003sw}%
  \BibitemOpen
  \bibfield  {author} {\bibinfo {author} {\bibfnamefont {S.}~\bibnamefont
  {Amoruso}} \emph {et~al.} (\bibinfo {collaboration} {ICARUS}),\ }\href
  {\doibase 10.1140/epjc/s2004-01597-7} {\bibfield  {journal} {\bibinfo
  {journal} {Eur. Phys. J. C}\ }\textbf {\bibinfo {volume} {33}},\ \bibinfo
  {pages} {233} (\bibinfo {year} {2004})},\ \Eprint
  {http://arxiv.org/abs/hep-ex/0311040} {arXiv:hep-ex/0311040 [hep-ex]}
  \BibitemShut {NoStop}%
\bibitem [{\citenamefont {Acciarri}\ \emph {et~al.}(2017)\citenamefont
  {Acciarri} \emph {et~al.}}]{Acciarri:2017sjy}%
  \BibitemOpen
  \bibfield  {author} {\bibinfo {author} {\bibfnamefont {R.}~\bibnamefont
  {Acciarri}} \emph {et~al.} (\bibinfo {collaboration} {MicroBooNE}),\ }\href
  {\doibase 10.1088/1748-0221/12/09/P09014} {\bibfield  {journal} {\bibinfo
  {journal} {JINST}\ }\textbf {\bibinfo {volume} {12}},\ \bibinfo {pages}
  {P09014} (\bibinfo {year} {2017})},\ \Eprint
  {http://arxiv.org/abs/1704.02927} {arXiv:1704.02927 [physics.ins-det]}
  \BibitemShut {NoStop}%
\bibitem [{\citenamefont {Winter}\ \emph {et~al.}(2006)\citenamefont {Winter},
  \citenamefont {Freedman}, \citenamefont {Rehm},\ and\ \citenamefont
  {Schiffer}}]{Winter:2004kf}%
  \BibitemOpen
  \bibfield  {author} {\bibinfo {author} {\bibfnamefont {W.~T.}\ \bibnamefont
  {Winter}}, \bibinfo {author} {\bibfnamefont {S.~J.}\ \bibnamefont
  {Freedman}}, \bibinfo {author} {\bibfnamefont {K.~E.}\ \bibnamefont {Rehm}},
  \ and\ \bibinfo {author} {\bibfnamefont {J.~P.}\ \bibnamefont {Schiffer}},\
  }\href {\doibase 10.1103/PhysRevC.73.025503} {\bibfield  {journal} {\bibinfo
  {journal} {Phys. Rev. C}\ }\textbf {\bibinfo {volume} {73}},\ \bibinfo
  {pages} {025503} (\bibinfo {year} {2006})},\ \Eprint
  {http://arxiv.org/abs/nucl-ex/0406019} {arXiv:nucl-ex/0406019 [nucl-ex]}
  \BibitemShut {NoStop}%
\bibitem [{\citenamefont {Bahcall}(1997)}]{Bahcall:1997eg}%
  \BibitemOpen
  \bibfield  {author} {\bibinfo {author} {\bibfnamefont {J.~N.}\ \bibnamefont
  {Bahcall}},\ }\href {\doibase 10.1103/PhysRevC.56.3391} {\bibfield  {journal}
  {\bibinfo  {journal} {Phys. Rev. C}\ }\textbf {\bibinfo {volume} {56}},\
  \bibinfo {pages} {3391} (\bibinfo {year} {1997})},\ \Eprint
  {http://arxiv.org/abs/hep-ph/9710491} {arXiv:hep-ph/9710491 [hep-ph]}
  \BibitemShut {NoStop}%
\bibitem [{\citenamefont {Konopinski}(1950)}]{Konopinski1950}%
  \BibitemOpen
  \bibfield  {author} {\bibinfo {author} {\bibfnamefont {E.~J.}\ \bibnamefont
  {Konopinski}},\ }\href
  {https://archive.org/details/TheTheoryOfBetaRadioactivity} {\emph {\bibinfo
  {title} {The Theory of Beta Radioactivity}}}\ (\bibinfo  {publisher} {Oxford
  University Press},\ \bibinfo {year} {1950})\BibitemShut {NoStop}%
\bibitem [{\citenamefont {Ormand}\ \emph {et~al.}(1995)\citenamefont {Ormand},
  \citenamefont {Pizzochero}, \citenamefont {Bortignon},\ and\ \citenamefont
  {Broglia}}]{Ormand:1994js}%
  \BibitemOpen
  \bibfield  {author} {\bibinfo {author} {\bibfnamefont {W.~E.}\ \bibnamefont
  {Ormand}}, \bibinfo {author} {\bibfnamefont {P.~M.}\ \bibnamefont
  {Pizzochero}}, \bibinfo {author} {\bibfnamefont {P.~F.}\ \bibnamefont
  {Bortignon}}, \ and\ \bibinfo {author} {\bibfnamefont {R.~A.}\ \bibnamefont
  {Broglia}},\ }\href {\doibase 10.1016/0370-2693(94)01605-C} {\bibfield
  {journal} {\bibinfo  {journal} {Phys. Lett. B}\ }\textbf {\bibinfo {volume}
  {345}},\ \bibinfo {pages} {343} (\bibinfo {year} {1995})},\ \Eprint
  {http://arxiv.org/abs/nucl-th/9405007} {arXiv:nucl-th/9405007 [nucl-th]}
  \BibitemShut {NoStop}%
\bibitem [{\citenamefont {Bhattacharya}\ \emph {et~al.}(1998)\citenamefont
  {Bhattacharya} \emph {et~al.}}]{Bhattacharya:1998hc}%
  \BibitemOpen
  \bibfield  {author} {\bibinfo {author} {\bibfnamefont {M.}~\bibnamefont
  {Bhattacharya}} \emph {et~al.},\ }\href {\doibase 10.1103/PhysRevC.58.3677}
  {\bibfield  {journal} {\bibinfo  {journal} {Phys. Rev. C}\ }\textbf {\bibinfo
  {volume} {58}},\ \bibinfo {pages} {3677} (\bibinfo {year}
  {1998})}\BibitemShut {NoStop}%
\bibitem [{\citenamefont {Beacom}\ and\ \citenamefont
  {Vogel}(1999)}]{Beacom:1998fj}%
  \BibitemOpen
  \bibfield  {author} {\bibinfo {author} {\bibfnamefont {J.~F.}\ \bibnamefont
  {Beacom}}\ and\ \bibinfo {author} {\bibfnamefont {P.}~\bibnamefont {Vogel}},\
  }\href {\doibase 10.1103/PhysRevD.60.033007} {\bibfield  {journal} {\bibinfo
  {journal} {Phys. Rev. D}\ }\textbf {\bibinfo {volume} {60}},\ \bibinfo
  {pages} {033007} (\bibinfo {year} {1999})},\ \Eprint
  {http://arxiv.org/abs/astro-ph/9811350} {arXiv:astro-ph/9811350 [astro-ph]}
  \BibitemShut {NoStop}%
\bibitem [{\citenamefont {Vogel}\ and\ \citenamefont
  {Beacom}(1999)}]{Vogel:1999zy}%
  \BibitemOpen
  \bibfield  {author} {\bibinfo {author} {\bibfnamefont {P.}~\bibnamefont
  {Vogel}}\ and\ \bibinfo {author} {\bibfnamefont {J.~F.}\ \bibnamefont
  {Beacom}},\ }\href {\doibase 10.1103/PhysRevD.60.053003} {\bibfield
  {journal} {\bibinfo  {journal} {Phys. Rev. D}\ }\textbf {\bibinfo {volume}
  {60}},\ \bibinfo {pages} {053003} (\bibinfo {year} {1999})},\ \Eprint
  {http://arxiv.org/abs/hep-ph/9903554} {arXiv:hep-ph/9903554 [hep-ph]}
  \BibitemShut {NoStop}%
\bibitem [{\citenamefont {Beacom}\ and\ \citenamefont
  {Parke}(2001)}]{Beacom:2001hr}%
  \BibitemOpen
  \bibfield  {author} {\bibinfo {author} {\bibfnamefont {J.~F.}\ \bibnamefont
  {Beacom}}\ and\ \bibinfo {author} {\bibfnamefont {S.~J.}\ \bibnamefont
  {Parke}},\ }\href {\doibase 10.1103/PhysRevD.64.091302} {\bibfield  {journal}
  {\bibinfo  {journal} {Phys. Rev. D}\ }\textbf {\bibinfo {volume} {64}},\
  \bibinfo {pages} {091302} (\bibinfo {year} {2001})},\ \Eprint
  {http://arxiv.org/abs/hep-ph/0106128} {arXiv:hep-ph/0106128 [hep-ph]}
  \BibitemShut {NoStop}%
\bibitem [{\citenamefont {Kurylov}\ \emph {et~al.}(2003)\citenamefont
  {Kurylov}, \citenamefont {Ramsey-Musolf},\ and\ \citenamefont
  {Vogel}}]{Kurylov:2002vj}%
  \BibitemOpen
  \bibfield  {author} {\bibinfo {author} {\bibfnamefont {A.}~\bibnamefont
  {Kurylov}}, \bibinfo {author} {\bibfnamefont {M.~J.}\ \bibnamefont
  {Ramsey-Musolf}}, \ and\ \bibinfo {author} {\bibfnamefont {P.}~\bibnamefont
  {Vogel}},\ }\href {\doibase 10.1103/PhysRevC.67.035502} {\bibfield  {journal}
  {\bibinfo  {journal} {Phys. Rev. C}\ }\textbf {\bibinfo {volume} {67}},\
  \bibinfo {pages} {035502} (\bibinfo {year} {2003})},\ \Eprint
  {http://arxiv.org/abs/hep-ph/0211306} {arXiv:hep-ph/0211306 [hep-ph]}
  \BibitemShut {NoStop}%
\bibitem [{\citenamefont {Gil~Botella}\ and\ \citenamefont
  {Rubbia}(2003)}]{GilBotella:2003sz}%
  \BibitemOpen
  \bibfield  {author} {\bibinfo {author} {\bibfnamefont {I.}~\bibnamefont
  {Gil~Botella}}\ and\ \bibinfo {author} {\bibfnamefont {A.}~\bibnamefont
  {Rubbia}},\ }\href {\doibase 10.1088/1475-7516/2003/10/009} {\bibfield
  {journal} {\bibinfo  {journal} {JCAP}\ }\textbf {\bibinfo {volume} {0310}},\
  \bibinfo {pages} {009} (\bibinfo {year} {2003})},\ \Eprint
  {http://arxiv.org/abs/hep-ph/0307244} {arXiv:hep-ph/0307244 [hep-ph]}
  \BibitemShut {NoStop}%
\bibitem [{\citenamefont {Kolbe}\ \emph {et~al.}(2003)\citenamefont {Kolbe},
  \citenamefont {Langanke}, \citenamefont {Martinez-Pinedo},\ and\
  \citenamefont {Vogel}}]{Kolbe:2003ys}%
  \BibitemOpen
  \bibfield  {author} {\bibinfo {author} {\bibfnamefont {E.}~\bibnamefont
  {Kolbe}}, \bibinfo {author} {\bibfnamefont {K.}~\bibnamefont {Langanke}},
  \bibinfo {author} {\bibfnamefont {G.}~\bibnamefont {Martinez-Pinedo}}, \ and\
  \bibinfo {author} {\bibfnamefont {P.}~\bibnamefont {Vogel}},\ }\href
  {\doibase 10.1088/0954-3899/29/11/010} {\bibfield  {journal} {\bibinfo
  {journal} {J. Phys. G}\ }\textbf {\bibinfo {volume} {29}},\ \bibinfo {pages}
  {2569} (\bibinfo {year} {2003})},\ \Eprint
  {http://arxiv.org/abs/nucl-th/0311022} {arXiv:nucl-th/0311022 [nucl-th]}
  \BibitemShut {NoStop}%
\bibitem [{\citenamefont {Bhattacharya}\ \emph {et~al.}(2009)\citenamefont
  {Bhattacharya}, \citenamefont {Goodman},\ and\ \citenamefont
  {Garcia}}]{Bhattacharya:2009zz}%
  \BibitemOpen
  \bibfield  {author} {\bibinfo {author} {\bibfnamefont {M.}~\bibnamefont
  {Bhattacharya}}, \bibinfo {author} {\bibfnamefont {C.~D.}\ \bibnamefont
  {Goodman}}, \ and\ \bibinfo {author} {\bibfnamefont {A.}~\bibnamefont
  {Garcia}},\ }\href {\doibase 10.1103/PhysRevC.80.055501} {\bibfield
  {journal} {\bibinfo  {journal} {Phys. Rev. C}\ }\textbf {\bibinfo {volume}
  {80}},\ \bibinfo {pages} {055501} (\bibinfo {year} {2009})}\BibitemShut
  {NoStop}%
\bibitem [{\citenamefont {Cheoun}\ \emph {et~al.}(2011)\citenamefont {Cheoun},
  \citenamefont {Ha},\ and\ \citenamefont {Kajino}}]{Cheoun:2011zza}%
  \BibitemOpen
  \bibfield  {author} {\bibinfo {author} {\bibfnamefont {M.-K.}\ \bibnamefont
  {Cheoun}}, \bibinfo {author} {\bibfnamefont {E.}~\bibnamefont {Ha}}, \ and\
  \bibinfo {author} {\bibfnamefont {T.}~\bibnamefont {Kajino}},\ }\href
  {\doibase 10.1103/PhysRevC.83.028801} {\bibfield  {journal} {\bibinfo
  {journal} {Phys. Rev. C}\ }\textbf {\bibinfo {volume} {83}},\ \bibinfo
  {pages} {028801} (\bibinfo {year} {2011})}\BibitemShut {NoStop}%
\bibitem [{\citenamefont {Suzuki}\ \emph {et~al.}(2014)\citenamefont {Suzuki},
  \citenamefont {Honma}, \citenamefont {Balantekin}, \citenamefont {Kajino},\
  and\ \citenamefont {Chiba}}]{Suzuki2014}%
  \BibitemOpen
  \bibfield  {author} {\bibinfo {author} {\bibfnamefont {T.}~\bibnamefont
  {Suzuki}}, \bibinfo {author} {\bibfnamefont {M.}~\bibnamefont {Honma}},
  \bibinfo {author} {\bibfnamefont {A.~B.}\ \bibnamefont {Balantekin}},
  \bibinfo {author} {\bibfnamefont {T.}~\bibnamefont {Kajino}}, \ and\ \bibinfo
  {author} {\bibfnamefont {S.}~\bibnamefont {Chiba}},\ }\href {\doibase
  10.1051/epjconf/20146607025} {\bibfield  {journal} {\bibinfo  {journal} {EPJ
  Web of Conferences}\ }\textbf {\bibinfo {volume} {66}},\ \bibinfo {pages}
  {07025} (\bibinfo {year} {2014})}\BibitemShut {NoStop}%
\bibitem [{\citenamefont {Hardy}\ and\ \citenamefont
  {Towner}(2015)}]{Hardy:2014qxa}%
  \BibitemOpen
  \bibfield  {author} {\bibinfo {author} {\bibfnamefont {J.~C.}\ \bibnamefont
  {Hardy}}\ and\ \bibinfo {author} {\bibfnamefont {I.~S.}\ \bibnamefont
  {Towner}},\ }\href {\doibase 10.1103/PhysRevC.91.025501} {\bibfield
  {journal} {\bibinfo  {journal} {Phys. Rev. C}\ }\textbf {\bibinfo {volume}
  {91}},\ \bibinfo {pages} {025501} (\bibinfo {year} {2015})},\ \Eprint
  {http://arxiv.org/abs/1411.5987} {arXiv:1411.5987 [nucl-ex]} \BibitemShut
  {NoStop}%
\bibitem [{\citenamefont {Karako{\c{c}}}\ \emph {et~al.}(2014)\citenamefont
  {Karako{\c{c}}} \emph {et~al.}}]{Karakoc:2014awa}%
  \BibitemOpen
  \bibfield  {author} {\bibinfo {author} {\bibfnamefont {M.}~\bibnamefont
  {Karako{\c{c}}}} \emph {et~al.},\ }\href {\doibase
  10.1103/PhysRevC.89.064313} {\bibfield  {journal} {\bibinfo  {journal} {Phys.
  Rev. C}\ }\textbf {\bibinfo {volume} {89}},\ \bibinfo {pages} {064313}
  (\bibinfo {year} {2014})}\BibitemShut {NoStop}%
\bibitem [{\citenamefont {Akimov}\ \emph {et~al.}(2017)\citenamefont {Akimov}
  \emph {et~al.}}]{Akimov:2017ade}%
  \BibitemOpen
  \bibfield  {author} {\bibinfo {author} {\bibfnamefont {D.}~\bibnamefont
  {Akimov}} \emph {et~al.} (\bibinfo {collaboration} {COHERENT}),\ }\href
  {\doibase 10.1126/science.aao0990} {\bibfield  {journal} {\bibinfo  {journal}
  {Science}\ }\textbf {\bibinfo {volume} {357}},\ \bibinfo {pages} {1123}
  (\bibinfo {year} {2017})},\ \Eprint {http://arxiv.org/abs/1708.01294}
  {arXiv:1708.01294 [nucl-ex]} \BibitemShut {NoStop}%
\bibitem [{\citenamefont {{National Nuclear Data Center}}(2016)}]{NNDC}%
  \BibitemOpen
  \bibfield  {author} {\bibinfo {author} {\bibnamefont {{National Nuclear Data
  Center}}},\ }\href {https://www.nndc.bnl.gov/nudat2/} {\enquote {\bibinfo
  {title} {{NuDat 2.7}},}\ } (\bibinfo {year} {2016}),\ \bibinfo {note}
  {[Online; accessed 2017-9-30]}\BibitemShut {NoStop}%
\bibitem [{\citenamefont {Bahcall}\ \emph {et~al.}(1995)\citenamefont
  {Bahcall}, \citenamefont {Kamionkowski},\ and\ \citenamefont
  {Sirlin}}]{Bahcall:1995mm}%
  \BibitemOpen
  \bibfield  {author} {\bibinfo {author} {\bibfnamefont {J.~N.}\ \bibnamefont
  {Bahcall}}, \bibinfo {author} {\bibfnamefont {M.}~\bibnamefont
  {Kamionkowski}}, \ and\ \bibinfo {author} {\bibfnamefont {A.}~\bibnamefont
  {Sirlin}},\ }\href {\doibase 10.1103/PhysRevD.51.6146} {\bibfield  {journal}
  {\bibinfo  {journal} {Phys. Rev. D}\ }\textbf {\bibinfo {volume} {51}},\
  \bibinfo {pages} {6146} (\bibinfo {year} {1995})},\ \Eprint
  {http://arxiv.org/abs/astro-ph/9502003} {arXiv:astro-ph/9502003 [astro-ph]}
  \BibitemShut {NoStop}%
\bibitem [{\citenamefont {Raghavan}\ \emph {et~al.}(1986)\citenamefont
  {Raghavan}, \citenamefont {Pakvasa},\ and\ \citenamefont
  {Brown}}]{Raghavan:1986fg}%
  \BibitemOpen
  \bibfield  {author} {\bibinfo {author} {\bibfnamefont {R.~S.}\ \bibnamefont
  {Raghavan}}, \bibinfo {author} {\bibfnamefont {S.}~\bibnamefont {Pakvasa}}, \
  and\ \bibinfo {author} {\bibfnamefont {B.~A.}\ \bibnamefont {Brown}},\ }\href
  {\doibase 10.1103/PhysRevLett.57.1801} {\bibfield  {journal} {\bibinfo
  {journal} {Phys. Rev. Lett.}\ }\textbf {\bibinfo {volume} {57}},\ \bibinfo
  {pages} {1801} (\bibinfo {year} {1986})}\BibitemShut {NoStop}%
\bibitem [{\citenamefont {Takeuchi}\ \emph {et~al.}(1999)\citenamefont
  {Takeuchi} \emph {et~al.}}]{Takeuchi:1999zq}%
  \BibitemOpen
  \bibfield  {author} {\bibinfo {author} {\bibfnamefont {Y.}~\bibnamefont
  {Takeuchi}} \emph {et~al.} (\bibinfo {collaboration} {SuperKamiokade}),\
  }\href {\doibase 10.1016/S0370-2693(99)00311-1} {\bibfield  {journal}
  {\bibinfo  {journal} {Phys. Lett. B}\ }\textbf {\bibinfo {volume} {452}},\
  \bibinfo {pages} {418} (\bibinfo {year} {1999})},\ \Eprint
  {http://arxiv.org/abs/hep-ex/9903006} {arXiv:hep-ex/9903006 [hep-ex]}
  \BibitemShut {NoStop}%
\bibitem [{\citenamefont {Blevis}\ \emph {et~al.}(2004)\citenamefont {Blevis}
  \emph {et~al.}}]{Blevis:2003ih}%
  \BibitemOpen
  \bibfield  {author} {\bibinfo {author} {\bibfnamefont {I.}~\bibnamefont
  {Blevis}} \emph {et~al.},\ }\href {\doibase 10.1016/j.nima.2003.10.103}
  {\bibfield  {journal} {\bibinfo  {journal} {Nucl. Instrum. Meth. A}\ }\textbf
  {\bibinfo {volume} {517}},\ \bibinfo {pages} {139} (\bibinfo {year}
  {2004})},\ \Eprint {http://arxiv.org/abs/nucl-ex/0305022}
  {arXiv:nucl-ex/0305022 [nucl-ex]} \BibitemShut {NoStop}%
\bibitem [{\citenamefont {Hardell}\ and\ \citenamefont
  {Beer}(1970)}]{Hardell1970}%
  \BibitemOpen
  \bibfield  {author} {\bibinfo {author} {\bibfnamefont {R.}~\bibnamefont
  {Hardell}}\ and\ \bibinfo {author} {\bibfnamefont {C.}~\bibnamefont {Beer}},\
  }\href {http://iopscience.iop.org/1402-4896/1/2-3/003} {\bibfield  {journal}
  {\bibinfo  {journal} {Physica Scripta}\ }\textbf {\bibinfo {volume} {1}},\
  \bibinfo {pages} {85} (\bibinfo {year} {1970})}\BibitemShut {NoStop}%
\bibitem [{\citenamefont {Nesaraja}\ and\ \citenamefont
  {McCutchan}(2016)}]{Nesaraja:2016ktw}%
  \BibitemOpen
  \bibfield  {author} {\bibinfo {author} {\bibfnamefont {C.~D.}\ \bibnamefont
  {Nesaraja}}\ and\ \bibinfo {author} {\bibfnamefont {E.~A.}\ \bibnamefont
  {McCutchan}},\ }\href {\doibase 10.1016/j.nds.2016.02.001} {\bibfield
  {journal} {\bibinfo  {journal} {Nucl. Data Sheets}\ }\textbf {\bibinfo
  {volume} {133}},\ \bibinfo {pages} {1} (\bibinfo {year} {2016})}\BibitemShut
  {NoStop}%
\bibitem [{\citenamefont {{National Nuclear Data Center}}(2013)}]{CapGam}%
  \BibitemOpen
  \bibfield  {author} {\bibinfo {author} {\bibnamefont {{National Nuclear Data
  Center}}},\ }\href {https://www.nndc.bnl.gov/capgam/index.html} {\enquote
  {\bibinfo {title} {{CapGam}},}\ } (\bibinfo {year} {2013}),\ \bibinfo {note}
  {[Online; accessed 2017-9-30]}\BibitemShut {NoStop}%
\bibitem [{\citenamefont {Wulandari}\ \emph {et~al.}(2004)\citenamefont
  {Wulandari}, \citenamefont {Jochum}, \citenamefont {Rau},\ and\ \citenamefont
  {von Feilitzsch}}]{Wulandari:2003cr}%
  \BibitemOpen
  \bibfield  {author} {\bibinfo {author} {\bibfnamefont {H.}~\bibnamefont
  {Wulandari}}, \bibinfo {author} {\bibfnamefont {J.}~\bibnamefont {Jochum}},
  \bibinfo {author} {\bibfnamefont {W.}~\bibnamefont {Rau}}, \ and\ \bibinfo
  {author} {\bibfnamefont {F.}~\bibnamefont {von Feilitzsch}},\ }\href
  {\doibase 10.1016/j.astropartphys.2004.07.005} {\bibfield  {journal}
  {\bibinfo  {journal} {Astropart. Phys.}\ }\textbf {\bibinfo {volume} {22}},\
  \bibinfo {pages} {313} (\bibinfo {year} {2004})},\ \Eprint
  {http://arxiv.org/abs/hep-ex/0312050} {arXiv:hep-ex/0312050 [hep-ex]}
  \BibitemShut {NoStop}%
\bibitem [{\citenamefont {de~Viveiros}(2010)}]{deViveiros:2010fnb}%
  \BibitemOpen
  \bibfield  {author} {\bibinfo {author} {\bibfnamefont {L.}~\bibnamefont
  {de~Viveiros}},\ }\emph {\bibinfo {title} {{Optimization of Signal versus
  Background in Liquid Xe Detectors Used for Dark Matter Direct Detection
  Experiments}}},\ \href {http://gradworks.umi.com/34/30/3430091.html} {Ph.D.
  thesis},\ \bibinfo  {school} {Brown U.} (\bibinfo {year} {2010})\BibitemShut
  {NoStop}%
\bibitem [{\citenamefont {Li}\ and\ \citenamefont {Beacom}(2014)}]{Li:2014sea}%
  \BibitemOpen
  \bibfield  {author} {\bibinfo {author} {\bibfnamefont {S.~W.}\ \bibnamefont
  {Li}}\ and\ \bibinfo {author} {\bibfnamefont {J.~F.}\ \bibnamefont
  {Beacom}},\ }\href {\doibase 10.1103/PhysRevC.89.045801} {\bibfield
  {journal} {\bibinfo  {journal} {Phys. Rev. C}\ }\textbf {\bibinfo {volume}
  {89}},\ \bibinfo {pages} {045801} (\bibinfo {year} {2014})},\ \Eprint
  {http://arxiv.org/abs/1402.4687} {arXiv:1402.4687 [hep-ph]} \BibitemShut
  {NoStop}%
\bibitem [{\citenamefont {Li}\ and\ \citenamefont
  {Beacom}(2015{\natexlab{a}})}]{Li:2015kpa}%
  \BibitemOpen
  \bibfield  {author} {\bibinfo {author} {\bibfnamefont {S.~W.}\ \bibnamefont
  {Li}}\ and\ \bibinfo {author} {\bibfnamefont {J.~F.}\ \bibnamefont
  {Beacom}},\ }\href {\doibase 10.1103/PhysRevD.91.105005} {\bibfield
  {journal} {\bibinfo  {journal} {Phys. Rev. D}\ }\textbf {\bibinfo {volume}
  {91}},\ \bibinfo {pages} {105005} (\bibinfo {year} {2015}{\natexlab{a}})},\
  \Eprint {http://arxiv.org/abs/1503.04823} {arXiv:1503.04823 [hep-ph]}
  \BibitemShut {NoStop}%
\bibitem [{\citenamefont {Li}\ and\ \citenamefont
  {Beacom}(2015{\natexlab{b}})}]{Li:2015lxa}%
  \BibitemOpen
  \bibfield  {author} {\bibinfo {author} {\bibfnamefont {S.~W.}\ \bibnamefont
  {Li}}\ and\ \bibinfo {author} {\bibfnamefont {J.~F.}\ \bibnamefont
  {Beacom}},\ }\href {\doibase 10.1103/PhysRevD.92.105033} {\bibfield
  {journal} {\bibinfo  {journal} {Phys. Rev. D}\ }\textbf {\bibinfo {volume}
  {92}},\ \bibinfo {pages} {105033} (\bibinfo {year} {2015}{\natexlab{b}})},\
  \Eprint {http://arxiv.org/abs/1508.05389} {arXiv:1508.05389
  [physics.ins-det]} \BibitemShut {NoStop}%
\bibitem [{\citenamefont {Behrens}\ and\ \citenamefont
  {J{\"a}necke}(1969)}]{Behrens1969}%
  \BibitemOpen
  \bibfield  {author} {\bibinfo {author} {\bibfnamefont {H.}~\bibnamefont
  {Behrens}}\ and\ \bibinfo {author} {\bibfnamefont {J.}~\bibnamefont
  {J{\"a}necke}},\ }\href {https://books.google.com/books?id=1EVsGQAACAAJ}
  {\emph {\bibinfo {title} {Numerical Tables for Beta-Decay and Electron
  Capture}}}\ (\bibinfo  {publisher} {Springer},\ \bibinfo {year}
  {1969})\BibitemShut {NoStop}%
\bibitem [{\citenamefont {Schenter}\ and\ \citenamefont
  {Vogel}(1983)}]{Schenter1983}%
  \BibitemOpen
  \bibfield  {author} {\bibinfo {author} {\bibfnamefont {G.~K.}\ \bibnamefont
  {Schenter}}\ and\ \bibinfo {author} {\bibfnamefont {P.}~\bibnamefont
  {Vogel}},\ }\href {\doibase 10.13182/NSE83-A17574} {\bibfield  {journal}
  {\bibinfo  {journal} {Nuclear Science and Engineering}\ }\textbf {\bibinfo
  {volume} {83}},\ \bibinfo {pages} {393} (\bibinfo {year} {1983})}\BibitemShut
  {NoStop}%
\bibitem [{\citenamefont {Hayes}\ and\ \citenamefont
  {Vogel}(2016)}]{Hayes:2016qnu}%
  \BibitemOpen
  \bibfield  {author} {\bibinfo {author} {\bibfnamefont {A.~C.}\ \bibnamefont
  {Hayes}}\ and\ \bibinfo {author} {\bibfnamefont {P.}~\bibnamefont {Vogel}},\
  }\href {\doibase 10.1146/annurev-nucl-102115-044826} {\bibfield  {journal}
  {\bibinfo  {journal} {Ann. Rev. Nucl. Part. Sci.}\ }\textbf {\bibinfo
  {volume} {66}},\ \bibinfo {pages} {219} (\bibinfo {year} {2016})},\ \Eprint
  {http://arxiv.org/abs/1605.02047} {arXiv:1605.02047 [hep-ph]} \BibitemShut
  {NoStop}%
\bibitem [{\citenamefont {Rogers}(1990)}]{rogers1990geology}%
  \BibitemOpen
  \bibfield  {author} {\bibinfo {author} {\bibfnamefont {H.}~\bibnamefont
  {Rogers}},\ }in\ \href
  {https://pubs.geoscienceworld.org/books/book/1824/chapter/107705612/geology-of-precambrian-rocks-in-the-poorman}
  {\emph {\bibinfo {booktitle} {Metallogeny of Gold in the Black Hills, South
  Dakota}}}\ (\bibinfo  {publisher} {Society of Economic Geologists},\ \bibinfo
  {year} {1990})\ p.\ \bibinfo {pages} {204}\BibitemShut {NoStop}%
\bibitem [{\citenamefont {Shultis}\ and\ \citenamefont
  {Faw}(2002)}]{shultis2002}%
  \BibitemOpen
  \bibfield  {author} {\bibinfo {author} {\bibfnamefont {J.~K.}\ \bibnamefont
  {Shultis}}\ and\ \bibinfo {author} {\bibfnamefont {R.~E.}\ \bibnamefont
  {Faw}},\ }\href {https://books.google.com/books?id=SO4Lmw8XoEMC} {\emph
  {\bibinfo {title} {Fundamentals of Nuclear Science and Engineering}}}\
  (\bibinfo  {publisher} {Taylor \& Francis},\ \bibinfo {year}
  {2002})\BibitemShut {NoStop}%
\bibitem [{\citenamefont {Ferrari}\ \emph {et~al.}(2005)\citenamefont
  {Ferrari}, \citenamefont {Sala}, \citenamefont {Fasso},\ and\ \citenamefont
  {Ranft}}]{Ferrari:2005zk}%
  \BibitemOpen
  \bibfield  {author} {\bibinfo {author} {\bibfnamefont {A.}~\bibnamefont
  {Ferrari}}, \bibinfo {author} {\bibfnamefont {P.~R.}\ \bibnamefont {Sala}},
  \bibinfo {author} {\bibfnamefont {A.}~\bibnamefont {Fasso}}, \ and\ \bibinfo
  {author} {\bibfnamefont {J.}~\bibnamefont {Ranft}},\ }\href@noop {} {\enquote
  {\bibinfo {title} {{FLUKA: A multi-particle transport code (Program version
  2005)}},}\ } (\bibinfo {year} {2005})\BibitemShut {NoStop}%
\bibitem [{\citenamefont {Battistoni}\ \emph {et~al.}(2007)\citenamefont
  {Battistoni}, \citenamefont {Muraro}, \citenamefont {Sala}, \citenamefont
  {Cerutti}, \citenamefont {Ferrari}, \citenamefont {Roesler}, \citenamefont
  {Fasso},\ and\ \citenamefont {Ranft}}]{Battistoni:2007zzb}%
  \BibitemOpen
  \bibfield  {author} {\bibinfo {author} {\bibfnamefont {G.}~\bibnamefont
  {Battistoni}}, \bibinfo {author} {\bibfnamefont {S.}~\bibnamefont {Muraro}},
  \bibinfo {author} {\bibfnamefont {P.~R.}\ \bibnamefont {Sala}}, \bibinfo
  {author} {\bibfnamefont {F.}~\bibnamefont {Cerutti}}, \bibinfo {author}
  {\bibfnamefont {A.}~\bibnamefont {Ferrari}}, \bibinfo {author} {\bibfnamefont
  {S.}~\bibnamefont {Roesler}}, \bibinfo {author} {\bibfnamefont
  {A.}~\bibnamefont {Fasso}}, \ and\ \bibinfo {author} {\bibfnamefont
  {J.}~\bibnamefont {Ranft}},\ }\href {\doibase 10.1063/1.2720455} {\bibfield
  {journal} {\bibinfo  {journal} {AIP Conf. Proc.}\ }\textbf {\bibinfo {volume}
  {896}},\ \bibinfo {pages} {31} (\bibinfo {year} {2007})}\BibitemShut
  {NoStop}%
\bibitem [{\citenamefont {Lesko}\ \emph {et~al.}(2011)\citenamefont {Lesko}
  \emph {et~al.}}]{Lesko:2011qk}%
  \BibitemOpen
  \bibfield  {author} {\bibinfo {author} {\bibfnamefont {K.~T.}\ \bibnamefont
  {Lesko}} \emph {et~al.},\ }\href@noop {} {\  (\bibinfo {year} {2011})},\
  \Eprint {http://arxiv.org/abs/1108.0959} {arXiv:1108.0959 [hep-ex]}
  \BibitemShut {NoStop}%
\bibitem [{\citenamefont {Chan}(2012)}]{Chan:rock}%
  \BibitemOpen
  \bibfield  {author} {\bibinfo {author} {\bibfnamefont {Y.-D.}\ \bibnamefont
  {Chan}},\ }\href
  {https://www.sanfordlab.org/article/low-background-construction-laboratories-4850-ft-level-davis-campus}
  {\enquote {\bibinfo {title} {{The Low-Background Construction of Laboratories
  at the 4850-ft Level Davis Campus}},}\ } (\bibinfo {year} {2012})\BibitemShut
  {NoStop}%
\bibitem [{\citenamefont {Ricci}\ \emph {et~al.}(2013)\citenamefont {Ricci},
  \citenamefont {Mantovani}, \citenamefont {Baldoncini}, \citenamefont
  {Esposito}, \citenamefont {Ludhova},\ and\ \citenamefont
  {Zavatarelli}}]{Ricci:2014qpa}%
  \BibitemOpen
  \bibfield  {author} {\bibinfo {author} {\bibfnamefont {B.}~\bibnamefont
  {Ricci}}, \bibinfo {author} {\bibfnamefont {F.}~\bibnamefont {Mantovani}},
  \bibinfo {author} {\bibfnamefont {M.}~\bibnamefont {Baldoncini}}, \bibinfo
  {author} {\bibfnamefont {J.}~\bibnamefont {Esposito}}, \bibinfo {author}
  {\bibfnamefont {L.}~\bibnamefont {Ludhova}}, \ and\ \bibinfo {author}
  {\bibfnamefont {S.}~\bibnamefont {Zavatarelli}},\ }\href@noop {} {\bibfield
  {journal} {\bibinfo  {journal} {PoS}\ }\textbf {\bibinfo {volume}
  {Neutel2013}},\ \bibinfo {pages} {077} (\bibinfo {year} {2013})},\ \Eprint
  {http://arxiv.org/abs/1403.4072} {arXiv:1403.4072 [hep-ex]} \BibitemShut
  {NoStop}%
\bibitem [{\citenamefont {Heise}(2015)}]{Heise:2015vza}%
  \BibitemOpen
  \bibfield  {author} {\bibinfo {author} {\bibfnamefont {J.}~\bibnamefont
  {Heise}},\ }\bibfield  {booktitle} {\emph {\bibinfo {booktitle}
  {{Proceedings, 2nd Workshop on Germanium-Based Detectors and Technologies:
  Vermillion, SD, USA, September 14-17, 2014}}},\ }\href {\doibase
  10.1088/1742-6596/606/1/012015} {\bibfield  {journal} {\bibinfo  {journal}
  {J. Phys. Conf. Ser.}\ }\textbf {\bibinfo {volume} {606}},\ \bibinfo {pages}
  {012015} (\bibinfo {year} {2015})},\ \Eprint
  {http://arxiv.org/abs/1503.01112} {arXiv:1503.01112 [physics.ins-det]}
  \BibitemShut {NoStop}%
\bibitem [{\citenamefont {Westerdale}\ and\ \citenamefont
  {Meyers}(2017)}]{Westerdale:2017kml}%
  \BibitemOpen
  \bibfield  {author} {\bibinfo {author} {\bibfnamefont {S.}~\bibnamefont
  {Westerdale}}\ and\ \bibinfo {author} {\bibfnamefont {P.~D.}\ \bibnamefont
  {Meyers}},\ }\href {\doibase 10.1016/j.nima.2017.09.007} {\bibfield
  {journal} {\bibinfo  {journal} {Nucl. Instrum. Meth. A}\ }\textbf {\bibinfo
  {volume} {875}},\ \bibinfo {pages} {57} (\bibinfo {year} {2017})},\ \Eprint
  {http://arxiv.org/abs/1702.02465} {arXiv:1702.02465 [physics.ins-det]}
  \BibitemShut {NoStop}%
\bibitem [{\citenamefont {Beacom}\ \emph {et~al.}(2017)\citenamefont {Beacom}
  \emph {et~al.}}]{JinpingNeutrinoExperimentgroup:2016nol}%
  \BibitemOpen
  \bibfield  {author} {\bibinfo {author} {\bibfnamefont {J.~F.}\ \bibnamefont
  {Beacom}} \emph {et~al.} (\bibinfo {collaboration} {Jinping}),\ }\href
  {\doibase 10.1088/1674-1137/41/2/023002} {\bibfield  {journal} {\bibinfo
  {journal} {Chin. Phys. C}\ }\textbf {\bibinfo {volume} {41}},\ \bibinfo
  {pages} {023002} (\bibinfo {year} {2017})},\ \Eprint
  {http://arxiv.org/abs/1602.01733} {arXiv:1602.01733 [physics.ins-det]}
  \BibitemShut {NoStop}%
\bibitem [{\citenamefont {Gaisser}\ and\ \citenamefont
  {Honda}(2002)}]{Gaisser:2002jj}%
  \BibitemOpen
  \bibfield  {author} {\bibinfo {author} {\bibfnamefont {T.~K.}\ \bibnamefont
  {Gaisser}}\ and\ \bibinfo {author} {\bibfnamefont {M.}~\bibnamefont
  {Honda}},\ }\href {\doibase 10.1146/annurev.nucl.52.050102.090645} {\bibfield
   {journal} {\bibinfo  {journal} {Ann. Rev. Nucl. Part. Sci.}\ }\textbf
  {\bibinfo {volume} {52}},\ \bibinfo {pages} {153} (\bibinfo {year} {2002})},\
  \Eprint {http://arxiv.org/abs/hep-ph/0203272} {arXiv:hep-ph/0203272 [hep-ph]}
  \BibitemShut {NoStop}%
\bibitem [{\citenamefont {Abe}\ \emph {et~al.}(2018{\natexlab{a}})\citenamefont
  {Abe} \emph {et~al.}}]{Abe:2017aap}%
  \BibitemOpen
  \bibfield  {author} {\bibinfo {author} {\bibfnamefont {K.}~\bibnamefont
  {Abe}} \emph {et~al.} (\bibinfo {collaboration} {Super-Kamiokande}),\ }\href
  {\doibase 10.1103/PhysRevD.97.072001} {\bibfield  {journal} {\bibinfo
  {journal} {Phys. Rev. D}\ }\textbf {\bibinfo {volume} {97}},\ \bibinfo
  {pages} {072001} (\bibinfo {year} {2018}{\natexlab{a}})},\ \Eprint
  {http://arxiv.org/abs/1710.09126} {arXiv:1710.09126 [hep-ex]} \BibitemShut
  {NoStop}%
\bibitem [{\citenamefont {Beacom}(2010)}]{Beacom:2010kk}%
  \BibitemOpen
  \bibfield  {author} {\bibinfo {author} {\bibfnamefont {J.~F.}\ \bibnamefont
  {Beacom}},\ }\href {\doibase 10.1146/annurev.nucl.010909.083331} {\bibfield
  {journal} {\bibinfo  {journal} {Ann. Rev. Nucl. Part. Sci.}\ }\textbf
  {\bibinfo {volume} {60}},\ \bibinfo {pages} {439} (\bibinfo {year} {2010})},\
  \Eprint {http://arxiv.org/abs/1004.3311} {arXiv:1004.3311 [astro-ph.HE]}
  \BibitemShut {NoStop}%
\bibitem [{\citenamefont {Bays}\ \emph {et~al.}(2012)\citenamefont {Bays} \emph
  {et~al.}}]{Bays:2011si}%
  \BibitemOpen
  \bibfield  {author} {\bibinfo {author} {\bibfnamefont {K.}~\bibnamefont
  {Bays}} \emph {et~al.} (\bibinfo {collaboration} {Super-Kamiokande}),\ }\href
  {\doibase 10.1103/PhysRevD.85.052007} {\bibfield  {journal} {\bibinfo
  {journal} {Phys. Rev. D}\ }\textbf {\bibinfo {volume} {85}},\ \bibinfo
  {pages} {052007} (\bibinfo {year} {2012})},\ \Eprint
  {http://arxiv.org/abs/1111.5031} {arXiv:1111.5031 [hep-ex]} \BibitemShut
  {NoStop}%
\bibitem [{\citenamefont {Abe}\ \emph {et~al.}(2011)\citenamefont {Abe} \emph
  {et~al.}}]{Abe:2011ts}%
  \BibitemOpen
  \bibfield  {author} {\bibinfo {author} {\bibfnamefont {K.}~\bibnamefont
  {Abe}} \emph {et~al.} (\bibinfo {collaboration} {Hyper-Kamiokande}),\
  }\href@noop {} {\  (\bibinfo {year} {2011})},\ \Eprint
  {http://arxiv.org/abs/1109.3262} {arXiv:1109.3262 [hep-ex]} \BibitemShut
  {NoStop}%
\bibitem [{\citenamefont {Abe}\ \emph {et~al.}(2018{\natexlab{b}})\citenamefont
  {Abe} \emph {et~al.}}]{Abe:2018uyc}%
  \BibitemOpen
  \bibfield  {author} {\bibinfo {author} {\bibfnamefont {K.}~\bibnamefont
  {Abe}} \emph {et~al.} (\bibinfo {collaboration} {Hyper-Kamiokande}),\
  }\href@noop {} {\  (\bibinfo {year} {2018}{\natexlab{b}})},\ \Eprint
  {http://arxiv.org/abs/1805.04163} {arXiv:1805.04163 [physics.ins-det]}
  \BibitemShut {NoStop}%
\bibitem [{\citenamefont {Andringa}\ \emph {et~al.}(2016)\citenamefont
  {Andringa} \emph {et~al.}}]{Andringa:2015tza}%
  \BibitemOpen
  \bibfield  {author} {\bibinfo {author} {\bibfnamefont {S.}~\bibnamefont
  {Andringa}} \emph {et~al.} (\bibinfo {collaboration} {SNO+}),\ }\href
  {\doibase 10.1155/2016/6194250} {\bibfield  {journal} {\bibinfo  {journal}
  {Adv. High Energy Phys.}\ }\textbf {\bibinfo {volume} {2016}},\ \bibinfo
  {pages} {6194250} (\bibinfo {year} {2016})},\ \Eprint
  {http://arxiv.org/abs/1508.05759} {arXiv:1508.05759 [physics.ins-det]}
  \BibitemShut {NoStop}%
\bibitem [{\citenamefont {Aalbers}\ \emph {et~al.}(2016)\citenamefont {Aalbers}
  \emph {et~al.}}]{Aalbers:2016jon}%
  \BibitemOpen
  \bibfield  {author} {\bibinfo {author} {\bibfnamefont {J.}~\bibnamefont
  {Aalbers}} \emph {et~al.} (\bibinfo {collaboration} {DARWIN}),\ }\href
  {\doibase 10.1088/1475-7516/2016/11/017} {\bibfield  {journal} {\bibinfo
  {journal} {JCAP}\ }\textbf {\bibinfo {volume} {1611}},\ \bibinfo {pages}
  {017} (\bibinfo {year} {2016})},\ \Eprint {http://arxiv.org/abs/1606.07001}
  {arXiv:1606.07001 [astro-ph.IM]} \BibitemShut {NoStop}%
\end{thebibliography}%



\begin{appendix}
\newpage
\clearpage
\setcounter{figure}{0}
\renewcommand{\thefigure}{A\arabic{figure}}
\renewcommand{\theHfigure}{A\arabic{figure}}

\titleformat{\section}[hang]{\bfseries\center\small}{}{1em}{Appendix \thesection.\quad#1}%
\titleformat{\subsection}[hang]{\bfseries\small\center}{}{1em}{\thesection.\thesubsection\quad#1}%

\onecolumngrid
\begin{center}
 \bf \large Supplemental Material
\end{center}
\vspace*{0.2cm}

\twocolumngrid

Here we provide additional details.  Appendix~\ref{sec:CC_sigma} focuses on the $\nu_e + \, ^{40}$Ar charged-current cross section, Appendix~\ref{sec:bck} on our calculation of detector backgrounds for DUNE, Appendix~\ref{sec:analysis} on our calculations of DUNE's sensitivity to solar neutrinos, and Appendix~\ref{sec:other_inputs} on examples of how our results change under different assumptions.


\section{SIGNAL CROSS SECTION}
\label{sec:CC_sigma}

To exploit the full potential of a solar-neutrino program in DUNE, the total cross section for
\begin{equation}
        \nu_e + \, ^{40}{\rm Ar} \rightarrow e^- + \, ^{40}{\rm K}^*,
\end{equation}
when convolved with the $^8$B spectrum and a detector threshold of $T_e = 5$~MeV, should be known to $\sim 1\%$.  With a larger uncertainty, DUNE alone can still precisely measure $\Delta m^2_{21}$ and $\phi(hep)$; DUNE in combination with other experiments can also still precisely measure $\sin^2\theta_{12}$ and $\phi(^8\text{B})$.  See Sec.~\ref{sec:detector}.

A cross-section uncertainty of $\lesssim 2.4\%$ has been claimed based on indirect measurements, though our assessment below suggests that $\lesssim 10\%$ is more realistic.  Importantly, the uncertainty is unlikely to be worse than that.  Although this may be surprising, based on the larger uncertainties for neutrino-nucleus interactions at GeV energies, the physics at these low energies is much simpler.  Our conclusions are robust to possible changes in the central value of the cross section.  For example, a 10\% change in the cross section and thus signal counts would change significances of measured parameters by $\simeq 5\%$.  Only the scale of its uncertainty is important.

A precise determination of the cross section is challenging but realistic.  There are ways to reduce the uncertainty on the cross section, including through a first direct measurement.


\subsection{Details of Calculations}
\label{sec:CC_sigma.1}

We first state the form of the cross section we use, which follows Ref.~\cite{Bhattacharya:2009zz}, which presents the most recent indirect measurements.
At leading order, the total cross section is
\begin{equation}
\sigma(E_\nu) = \sum_i \frac{G_{F, \beta}^2 \left| V_{ud} \right|^2}{\pi}
\left| \mathcal{M}_{o \rightarrow i} \right|^2
E_e^i \, p_e^i \, F(Z,E_e^i),
\label{eq:CC_cross_section}
\end{equation}
where $i$ indexes transitions from the ground state of $^{40}$Ar to distinct relevant nuclear excited states in $^{40}{\rm K}$, collectively denoted with a * (transitions to its ground state are forbidden by selection rules).  The $i$-dependent terms are the amplitude squared, phase space ($E_e$ is the electron's total energy and $p_e = v_e E_e \simeq E_e$ its momentum), and Fermi function $F$ (to account for Coulomb effects).  $G_{F, \beta}$ is the Fermi constant for beta decay (see below) and $V_{ud}$ is the quark mixing matrix element.  For each nuclear transition, which could, in principle, be identified by the total gamma-ray energy, the neutrino spectrum is sampled faithfully, weighted by cross section and shifted by the nuclear threshold.

For the final states we consider, the transition amplitudes squared can be expressed as
\begin{equation}
\left|\mathcal{M}_{o\rightarrow i}\right|^2 = B_i(\text{F})+ B_i(\text{GT}),
\end{equation}
where the transitions are of the Fermi or Gamow-Teller type, or mixed~\cite{Konopinski1950}.  For allowed Fermi transitions, which correspond to the vector part of the weak current, the leptons have total spin zero and the change in the nuclear total angular momentum is zero.  For allowed Gamow-Teller transitions, which correspond to the axial-vector part of the weak current, the leptons have total spin one and the change in the nuclear total angular momentum is one or zero (but not $0 \rightarrow 0$).  For $\nu_e + \, ^{40}$Ar, the transitions are seemingly not mixed and the only relevant Fermi transition is super-allowed, with its strength given by a sum rule, $B(\text{F}) = N - Z = 4$ (defined for the parent nucleus)~\cite{Ormand:1994js, Bhattacharya:2009zz}.  For the Gamow-Teller strengths, we use those based on measurements of $^{40}\text{Ar}(p, n)^{40}\text{K}^*$ in forward-angle kinematics~\cite{Bhattacharya:2009zz}.  See Table~\ref{table:CC_cross_section}.

\begin{table}[t]
\centering
\vspace{0.8em}
\begin{tabular}{||C{1.0cm}|C{2.0cm}|C{1.4cm}|C{1.4cm}||}
  \hline\hline
   i & $\Delta E_i$ [MeV] & $B_i({\rm F})$ & $B_i({\rm GT})$\\[-1pt] \hline
  \hline
   1 & 2.333 &      & 1.64 \\[-1pt] \hline
   2 & 2.775 &      & 1.49 \\[-1pt] \hline
   3 & 3.204 &      & 0.06 \\[-1pt] \hline
   4 & 3.503 &      & 0.16 \\[-1pt] \hline
   5 & 3.870 &      & 0.44 \\[-1pt] \hline
   6 & 4.384 & 4.00 &      \\[-1pt] \hline
   7 & 4.421 &      & 0.86 \\[-1pt] \hline
   8 & 4.763 &      & 0.48 \\[-1pt] \hline
   9 & 5.162 &      & 0.59 \\[-1pt] \hline
  10 & 5.681 &      & 0.21 \\[-1pt] \hline
  11 & 6.118 &      & 0.48 \\[-1pt] \hline
  12 & 6.790 &      & 0.71 \\[-1pt] \hline
  13 & 7.468 &      & 0.06 \\[-1pt] \hline
  14 & 7.795 &      & 0.14 \\[-1pt] \hline
  15 & 7.952 &      & 0.97 \\[-1pt] \hline\hline
  total & & 4.00 & 8.29 \\[-1pt] \hline\hline
\end{tabular}
\vspace{1em}
\caption{Transition strengths for $\nu_e + \, ^{40}{\rm Ar} \rightarrow e^- + \, ^{40}{\rm K}^*$~\cite{Bhattacharya:2009zz}.  Here we have multiplied the $B_i({\rm GT})$ values by the axial coupling constant squared (with $g_A = -1.26$), so that they have the same normalization as $B_i({\rm F})$, as appropriate for Ref.~\cite{Bhattacharya:2009zz}.  In the literature, it is not always clearly noted if $B_i({\rm GT})$ values are normalized with or without this factor. The energy of the Fermi state is taken from Ref.~\cite{Bhattacharya:1998hc}.}
\label{table:CC_cross_section}
\end{table}

Because the interaction is a 2-to-2 process and the nuclear transitions are between discrete states, the kinematics and the phase-space factor are simple.  (Recoil-order corrections are $\lesssim 5$ keV, and can be neglected.)  The electron kinetic energy is $T_e = E_\nu - Q_i$, where $Q_i = Q_\text{gs} + \Delta E_i$, with $Q_\text{gs} = 1.504$~MeV the reaction threshold to reach the ground state of $^{40}$K (including creation of the electron) and $\Delta E_i$ the excitation energy above that~\cite{NNDC, Bhattacharya:2009zz}.  We assume a detection threshold of $T_e = 5$~MeV, conservatively neglecting the detectability of nuclear de-excitation gamma rays (of total energy $\Delta E_i$), which primarily undergo Compton scattering, and may be detectable in coincidence, improving particle and reaction identification.  For a neutrino interaction to register, its energy must exceed $E_\nu^{thr,i} = (Q_\text{gs} + \Delta E_i) + 5 {\rm \ MeV}$, which is 8.837~MeV for the lowest allowed transition and higher for others.  The electron angular distribution, $d\sigma/d\cos\theta$, is $\propto 1 + \cos\theta$ for Fermi transitions and $\propto 1 - \frac{1}{3} \cos\theta$ for Gamow-Teller transitions~\cite{Beacom:1998fj, Vogel:1999zy}.  For an isotropic angular distribution, the fraction of $\nu_e + \, ^{40}{\rm Ar}$ events outside the forward cone is 88\%.  Taking into account the weighting of different transitions does not change this number appreciably.

\begin{figure}[t]
\begin{center}
\includegraphics[width=\columnwidth]{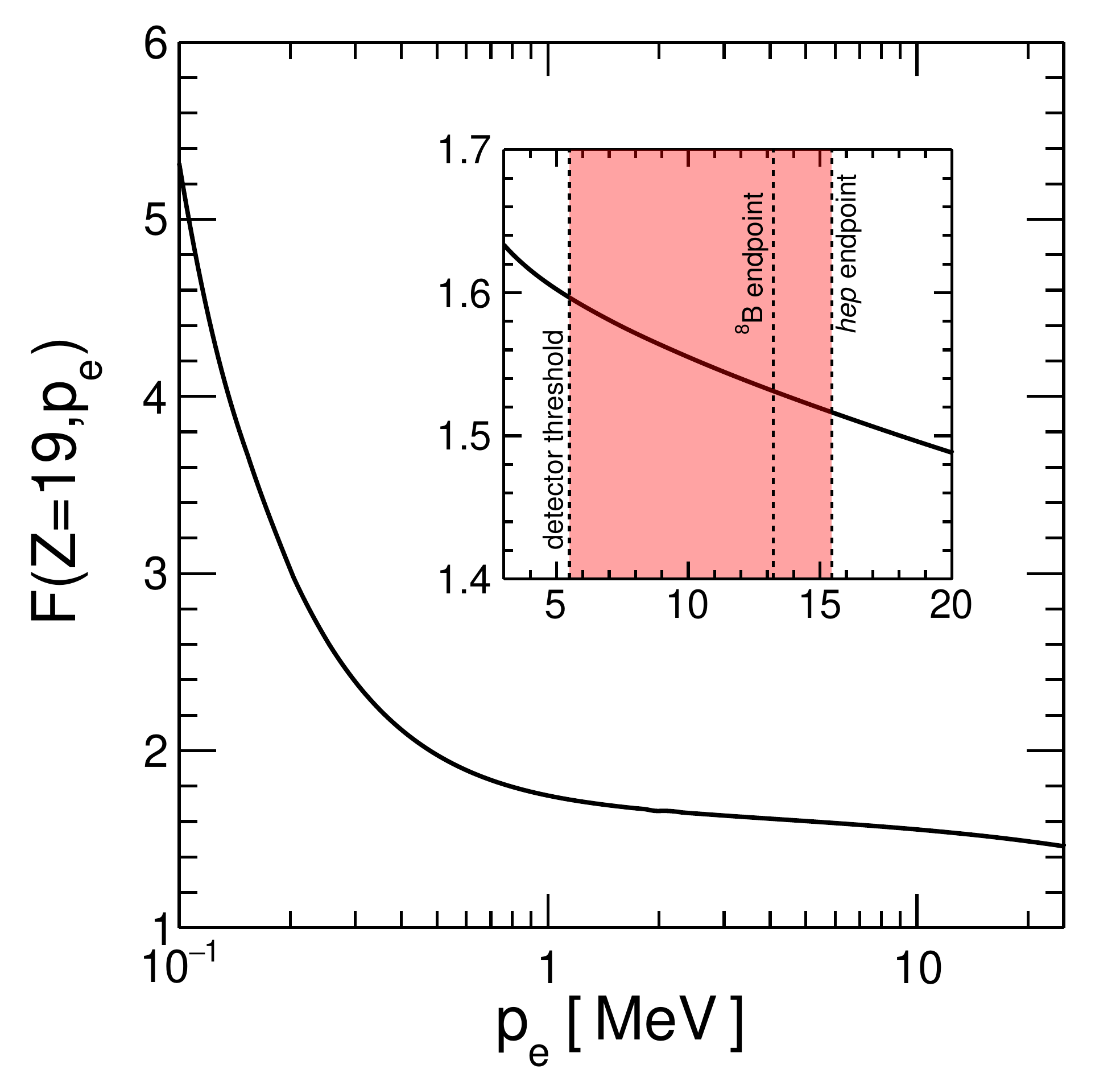}
\caption{Fermi function for $\nu_e + \, ^{40}{\rm Ar} \rightarrow e^- + \, ^{40}{\rm K}^*$ in terms of electron momentum, interpolated from data in Ref.~\cite{Behrens1969, Bhattacharya:2009zz}.  The shaded region indicates the approximate range of interest.}
\label{fig:fermi_function}
\end{center}
\end{figure}

The Fermi function $F$ accounts for the distortion of the outgoing electron wave function due to its Coulomb interaction with the nuclear charge $Z$ (defined for the daughter nucleus)~\cite{Schenter1983, Behrens1969, Hayes:2016qnu}.  For an outgoing electron, the interaction is attractive, making $F > 1$, and, in this case, the correction is substantial, a factor $\simeq 1.6$.  A commonly used analytic estimate, $F \simeq 2\pi \nu / [1 - \exp(- 2 \pi \nu)]$, with $\nu = Z \alpha /v_e$, based on the solution to the Schr\"{o}edinger equation for an electron in the potential of a point-like nucleus, is only suitable for small Z~\cite{Schenter1983}.  At high momentum, this is a constant; at low momentum, it varies as $1/v_e$.  A detailed calculation of $F$, based on the solution of the Dirac equation for an electron in the potential of a finite-sized nucleus, is tabulated in Table II of Ref.~\cite{Behrens1969, Bhattacharya:2009zz}.  Figure~\ref{fig:fermi_function} shows our interpolated result, which differs from the simple analytic expression~\cite{Schenter1983} by being $\sim 10\%$ larger and by varying at large momenta, both of which are important.

\begin{figure}[t]
\begin{center}
\includegraphics[width=\columnwidth]{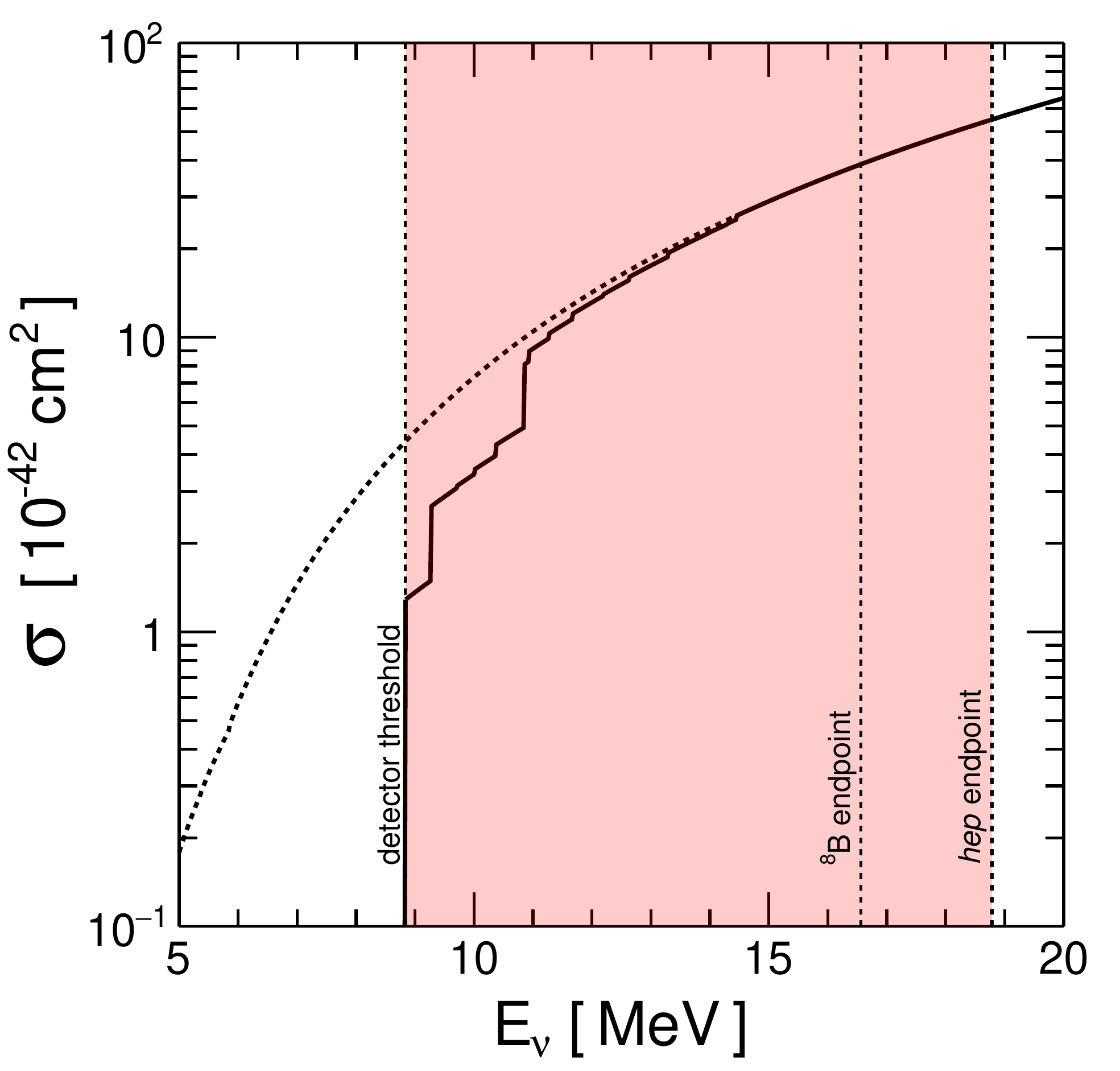}
\caption{Total cross section for $\nu_e + \, ^{40}{\rm Ar} \rightarrow e^- + \, ^{40}{\rm K}^*$.  The solid line takes into account the 5-MeV threshold of DUNE, while the dotted line neglects it.}
\label{fig:CC_cross_section}
\end{center}
\end{figure}

Figure~\ref{fig:CC_cross_section} shows the total cross section for $\nu_e + \, ^{40}{\rm Ar}$, combining the factors above, including the detector threshold of $T_e = 5$ MeV.  At $E_\nu = 10 $~MeV, a typical energy, the relevant cross section is $\sigma \approx 3 \times 10^{-42}$ cm$^2$, a factor $\sim 30$ larger than that for $\nu_e + e^-$, though the latter's density of targets is 18 times higher.  The $\nu_e + \, ^{40}{\rm Ar}$ cross section grows rapidly, faster than $\sigma \propto E_e^2 \propto (E_\nu - E_\nu^{thr,i} + 5 {\rm\ MeV} + m_e)^2$, because with increasing neutrino energy, more nuclear thresholds are surpassed, and because of the strong effect of the detector threshold.  In our calculations, we do not use this total cross section; instead, we sum the partial cross sections for each independent transition.

\begin{figure}[t]
\begin{center}
\includegraphics[width=\columnwidth]{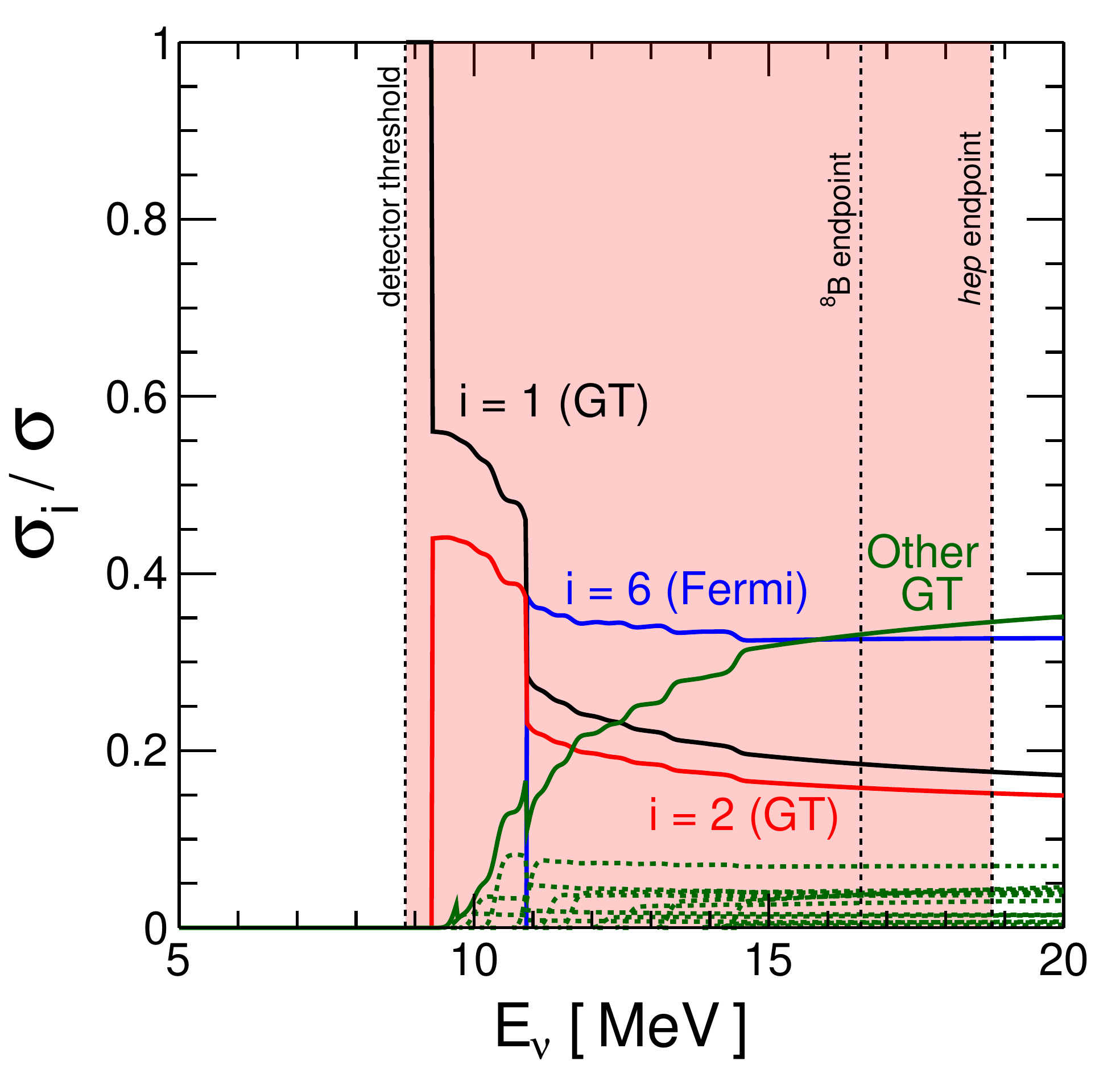}
\caption{Relative contributions to the total cross section for $\nu_e + \, ^{40}{\rm Ar} \rightarrow e^- + \, ^{40}{\rm K}^*$, with the minor transitions smoothed.  The detector threshold is taken into account.}
\label{fig:CC_components}
\end{center}
\end{figure}

Figure~\ref{fig:CC_components} shows how different transitions contribute to the total cross section, now showing only the case where the detector threshold of $T_e = 5$~MeV is applied to each transition separately.  The relative contributions vary depending on their strengths compared to those of all other kinematically accessible transitions.  Below $E_\nu = 10.888$ MeV, corresponding to the threshold for the super-allowed Fermi transition, the cross section is dominated by the two lowest-energy Gamow-Teller transitions.  For higher $E_\nu$, there are four comparable contributions: each of the two lowest-energy Gamow-Teller transitions, the Fermi transition, and the sum of all other kinematically accessible Gamow-Teller transitions.  (Though we do not exploit the angular distribution, doing so~\cite{Beacom:1998fj, Vogel:1999zy} would increase sensitivity.)

In principle, there are additional contributions to the $\nu_e + \, ^{40}$Ar total cross section from particle-unbound final states, e.g., $e^- + \, ^{39}{\rm Ar} + p$ and $e^- + \, ^{39}{\rm K} + n$.  However, even if the transition strengths are appreciable, the nuclear thresholds are above 9~MeV~\cite{NNDC}, making the neutrino thresholds above 14~MeV, and even higher if we account for the energy of the outgoing nucleons.  For a given neutrino energy, $T_e$ will be substantially lower than for the particle-bound transitions, burying any additional contributions in the falling total spectra of Fig.~3.


\subsection{Assessment of Uncertainties}
\label{sec:CC_sigma.2}

To assess the uncertainty on the cross section, we first consider two data-based evaluations.  The $B_i({\rm GT})$ values we use are from the $^{40}\text{Ar}(p,n)^{40}$K$^*$ data of Ref.~\cite{Bhattacharya:2009zz}.  For the uncertainties, we focus on the strengths themselves, though there may also be uncertainties in the excitation energies (through identification of the relevant transitions).  Though the uncertainties on individual $B_i$ are $\sim$ 5--20\%, the uncertainty on the summed strength is 2.4\%.  This reduction in uncertainty is because the strength for the super-allowed Fermi transition is considered known from the sum rule and because the uncertainties of independent Gamow-Teller transitions combine in a central-limit fashion for the total strength.  The uncertainty on the total cross section (convolved with the $^8$B spectrum and with a detector threshold of $T_e = 5$~MeV) is comparable to that on the summed strength.  In a paper preceding Ref.~\cite{Bhattacharya:2009zz}, a related group of authors obtained $B_i({\rm F + GT})$ values based on measurements of $^{40}$Ti $\beta^+$ decay to $^{40}$Sc$^*$, the isospin mirror process for $\nu_e + \, ^{40}{\rm Ar} \rightarrow e^- + \, ^{40}{\rm K}^*$, with an uncertainty of 2.1\% on the summed strength~\cite{Bhattacharya:1998hc}.  However, there are discrepancies between the two techniques, with the convolved cross sections at the most important energies differing at the $\sim 10\%$ level.  It is difficult to assess the systematics, as the two techniques access somewhat different transitions, and both rely on some assumptions: for the $(p,n)$ data, that the weak transitions are in the same proportions as the strong transitions; for the beta-decay data, that isospin symmetry holds.  A shell-model study~\cite{Karakoc:2014awa} suggests that the $(p,n)$ data are preferred over the beta-decay data, which is one of the reasons we adopted the strengths from Ref.~\cite{Bhattacharya:2009zz}.  Further work is needed to resolve  differences between the two techniques. 

The cross section can also be evaluated using $B_i$ values calculated with nuclear theory~\cite{Ormand:1994js, GilBotella:2003sz, Kolbe:2003ys, Cheoun:2011zza, Suzuki2014}.  For the low energies of solar neutrinos, the preferred technique is the nuclear shell model, treating most of the nucleons as belonging to a closed core, and treating the remaining valence nucleons as subject to an effective potential from the core as well as to a residual nucleon-nucleon interaction.  For the higher energies of supernova neutrinos, the preferred technique is the random phase approximation (RPA), which describes collective states of nuclei in a basis of particle-hole excitations.  A hybrid approach is also possible, with RPA used to calculate transitions not well described by the shell model.  Compared to our calculation, the hybrid calculation of Ref.~\cite{Suzuki2014}, which we consider to be the most reliable, is $\simeq 10\%$ smaller.  However, its $B_i$ values for known important low-lying transitions fall well below the data of Ref.~\cite{Bhattacharya:2009zz}, allowing its cross section to be smaller, despite the additional contributions of RPA-calculated transitions ($\lesssim 10\%$ at solar-neutrino energies).  We neglect the results of some other calculations: the early shell-model calculation of Ref.~\cite{Ormand:1994js} ($\simeq 30\%$ smaller than ours, but it omits relevant nuclear operators) and the RPA calculations of Refs.~\cite{GilBotella:2003sz, Kolbe:2003ys, Cheoun:2011zza} (a factor of a few difference, but they are not suited for the solar-neutrino energy range).  New work is needed on all of these techniques.

Thus, based on both experimental and theoretical evaluations of the cross section, its uncertainty (for $^8$B neutrinos in DUNE) is likely $\lesssim 10\%$.  This is small enough that our conclusions are robust, but large enough that new efforts to reduce uncertainties are needed.


\subsection{Towards Reducing Uncertainties}
\label{sec:CC_sigma.3}

Even without new experimental data, it seems likely that the cross-section uncertainty can be reduced through new theoretical work: calculations of the $B_i$, calculations to help reconcile differences in experimental inputs, and the development of a framework to consistently include all known effects.  These effects include corrections that, while not all uncertainties per se, cause discrepancies between different results if they are not applied uniformly, of which we give several examples, following e.g., Refs.~\cite{Vogel:1999zy, Beacom:2001hr, Kurylov:2002vj, Hardy:2014qxa, Hayes:2016qnu}.  The inner radiative corrections for charged-current semi-leptonic processes with nucleons increase the cross section by $\simeq 2.4\%$, which can be absorbed into a change in the Fermi constant for beta decay ($G_{F, \beta}$) relative to that measured from muon decay ($G_F$).  The outer radiative corrections can also increase the cross section by $\sim$ 1--2\%, depending on the nucleus and how the energy deposition by bremsstrahlung affects the detectability of the electron.  In principle, updating the value of the axial coupling constant from $g_A = -1.26$ to the contemporary $-1.27$ would also increase the cross section by $\simeq 2\%$, but $g_A$ may be quenched in nuclei.  In addition, the effects of the following should also be considered: isospin-violation corrections, subdominant argon isotopes (0.4\%), particle-unbound transitions, forbidden transitions, more accurate identifications of the energies of the relevant nuclear transitions, and so on.

To reduce the uncertainty to $< 1\%$, new data are needed, starting with comprehensive new measurements of auxiliary data to evaluate $B_i$, e.g., from $^{40}\text{Ar}(p, n)^{40}\text{K}^*$ and $^{40}$Ti $\beta^+$ decay.  It seems likely that new measurements, supported by theoretical efforts, could ensure that these techniques reach their intended precision.

Ultimately, the $\nu_e + \, ^{40}$Ar cross section must be directly measured with a laboratory source of neutrinos, which has never been done.  Achieving an uncertainty $< 1\%$ will require the statistics of $\gtrsim 10^4$ events and commensurate control of systematics.  The neutrino source could be accelerator-produced $\mu^+$ decay at rest, for which the $\nu_e$ spectrum, before weighting with the rising cross section, peaks at $\simeq 35$~MeV.  At this energy, the total charged-current cross section is $\simeq 300 \times 10^{-42}$ cm$^2$, taking into account only the nuclear transitions noted above; the true cross section will be larger.  At the location of the suite of COHERENT neutrino detectors at the Spallation Neutron Source at Oak Ridge, the time-averaged $\nu_e$ flux is $\simeq 10^7$ cm$^{-2}$ s$^{-1}$~\cite{Akimov:2017ade}.  We thus estimate that an exposure of $\sim 10$ ton-year is needed.  With a low detection threshold and good collection of scintillation light, the nuclear transition of each interaction could be identified by the total energy of its de-excitation gamma rays, which would reduce backgrounds.  If systematics are more challenging than statistics, then identifying specific nuclear transitions by their gamma rays could be used to measure relative strengths (including compared to the super-allowed Fermi transition) instead of the absolute cross section.  There would be numerous technical challenges, but the importance of the problem encourages significant investments.


\section{BACKGROUNDS}
\label{sec:bck}

For the detection of MeV neutrinos, backgrounds must be seriously considered.  Standard techniques developed for previous solar-neutrino and other experiments can powerfully reduce backgrounds.  These include defining a fiducial volume, removing U/Th from liquids and Rn from air, selecting low-background materials, applying a short deadtime after high-energy events, and so on.  There are some special aspects of MeV backgrounds in DUNE due to its target material, its readout technology, its unusual depth, and the fact that it is a target for GeV accelerator-produced neutrinos.  We review some backgrounds that will be unimportant after standard cuts, then provide details on three that will remain important.

We first discuss the expected livetime fraction for DUNE for MeV neutrinos, i.e., when the detector is quiet from high-energy events or their aftermaths.  Following a high-energy event, the produced charge takes $\simeq$ 2--3~ms to drift out of the volume~\cite{Acciarri:2015uup, Strait:2016mof, Acciarri:2016ooe}.  High-energy events can be created by atmospheric muons (0.05 Hz per 10~kton~\cite{Acciarri:2016ooe, Zhu:2018rwc}) and neutrinos from the Fermilab beam ($\simeq 1$ Hz, with duration $\simeq 10\, \mu$s~\cite{Strait:2016mof}, though the expected number of interactions per spill is tiny).  In a conservative case, in which we apply a holdoff of 10~ms every 1~s, this induces a detector deadtime of only $\simeq 1\%$.  As discussed below, this is enough time that any neutrons created have escaped or captured; long-lived beta-decaying isotopes must be treated separately.  The expected exposure of DUNE for MeV events, i.e., when the detector is quiet, is thus nearly the same as the calendar time.

Other neutrino fluxes can be ignored.  The fluxes of MeV atmospheric neutrinos~\cite{Gaisser:2002jj, Abe:2017aap} and the Diffuse Supernova Neutrino Background~\cite{Beacom:2010kk, Bays:2011si} are even smaller than that of $hep$ neutrinos.  The reactor flux~\cite{Ricci:2014qpa, JinpingNeutrinoExperimentgroup:2016nol} is comparable to the $^8$B flux, but the high threshold~\cite{NNDC} for $\bar{\nu}_e$ events on $^{40}$Ar, combined with the low energies of reactor neutrinos, make this irrelevant.  For GeV atmospheric neutrinos, they can be vetoed as above.

\begin{figure*}[t]
\begin{center}
\includegraphics[width=1.9\columnwidth]{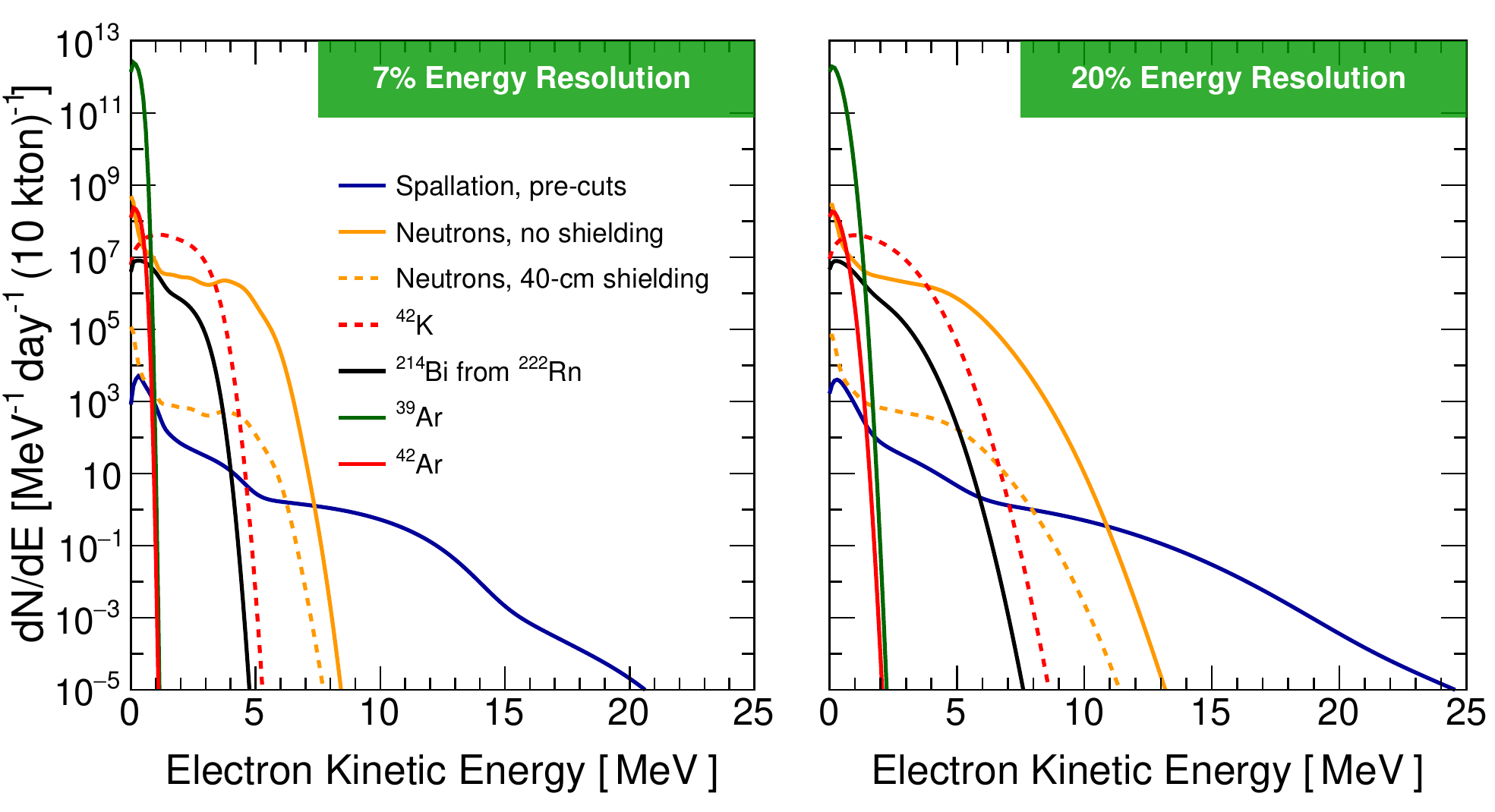}
\caption{Spectra of major possible backgrounds.  Note the large range on the y-axes.  {\bf Left:}  Assuming 7\% energy resolution (as in our main results).  {\bf Right:} Assuming 20\% energy resolution (as in the alternate cases considered below).  The spallation backgrounds are taken from Ref.~\cite{Zhu:2018rwc} and the neutron backgrounds are calculated below.  For $^{39}$Ar and $^{42}$Ar ($^{42}$K) decays, we assume rates of $10^7$ Hz and $10^3$ Hz per module, respectively~\cite{Acciarri:2015uup, Strait:2016mof, Acciarri:2016ooe, Zhu:2018rwc}.  For the $^{222}$Rn activity, we assume that $\sim 10$ mBq/m$^3$ can be achieved in steady state (for comparison, in Super-K the pre-purification level was about $\sim 50$ times worse than that; the post-purification level is a few to several times better than that, depending on depth in the detector~\cite{Takeuchi:1999zq}).}
\label{fig:all_backgrounds}
\end{center}
\end{figure*}

We expect that betas and gammas from intrinsic radioactivities~\cite{Arneodo:2000fa} can be mitigated effectively, leaving a negligible rate inside the fiducial volume and above 5~MeV.  These long-lived radioactivities have low energies, but can effectively reach higher energies if their rate is high and/or the energy resolution is poor.  Standard MeV techniques such as U/Th removal from the liquid, Rn removal from the air, and selection of low-radioactivity material can significantly lower background rates.  Good energy resolution (7\%) will ensure that the end-point energies of both $^{42}$K (3.53~MeV, from the $^{42}$Ar decay chain~\cite{NNDC}) and $^{214}$Bi (3.27~MeV, from the $^{222}$Rn decay chain~\cite{NNDC}) are too low to affect the solar analysis.  We also considered backgrounds due to pileup, i.e., time and space coincidences of multiple events that could mimic a single event of higher energy.  For the decays of $^{42}$Ar ($^{42}$K) or $^{39}$Ar (endpoint 0.57~MeV)\protect{~\cite{Acciarri:2016ooe, Arneodo:2000fa}}, or the capture of external neutrons (even without shielding), the rates are negligible.  Pileup for these and other cases is discussed in detail in Ref.~\cite{Zhu:2018rwc}.

Figure~\ref{fig:all_backgrounds} shows, for initial orientation, the spectra of several major possible backgrounds under two assumptions about energy resolution.  These results show that the most important background above 5 MeV electron energy is due to neutron radiative captures.  These results also show that low-energy radioactivities are much less important, even with poor energy resolution.  If their rates were an order of magnitude higher than our estimates, they still would not have an impact on our analysis. We thus move on to a more detailed discussion of the backgrounds above 5 MeV electron energy.

\begin{figure}[t]
\begin{center}                  
\includegraphics[width=\columnwidth]{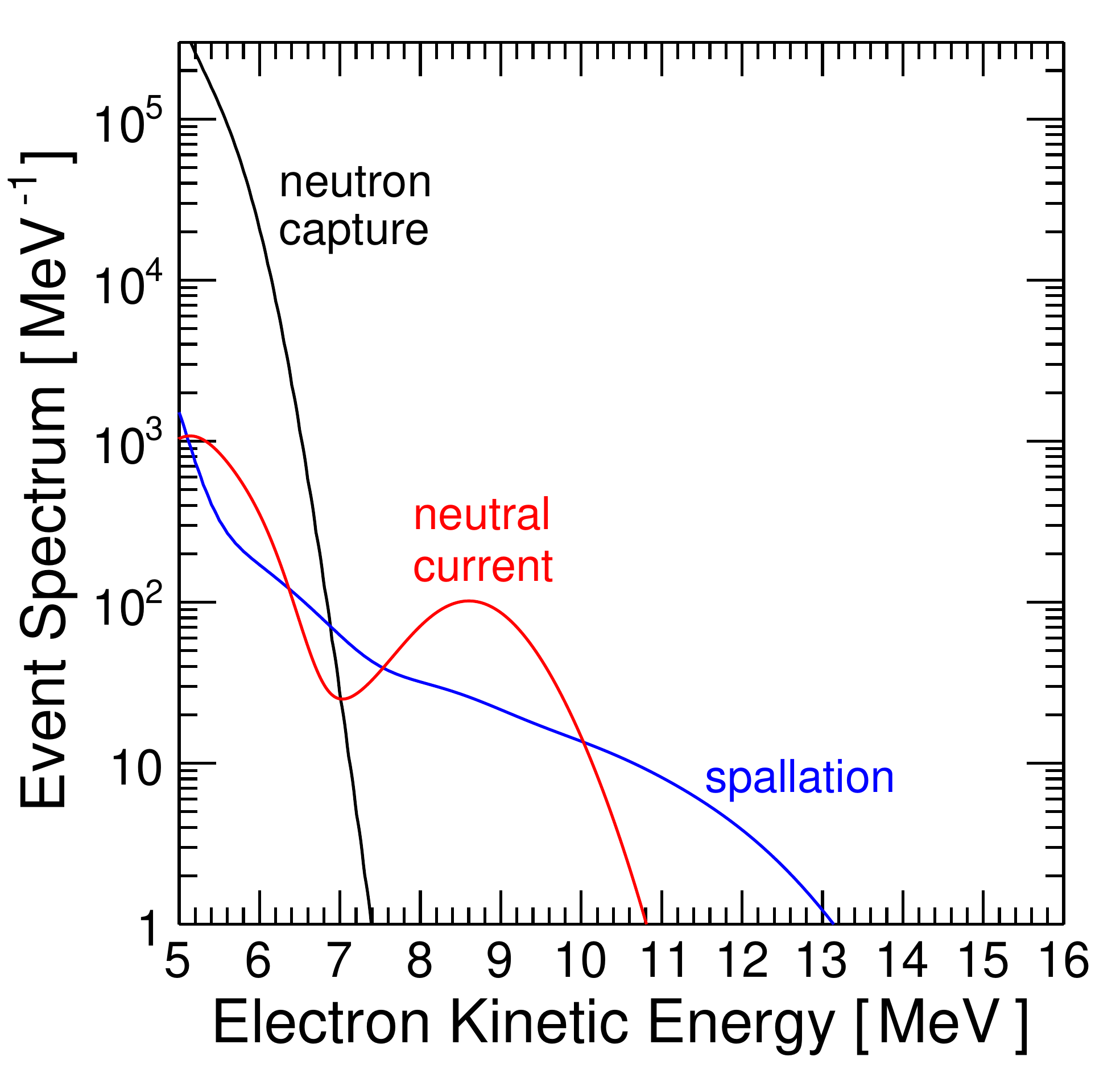}
\caption{Contributions of different sources of background after cuts, for an exposure of 100~kton-year, assuming 40 cm of shielding.}
\label{fig:background_components}
\end{center}
\end{figure}

Figure~\ref{fig:background_components} shows the three components of the background that will remain significant after cuts.  In Fig.~3, the total background yields (after cuts) outside and inside the forward cone are $14.0 \times 10^4$ and $1.9 \times 10^4$, respectively.  As in the main text, we discuss them in order of increasing energy and decreasing importance.  At the end of this section, we comment on the potential impact of particle-identification techniques.

\subsection{Neutron Backgrounds From Radioactivities}
\label{sec:neutron_bck}

The dominant background is due to MeV neutrons originating in the surrounding rock.  (In Ref.~\cite{Zhu:2018rwc}, we show that neutrons produced by muons are much less important; see Fig.~12 there.)  The neutrons are produced in $(\alpha, n)$ reactions generated by $^{238}$U or $^{232}$Th daughter decays as well as through the spontaneous fission of $^{238}$U.  Once neutrons enter the active volume, they mostly elastically scatter with argon.  Within a few ms, they either exit the detector or undergo radiative capture on argon, producing multiple gamma rays with energy up to 6.1~MeV, the total energy release~\cite{Hardell1970, Nesaraja:2016ktw, CapGam}.  Gamma rays then Compton-scatter or pair-produce electrons with energies up to the maximum gamma-ray energy, creating signal-like events.  (External gamma rays from the same processes, which we calculate, are significantly less important, due to the fiducial-volume cut.)

These neutrons cannot be rejected by the fiducial volume cut because neutrons below 5--10 MeV (above which inelastic cross sections become large) do not lose much energy through elastic scattering on argon: neutrons can travel $\sim$ 100~meters in LAr before they capture.  (In the Sudbury Neutrino Observatory phase with pure D$_2$O, the neutron-capture distance was long because capture was inefficient while energy loss through elastic scattering was efficient; here, it is the opposite.)  In our nominal analysis, we assume 40~cm of water or other hydrogenous shielding, which must be kept radiopure but which can be passive.  Roughly, each 10~cm of water reduces the neutron background by somewhat less than an order of magnitude.  Note that comparable sensitivity can be achieved with no shielding and twice the exposure.

This background rate depends on the type of rock surrounding the detector.  The DUNE far detector will be placed at the 4850 feet level in the Homestake Gold Mine~\cite{Strait:2016mof}, located in Lead, South Dakota.  At this level, the rock belongs mostly to the Poorman formation~\cite{Lesko:2011qk, Chan:rock, Heise:2015vza, Strait:2016mof}.  The Poorman rock has an average concentration of $^{238}$U of 3.43~ppm and $^{232}$Th of 7.11~ppm~\cite{rogers1990geology}, which we adopt for our calculation.  This is a conservative choice, because some samples of the Poorman rock, in particular those in the Yates unit, are known to possess a sub-ppm concentration of U/Th~\cite{rogers1990geology}, as is the case for Davis Campus where LUX is hosted~\cite{deViveiros:2010fnb}.  Based on a survey reported in Ref.~\cite{Strait:2016mof}, we neglect the presence of rhyolite, which might increase the radioactive contamination.  Regarding the concentration of other elements, we adopt the average values for the Poorman formation reported in Table 1 in Ref.~\cite{rogers1990geology}. This is also a conservative assumption, because the average water (OH radicals and actual water) content is $\sim 4$\%, whereas some samples show fluctuations up to 10\%, which would lower the neutron level.  The rock composition we adopt is reported in Table~\ref{table:rock_composition}.

\begin{table}[t]
\centering
\vspace{0.8em}
\begin{tabular}{||C{1.5cm}|C{3cm}||}
  \hline\hline
   Element & weight percentage \\[-1pt] \hline
  \hline
   O   &  46.0   \\[-1pt] \hline
   Si  &  24.6   \\[-1pt] \hline
   Fe  &  9.1    \\[-1pt] \hline
   Al  &  5.4    \\[-1pt] \hline
   C   &  3.6    \\[-1pt] \hline
   Mg  &  2.9    \\[-1pt] \hline
   Ca  &  2.7    \\[-1pt] \hline
   K   &  2.7    \\[-1pt] \hline
   S   &  1.9    \\[-1pt] \hline
   Na  &  0.4    \\[-1pt] \hline
   H   &  0.2    \\[-1pt] \hline
   Mn  &  0.2    \\[-1pt] \hline
   Ti  &  0.2    \\[-1pt] \hline
   F   &  0.1    \\[-1pt] \hline
   Th  & trace (7.11 ppm) \\[-1pt] \hline
   U   & trace (3.43 ppm) \\[-1pt] \hline\hline
  total& 100.0 \\[-1pt] \hline
    \hline
\end{tabular}
\vspace{1em}
\caption{The rock composition near the detector~\cite{rogers1990geology}.}
\label{table:rock_composition}
\end{table}

We prepared dedicated simulations to assess the neutron backgrounds.  For neutrons produced through $(\alpha, n)$ interactions, which are the most important, we use the simulation package NeuCBOT~\cite{Westerdale:2017kml} to get the neutron yield and spectrum.  To calculate the neutron rate from spontaneous fission, we approximate the neutron spectrum using using the Watt distribution~\cite{Arneodo:2000fa, shultis2002, Wulandari:2003cr, deViveiros:2010fnb}, and take the neutron yield per fission to be 2~\cite{shultis2002}.  We propagate neutrons through rock, water shielding, the active volume, and finally the fiducial volume of the detector using the simulation package FLUKA~\cite{Ferrari:2005zk, Battistoni:2007zzb}.  We then record all neutron captures in the fiducial volume.  Last, we convert the gamma ray spectrum from neutron captures~\cite{Hardell1970, Nesaraja:2016ktw, CapGam} to an electron spectrum by simulating gamma rays in FLUKA.  We then record the electron kinetic energy for Compton scattering and the total kinetic energy for pair production.  The most important contribution to this background is neutrons of a few MeV.

There are also neutrons produced inside the detector, from $(\alpha,n)$ reactions generated the decay chain of $^{222}$Rn that has diffused into the LAr from the rock.  For a $^{222}$Rn activity of $\sim 1$ mBq/m$^3$ in the liquid, which has been achieved or improved upon in other detectors following air-purification procedures~\cite{Takeuchi:1999zq, Blevis:2003ih}, we find that this background is negligible.  We neglect backgrounds that follow from the decay chains of U/Th in the detector, as LAr can be easily purified~\cite{Franco:2015pha}.


\subsection{Neutrino-Argon Neutral-Current Interactions}
\label{sec:NC_sigma}

Neutrino neutral-current interactions with argon produce an excited state of the argon nucleus:
\begin{equation}
\nu_{e,\mu,\tau} + \, ^{40}\text{Ar} \rightarrow \nu_{e,\mu,\tau} + \, ^{40}\text{Ar}^* \ ,
\label{NC}
\end{equation}
where the de-excitation gamma rays might be tagged.  Because there is only one shell-model calculation of neutral-current cross section~\cite{Raghavan:1986fg}, that is what we adopt.  We calculate the cross section based on Eqs.~(1, 2) and Table~I in Ref.~\cite{Raghavan:1986fg}.  In this case, there are three excited states of argon, at 6.1, 9.6, and 10.42~MeV.  The first two states emit gamma rays at the corresponding excitation energies, whereas the 10.42 MeV state leads to emission of a 6.1-MeV gamma ray.

It is difficult to access the uncertainty on the cross section.  The only shell-model calculation~\cite{Raghavan:1986fg} differs significantly from two RPA calculations~\cite{Kolbe:2003ys, Cheoun:2011zza}.  One of those~\cite{Kolbe:2003ys} has a cross section that is almost a factor of 10 larger than the shell-model calculation.  If that were correct, there would be a chance to turn the neutral-current channel into a signal, which would be valuable because it would detect the $\nu_{\mu,\tau}$ component of the solar flux at full strength, unlike the elastic-scattering channel.  A background to the neutral-current channel would be charged-current events with the electron below threshold but the gamma-ray above.  Further theoretical and experimental investigation is needed.


\subsection{Cosmic-Ray Muon-Induced Backgrounds}

When a cosmic-ray muon passes through the detector (for each module, the total rate is 0.05~Hz~\cite{Acciarri:2016ooe, Zhu:2018rwc}), it can rarely produce a large shower along its track.  Secondary particles, especially pions and neutrons, may break argon nuclei and produce beta-unstable isotopes.  These isotopes subsequently beta decay and produce electrons, mimicking neutrino signals.  They are called spallation backgrounds (see, e.g., Refs~\cite{Li:2014sea, Li:2015kpa, Li:2015lxa}).

\begin{figure*}[t]
\begin{center}
\includegraphics[width=1.9\columnwidth]{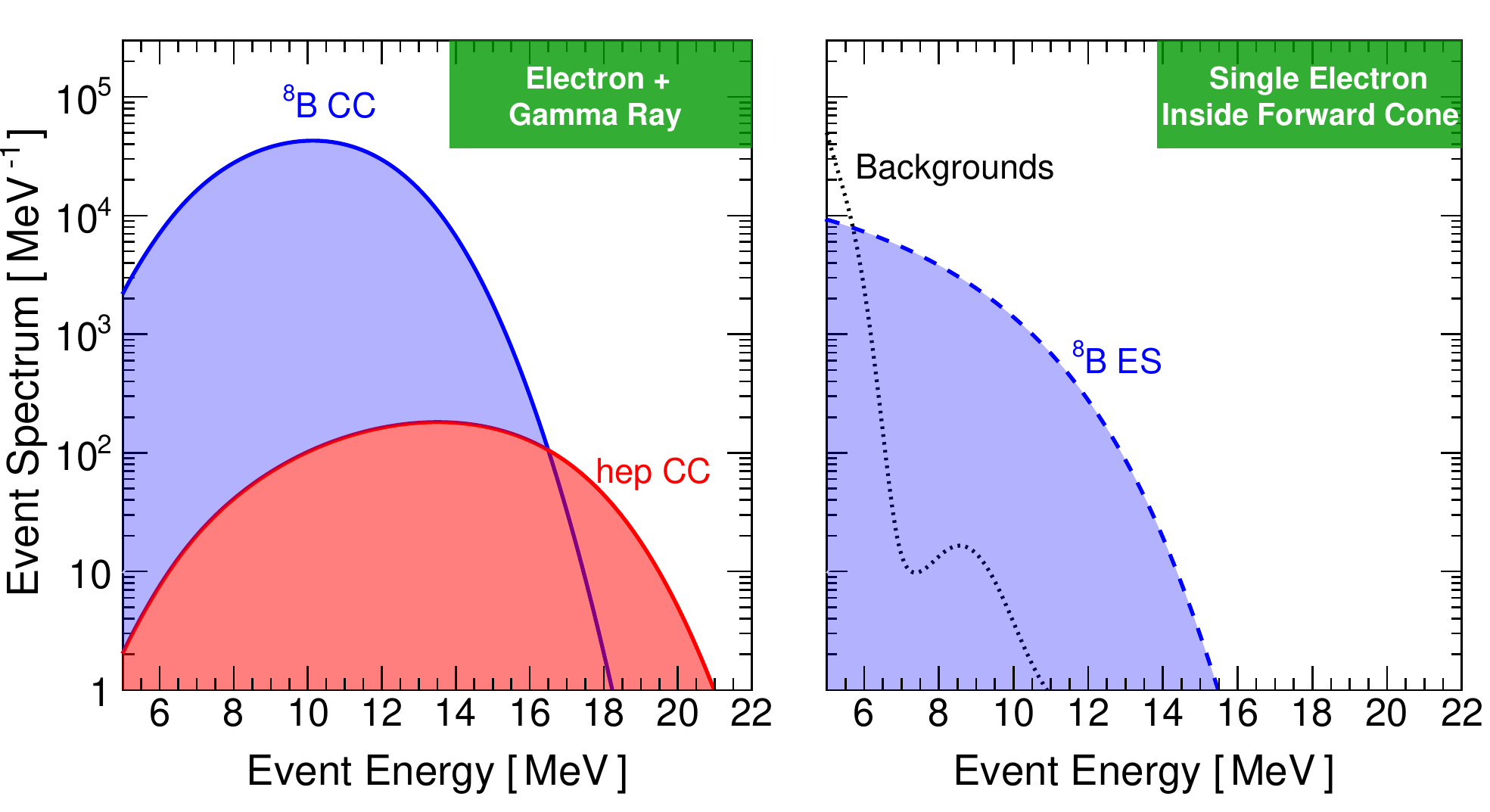}
\caption{Same as Fig.~3, but assuming that the full neutrino energy for $\nu_e + \, ^{40}{\rm Ar}$ events could be reconstructed instead of only the energy of the leading electron, which would allow separating $\nu_e + \, ^{40}{\rm Ar}$ events (left panel) from backgrounds and $\nu_{e,\mu,\tau} + e^-$ events (right panel).}
\label{fig:event_spectrum_ideal_7percent}
\end{center}
\end{figure*}

\begin{figure*}
\vspace{-2em}
\begin{center}
\includegraphics[width=1.9\columnwidth]{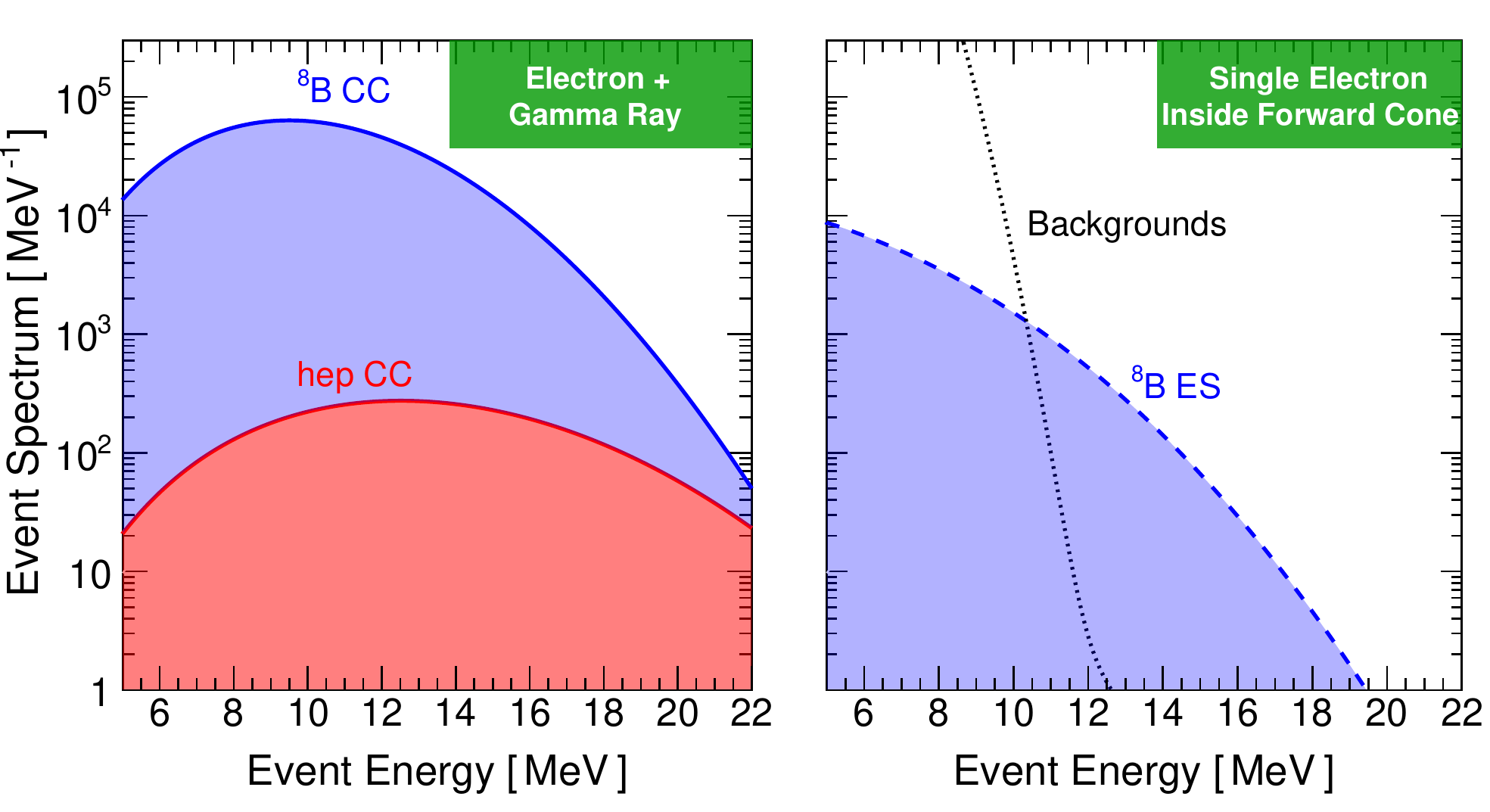}
\caption{Same as Fig.~\ref{fig:event_spectrum_ideal_7percent}, but with no shielding and with an energy resolution of 20\% instead of 7\%.}
\label{fig:event_spectrum_ideal_20percent}
\end{center}
\end{figure*}

Due to the low muon rate in DUNE, spallation backgrounds can be efficiently reduced by simple cuts to reject delayed events following muons.  A per-module cut of duration 250~ms, followed by a cylinder cut of radius 2.5~m and duration 40~s, induces a deadtime of only $5\%$.  The first cut decimates short-lived radioactivities (and neutrons, which capture in a few ms), and the second greatly reduces isotopes that are longer-lived.  Above 10~MeV, where spallation is most relevant, these cuts reduce the spallation background by $\simeq 99.5\%$, which is taken into account in Fig.~3.  More sophisticated cuts can further reduce the backgrounds while maintaining low deadtime~\cite{Zhu:2018rwc}.

Muons that interact only in the nearby rock do not pose a problem.  The neutrons they produce are subdominant to those produced by radioactivities in the rock~\cite{Zhu:2018rwc}.  Showers induced by the muons may enter the detector, producing spallation isotopes, but these instances will be reduced by the fiducial-volume cut and our simulations show that they can be identified by the presence of charged shower particles entering the detector.  Simple cuts, similar to those above, will be sufficient.


\subsection{Particle-Identification Techniques}
If signals and backgrounds could be separated by particle-identification techniques, that would increase the sensitivity of the solar neutrino program.  As noted above, $\nu_{e,\mu,\tau} + e^-$, $\nu_e + \, ^{40}{\rm Ar}$, and neutron-capture events would have one electron, an electron with gammas, and multiple gammas, respectively.  For gamma rays, the radiation length is 14~cm and Compton-scattering dominates~\cite{Patrignani:2016xqp}, so these event classes should be distinct.

Once a single electron from a $\nu_e + \, ^{40}{\rm Ar}$ event above 5~MeV has triggered the detector, it should be possible to reconstruct, using time and space coincidences, at least some of the lower-energy electrons produced by gamma rays in association.  These gamma rays carry a substantial amount of energy;  see Table ~\ref{table:CC_cross_section}, where the $\Delta E_i$ values are 2.3 MeV or larger.  In principle, this technique could be even employed for events in which the primary and secondary electrons are all individually below 5~MeV, but above in total.  The presence of these gamma rays would strongly favor that an event is due to $\nu_e + \, ^{40}{\rm Ar}$.  For $\nu_{e,\mu,\tau} + e^-$ events, gamma rays can be produced through bremsstrahlung, but that is suppressed because the event energies are well below the critical energy of 32 MeV~\cite{Patrignani:2016xqp, Amoruso:2003sw, Acciarri:2017sjy}.  And for neutron capture, it typically produces several gamma rays, with total energy 6.1 MeV ~\cite{Hardell1970, Nesaraja:2016ktw, CapGam}.  However, induced electrons above 5 MeV are dominantly produced by a single 5.6-MeV gamma ray; we assume that the remaining 0.5 MeV in gamma rays is undetectable, even in coincidence.  Further study is needed to assess the the potential to separate these event classes from each other (as well as from the less-important backgrounds due to $\nu_{e,\mu,\tau} + \, ^{40}\text{Ar}$ events and spallation beta decays).

The impact of particle-identification techniques for $\nu_e + \, ^{40}{\rm Ar}$ events would be even greater if the energy deposited via gamma rays could be estimated.  In our main analysis, we conservatively assumed that the only detectable energy in an event is due to single electrons above 5~MeV.   Figure~\ref{fig:event_spectrum_ideal_7percent} shows how Fig.~3 would be improved if, for $\nu_e + \, ^{40}{\rm Ar}$ events, the measurable energy was the full neutrino energy instead of only the leading electron energy.  This would allow separation of $\nu_e + \, ^{40}{\rm Ar}$ events from $\nu_{e,\mu,\tau} + e^-$ events and backgrounds.  Perfect separation is an unrealistic ideal, but this figure gives a sense of the benefits that would follow from being able to reconstruct at least some of the energy deposited in gamma rays.  We use the full neutrino energy to account for both the measurable energy deposited by gamma rays as well for a correction of $Q_\text{gs} = 1.504$~MeV for the nuclear threshold; once it is known that the event is due to $\nu_e + \, ^{40}{\rm Ar}$, this correction is appropriate.

This capability would lead to multiple benefits.  First, it would improve the precision of the measurements.  Second, it would allow testing the survival probability over a wider energy range (including detecting the upturn at low energies), increasing sensitivity to new physics.  Third, it would increase the robustness of the measurements to less favorable detector properties than we have assumed.  For example, Fig.~\ref{fig:event_spectrum_ideal_20percent} shows how Fig.~\ref{fig:event_spectrum_ideal_7percent} would be altered by an energy resolution of 20\% instead of 7\% and without any shielding.  In that case, the $\nu_e + \, ^{40}{\rm Ar}$ results would be better than we have assumed, though the $\nu_{e,\mu,\tau} + e^-$ results would be worse.  The latter could be repaired by improving either energy resolution or shielding or by combining DUNE results on $\nu_e + \, ^{40}{\rm Ar}$ with Hyper-Kamiokande results on $\nu_{e,\mu,\tau} + e^-$.


\section{ANALYSIS DETAILS}
\label{sec:analysis}

We provide technical details on our calculations of the neutrino fluxes at the detector and how the fitted parameters are deduced from the observables.


\subsection{Main Analysis}

As shown in the main text, we separate the events into two categories: inside and outside the forward cone, which is centered with the direction away from the Sun and has a half-angle of 40$^\circ$.  Each category is further divided into day and night events, each having an exposure of 50 kton-year. The day and night events are further divided into 13 energy bins, with the following extrema: 5.0, 5.5, 6.0, 6.5, 7.0, 8.0, 8.5, 9.0, 10.0, 11.0, 12.0, 13.0, 14.0, 20.0 MeV.  The night events are also divided into 10 equally spaced $\cos\theta_z$ bins with a width of 0.1, where $\theta_z$ is the zenith angle of the Sun.  We checked that variations in the binning do not significantly change our results.

The survival probability of $\nu_e$ for the day events is calculated using the equation
\begin{equation}
P_{ee}^{\text{day}}(E_\nu)=\int dr\,\phi_x(r)\sum_{i=1}^3\left|U_{ei}\right|^2\left|U_{ei}^{\odot}(E_\nu,r)\right|^2\,,
\label{eq:prob_day}
\end{equation}
where $U_{ei}$ and $U_{ei}^{\odot}$ are the mixing matrix elements in vacuum and in matter respectively, $r$ is the radial distance from the center of the Sun and $\phi_x(r)$ is the radial distribution of neutrinos produced in the Sun ($x=$ {\it hep}, $^8$B).  We take the electron density and radial distribution of $^8$B and {\it hep} neutrinos from the solar model BS05(OP)~\cite{Bahcall:2004pz, Bahcall_website}.    For night events we use the equation
\begin{equation}
P_{ee}^{\text{night}}(E_\nu,\theta_z)=\int dr\,\phi_x(r)\sum_{i=1}^3\left|U_{ei}^\odot(E_\nu,r)\right|^2P^{\oplus}_{ie}(E_\nu,\theta_z)\,,
\label{eq:prob_night}
\end{equation}
where $P^{\oplus}_{ie}$ is the probability for the $i-$th mass eigenstate to be detected as an electron neutrino after propagating through Earth. $P^{\oplus}_{ie}$ is calculated according to the method proposed in Ref.~\cite{Lisi:1997yc}. In particular, for each zenith angle bin we convolve $P_{ie}^{\oplus}(\theta_z)$ with the solar exposure $W(\theta_z)$ calculated for the latitude of DUNE site
\begin{equation}
\langle P^{\oplus}_{ie}(E_\nu)\rangle=\frac{\int_{\theta_{z,1}}
^{\theta_{z,2}}W(\theta_z)P^{\oplus}_{ie}(E_\nu,\theta_z)d\theta_z}{\int_{\theta_{z,1}}
^{\theta_{z,2}}W(\theta_z)d\theta_z}\,,
\label{average_Pnight}
\end{equation}
where we take the analytical expression for $W(\theta_z)$ from Ref.~\cite{Lisi:1997yc}.

We generate our simulated number of data events in the $i-$th bin $N_i^{\text{exp}}=N_i^{\text{exp},^8\text{B}}+N_i^{\text{exp},\text{hep}}$ assuming the best fit of solar experiments, i.e. $\Delta m^2_{21} = 4.85 \times 10^{-5}$ eV$^2$ and $\sin^2\theta_{12}=0.308$.  We perform a scan of the parameter space identified by the following parameters: $\sin^2\theta_{12}$, $\Delta m^2_{21}$, $\phi(^8$B), $\phi(hep)$.  For each point of this four-dimensional parameter space we calculate the theoretical expectation of events in the $i-$th bin
\begin{equation}
\begin{split}
N_i^{\text{th}}(\sin^2\theta_{12},\Delta m^2_{21},\alpha_{^8\text{B}},\alpha_{hep})=\\
N_i^{\text{th},^8\text{B}}(\sin^2\theta_{12},\Delta m^2_{21})[1+\alpha_{^8\text{B}}]\,+\,\\
N_i^{\text{th},hep}(\sin^2\theta_{12},\Delta m^2_{21})[1+\alpha_{hep}]\,,
\end{split}
\label{eq:Nth}
\end{equation}
where $\alpha_{^8\text{B}}$ and $\alpha_{hep}$ represent the deviations from the standard values of $\phi(^8\text{B})$ and $\phi(hep)$, and we calculate the $\chi^2$ as
\begin{equation}
\begin{split}
\chi^2(\sin^2\theta_{12},\Delta m^2_{21},\phi(^8\text{B}),\phi(hep))=\\
\sum_i^{N_{\text{bins}}}\left(\frac{N_i^{\text{exp}}-N_i^{\text{th}}(\sin^2\theta_{12},\Delta m^2_{21},\alpha_{^8\text{B}},\alpha_{hep})}{\sqrt{N_i^{\text{exp}}+B_i}}\right)^2\,,
\end{split}
\label{eq:chi2}
\end{equation}
where $B_i$ is the background contribution.  We do not use any priors on the four parameters and we do not include energy or zenith-angle dependent systematics.  The allowed regions in the space ($\sin^2\theta_{12},\Delta m^2_{21}$) are obtained by marginalizing over the other parameters and by adopting the 1, 2, and 3-$\sigma$ levels for 2 d.o.f., corresponding to $\Delta \chi^2=$ 2.3, 6.2, and 11.8.


\subsection{Day-night Asymmetry}
\label{sec:day-night}

\begin{figure}[t]
\begin{center}
\includegraphics[width=\columnwidth]{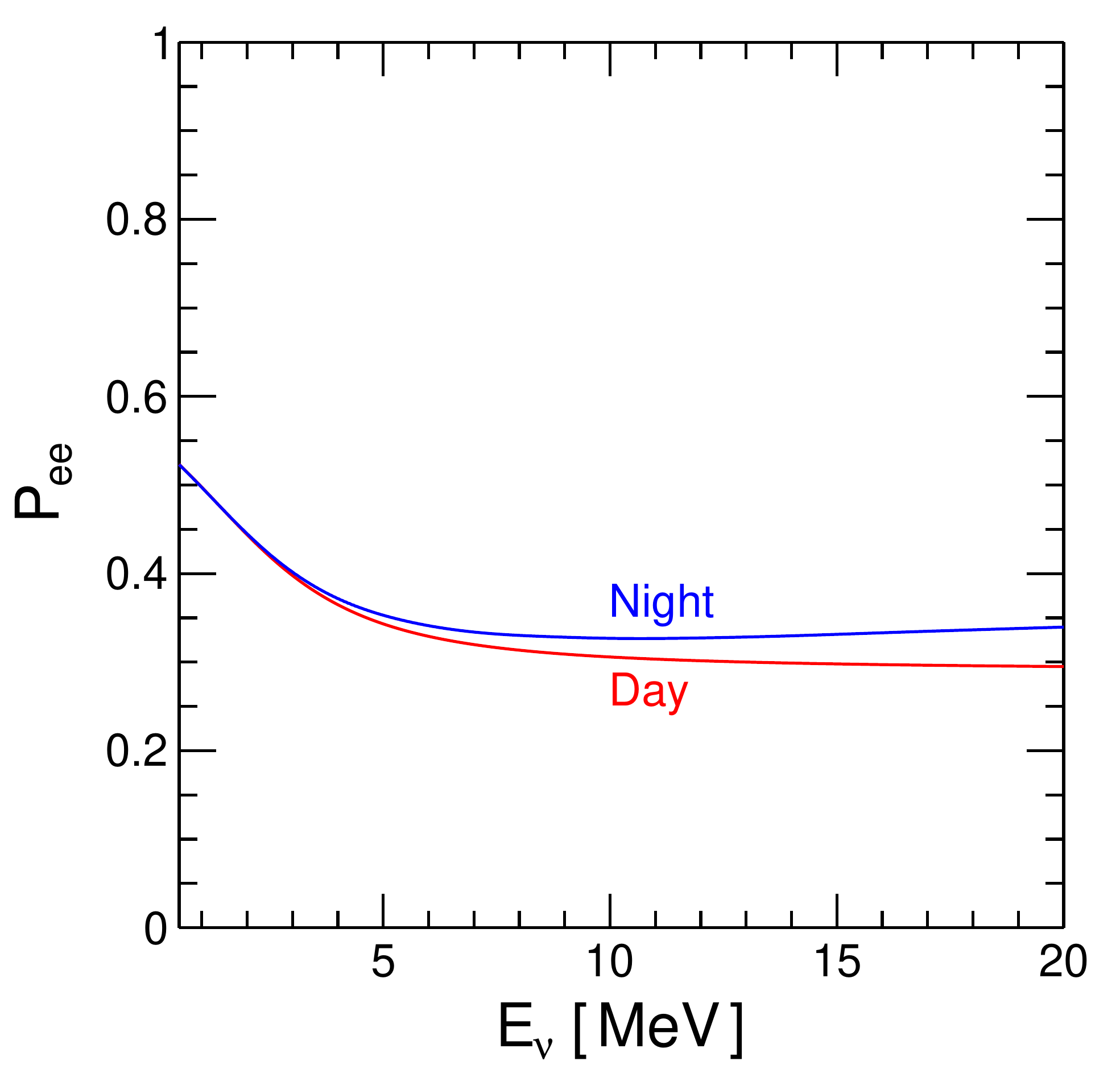}
\caption{The day and night oscillation probabilities as a function of energy, calculated for the solar best fit $\Delta m^2_{21}=4.85\times10^{-5}$ eV$^2$.}
\label{fig:daynight_prob}
\end{center}
\end{figure}

\begin{figure}[t]
\begin{center}
\includegraphics[width=\columnwidth]{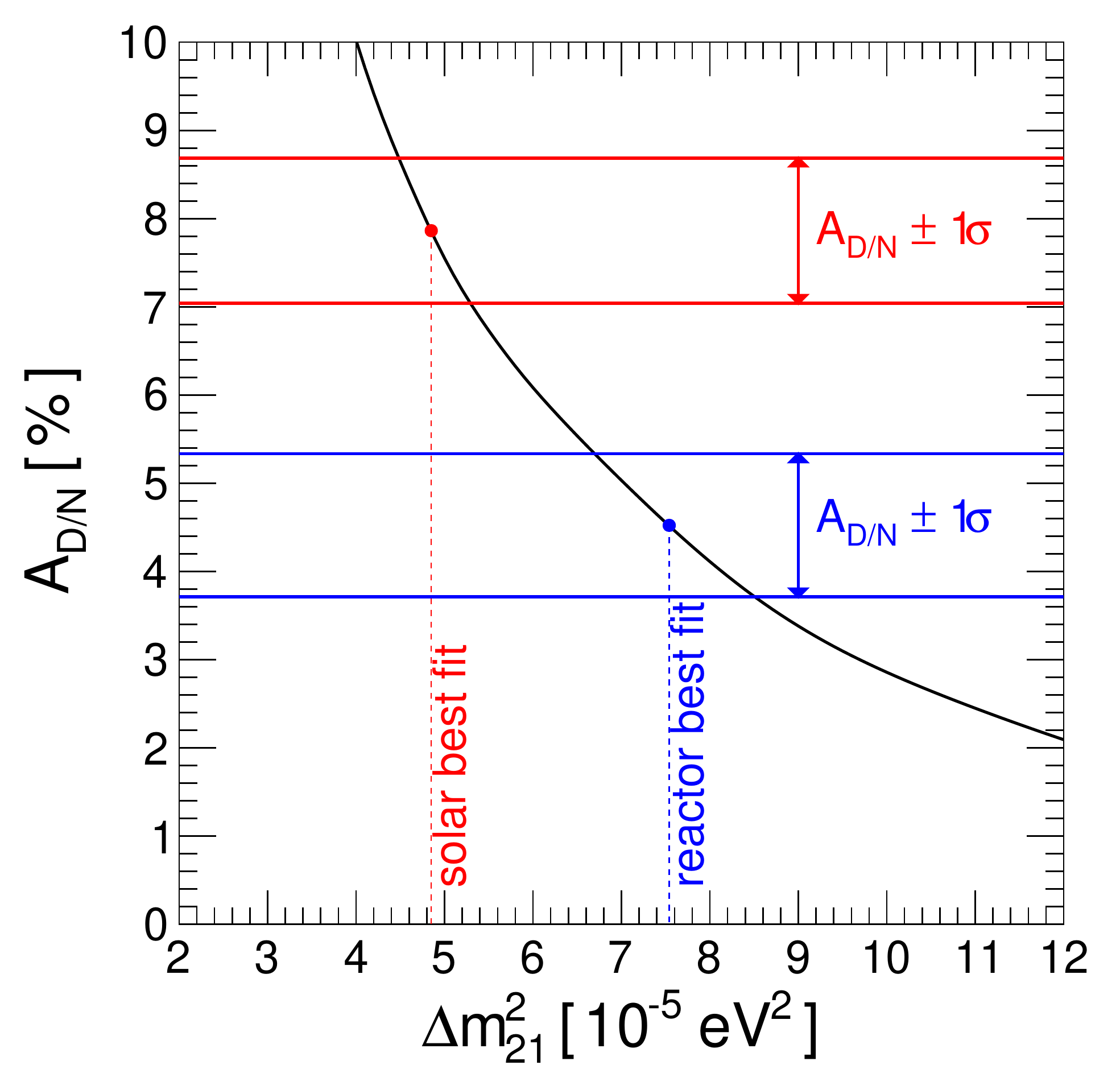}
\caption{The day-night asymmetry as a function of $\Delta m^2_{21}$.  The red (blue) band represents the 1-$\sigma$ statistical uncertainties on $A_{D/N}$ for the solar (KamLAND) best fit.}
\label{fig:daynight_vs_dm2}
\end{center}
\end{figure}

Figure~\ref{fig:daynight_prob} shows the survival probability of electron neutrinos for both day and night assuming the solar best fit, i.e. $\Delta m^2_{21}=4.85\times 10^{-5}$ eV$^2$. The former is obtained using Eq.~\ref{eq:prob_day}, averaging over the production point distribution of $^8$B neutrinos. The latter is calculated using Eq.~\ref{average_Pnight} and smearing in energy, in order to remove some residual fast variations. The difference between the probabilities increases linearly with energy and it reaches $\sim15\%$ at 16~MeV.  However, most of the detected neutrinos have an energy around 10~MeV, for which the difference is $\sim7\%$.  The upturn, i.e., the rise of $P_{ee}$ going below $E_\nu<5$ MeV, is hardly observable in DUNE.  In the next paragraph we perform a more precise evaluation of the day-night effect and its significance. A more detailed calculation and discussion can be found in Ref.~\cite{Ioannisian:2017dkx}, where similar results are obtained.

The significance of the day-night asymmetry is evaluated separately from the main analysis.  In each of the 13 energy bins, we have the expected signal rate (charged-current events outside the forward cone only) during the day $s^i_D$, during the night $s^i_N$, and the background rate (other events outside the forward cone, including elastic-scattering) during the day and night $b^i_D = b^i_N$.  Then, in each bin, we evaluate the day-night asymmetry
\begin{equation}
A_{D/N, i}=\frac{\phi_{D,i}-\phi_{N,i}}{\frac{1}{2}(\phi_{D,i}+\phi_{N,i})} = \frac{s_{D,i}-s_{N,i}}{\frac{1}{2}(s_{D,i}+s_{N,i}+2b_{D,i})} .
\end{equation}
We calculate the total asymmetry $A_{D/N}$ through a weighted average,
\begin{equation}
A_{D/N} = \frac{\sum^{13}_{i=1} A_{D/N,i}/\sigma ^2_{A,i}}{\sum^{13}_{i=1}1/\sigma ^2_{A,i}} ,
\end{equation}
where the weight of each bin is its statistical uncertainty, which includes both signal and background.  This gives the total uncertainty on the day-night asymmetry as
\begin{equation}
\sigma_A = \frac{1}{\sqrt{\sum^{13}_{i=1}1/\sigma ^2_{A,i}}} .
\end{equation}
For the solar best-fit parameters, this gives
\begin{equation}
A_{D/N} = -(7.67 \pm 0.74) \%.
\end{equation}

At the location of DUNE, the day and night exposures differ by a few percent~\cite{daytime_models}, which would have to be taken into account in an analysis of real data (along with shutdown periods).  In effect, we correct for this by assuming equal exposures, though we do use the correct location of DUNE to determine neutrino trajectories through Earth.  Once corrected for, this has a negligible effect on the statistical uncertainties.  Similarly for correcting for the eccentricity of Earth's orbit.

Figure~\ref{fig:daynight_vs_dm2} shows the calculated asymmetry as a function of $\Delta m^2_{21}$.  We note that $A_{D/N} \propto E_\nu/\Delta m^2_{21}$, so that the effect is enhanced at high energies, though its significance depends also on the falling statistics.  Further, that symmetric uncertainties on $A_{D/N}$ lead to somewhat asymmetric uncertainties on $\Delta m^2_{21}$.

Finally, we emphasize that the day-night sensitivity of DUNE is greater than that of Super-K~\cite{Abe:2016nxk}.  Part of the reason is the increased statistics, due to the much larger cross section, but the asymmetry itself is also larger.  In DUNE, higher neutrino energies are emphasized, due to the stronger energy dependence of the cross section, as well as the larger difference between neutrino and electron energy.  Also in DUNE, the day-night asymmetry will be measured through a charged-currrent channel, avoiding dilution due to the $\nu_{\mu, \tau}$ component of the flux.  Last, the tighter relation between neutrino and electron energy reduces the smearing of the day-night signal between different energy bins.


\section{EFFECTS OF INPUT CHOICES}
\label{sec:other_inputs}

We report how our results change with different assumptions about the inputs.

\begin{figure}[t]
\begin{center}                
\includegraphics[width=\columnwidth]{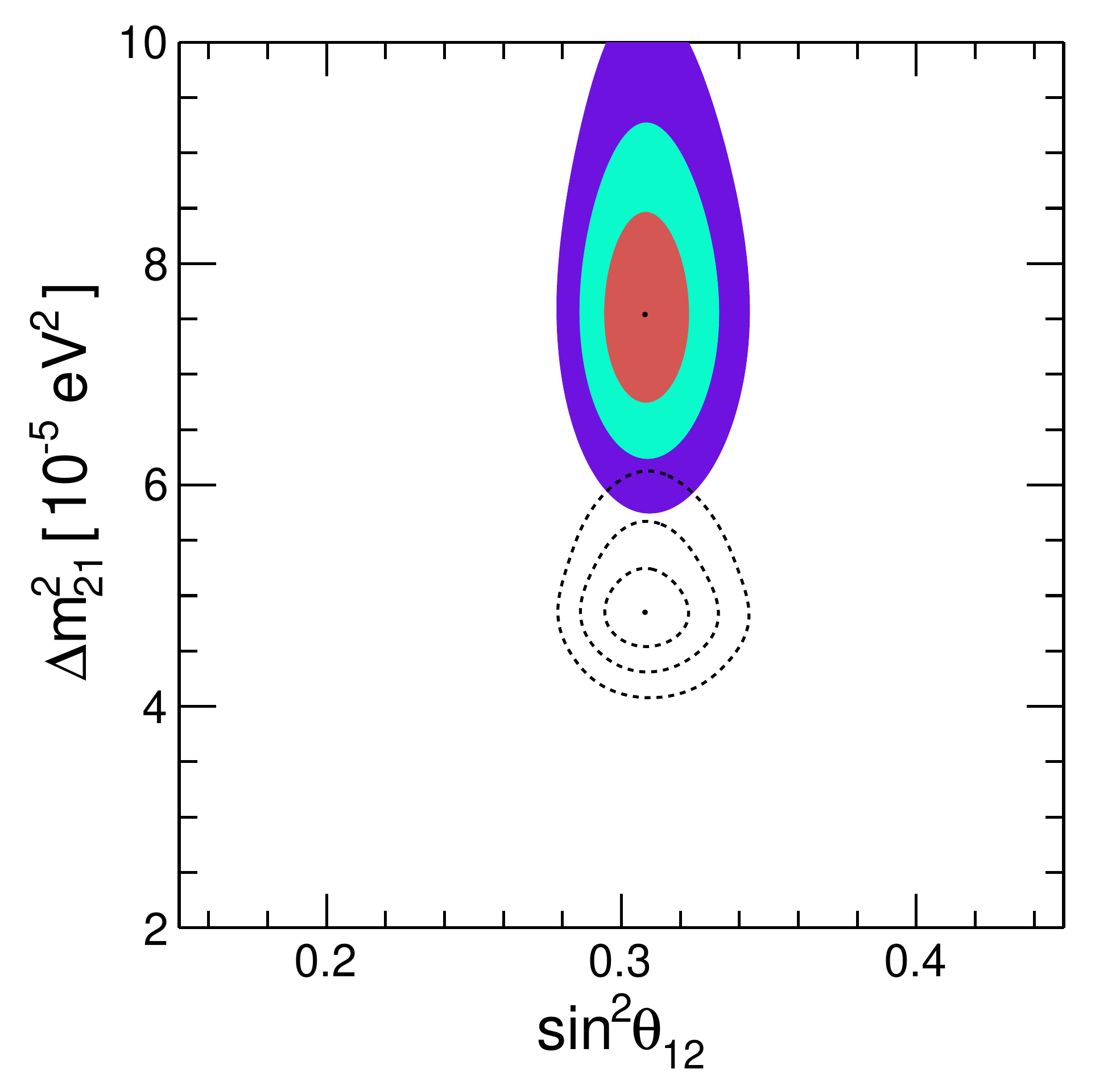}        
\caption{Allowed regions at 1, 2, 3 $\sigma$ obtained assuming $\Delta m^2_{21} = 7.54 \times 10^{-5}$ eV$^2$, the global best fit value, including reactor data.  The dashed contours are for the solar-only best fit value, as in Fig.~2.}
\label{fig:global_parameters}
\end{center}
\end{figure}

\begin{figure*}[t]
\begin{center}
\includegraphics[width=\columnwidth]{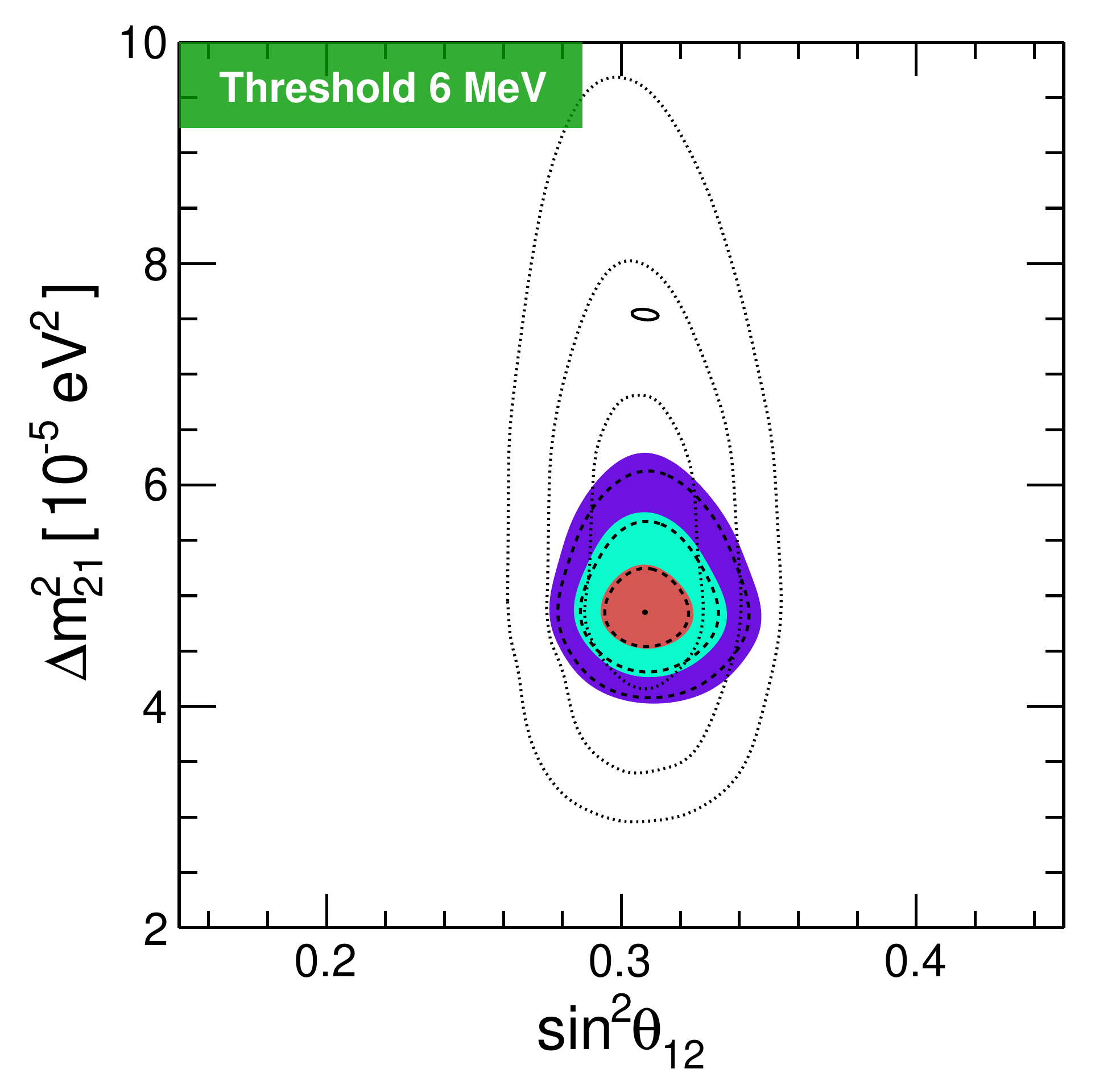}
\hspace{0.25cm}
\includegraphics[width=\columnwidth]{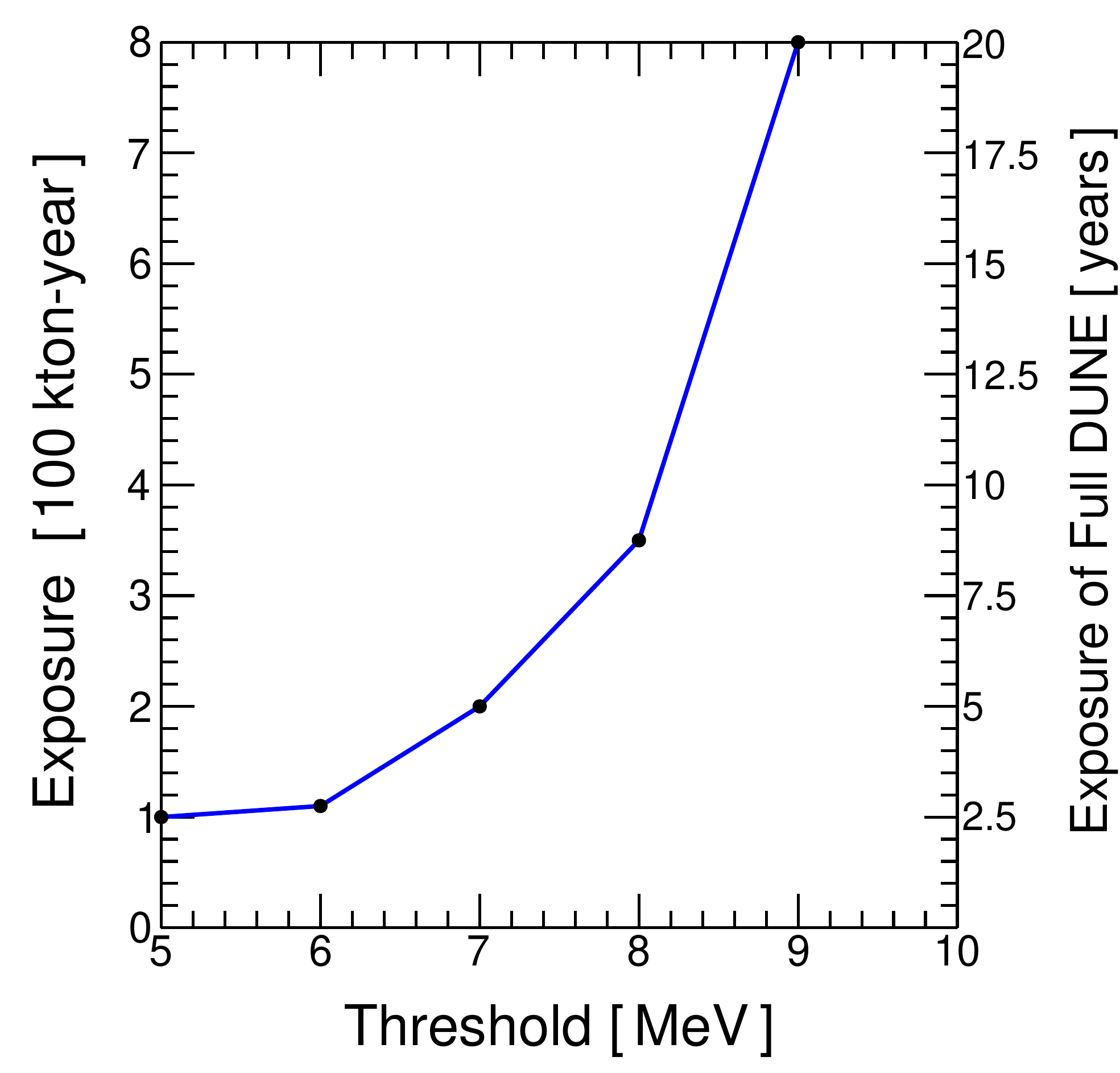}
\caption{{\bf Left:} Allowed regions at 1, 2, 3-$\sigma$ for a detection threshold of 6 MeV.  The dashed lines are for our nominal case, with a threshold of 5~MeV (effectively $\simeq 5.8$~MeV), for which results are shown in Fig.~2.  The dotted contours refer to the present sensitivity and the solid contour to the expected sensitivity of JUNO (see Figs.~1 and~2).  {\bf Right:} With varying analysis threshold, the exposure required to obtain comparable mixing-parameter sensitivity to that of our nominal case.}
\label{fig:higher_threshold}
\end{center}
\end{figure*}


\subsection{Choice of Best-Fit Mixing Parameters}
\label{sec:parameters}

In the main analysis, we generated our simulated data set assuming the best-fit value of $\Delta m^2_{21}$ from solar experiments, to study the capability of DUNE to resolving the long-standing tension with KamLAND.  If we assume that there is no new effect in the propagation of solar neutrinos, the best fit of DUNE is expected to be the same as that of KamLAND.  We thus also generate a simulated data set for $\Delta m^2_{21} = 7.54 \times 10^{-5}$ eV$^2$.

Figure~\ref{fig:global_parameters} shows the allowed regions at 1, 2, 3-$\sigma$ in the mixing-parameter plane.  The precision on $\sin^2\theta_{12}$ is basically unchanged with respect to Fig.~2.  However, now the 3-$\sigma$ constraint on $\Delta m^2_{21}$ is about three times larger than what reported in Fig.~2, due to the smaller day-night asymmetry.


\subsection{Assumed Detector Properties}
\label{sec:detector}

The detection threshold, nominally 5~MeV, depends on two factors: the ability of the data-acquisition system to register, recognize, and record signals in the presence of noise, and the rate of background events in this energy range.  The first  seems achievable, but remains to be demonstrated~\cite{Acciarri:2016ooe}.  Regarding backgrounds, we show in Fig.~3 that even with shielding, neutrons are a significant background in the energy range 5--6 MeV.  Figure~\ref{fig:higher_threshold} (left panel) shows that if we increase the analysis threshold to 6 MeV, the sensitivity barely changes.

To assess the impact of reduced shielding, we first consider how the mixing-parameter sensitivity depends on changes to the analysis threshold for any reason, then discuss shielding specifically.  Figure~\ref{fig:higher_threshold} (right panel) shows how much the runtime must be increased to obtain mixing-parameter sensitivity comparable to that of our nominal case, with a 5-MeV threshold, 40-cm shielding, and an exposure of 100~kton-year (2.5 years of four modules, or 5~years of two modules).  For increases of up to a few MeV, for which the spectra shown in Fig.~3 fall off slowly, the increases in required runtime are reasonable.  Beyond that, the spectra fall steeply, and the required runtime increases quickly.

We stress that reasonable changes in the analysis threshold only affect the sensitivity through statistics, which can be restored with a longer runtime.  This is because $\sin^2\theta_{12}$ is determined from the electron-neutrino survival probability, which is near-constant at these energies, and because $\Delta m^2_{21}$ is determined from the day-night asymmetry, which increases with energy.

Reductions in the assumed shielding would increase the required analysis threshold.  Because the neutron background spectrum is steep (see Fig.~\ref{fig:background_components} for details), this correspondence is sharp.  With 40, 30, 20, 10, or 0~cm of shielding, the effective analysis threshold is $\simeq$ 5.8, 6.2, 6.5, 6.9, or 7.2~MeV.  Even without shielding, DUNE could make world-leading measurements of solar neutrinos; to achieve comparable sensitivity to our nominal case, the required exposure would have to increase by only a factor $\simeq 2$.  A low threshold is also preferred to allow tests of shape distortions in the spectrum and to enhance particle-identification techniques.

A closely related point is energy resolution, because it smears background spectra from lower to higher energies.  If the energy resolution could be made better than the expected 7\%, it would help the sensitivity somewhat and might help particle-identification techniques more.  If the energy resolution is worse than expected, the sensitivity will be degraded.  In Refs.~\cite{Acciarri:2015uup, Strait:2016mof, Acciarri:2016ooe}, an energy resolution of $\simeq 20\%$ is considered if, contrary to expectations, the electron lifetime in LAr is short and corrections for this cannot be made.  With this resolution and 40 cm of shielding to reduce neutron backgrounds, sensitivity compared to that of our nominal case would be obtained if the exposure were increased by only a factor $\simeq 2$.  However, with this resolution and no shielding, the ability to measure solar neutrinos would largely be lost.  At least one, and ideally both, of resolution and shielding, must be prioritized.

An enhanced light-detection system could help in several ways, including refining triggering; better determining the $t_0$ of events, important for energy resolution; and particle identification.  This would also benefit supernova neutrino detection and perhaps other programs.

For angular resolution, though we assumed a smearing of $25^\circ$, following ICARUS~\cite{Arneodo:2000fa}, we conservatively defined the forward cone to have a half-angle of $40^\circ$.  Further, we find that the results are not sensitive to reasonable variations from this. 

Calibration of the full detector in the MeV range will be challenging, but is important to control systematics.  On the experimental side, this will be aided by the extensive knowledge of techniques built up by earlier MeV solar and reactor neutrino experiments.  Calibration sources will likely include muon decays at rest, spallation decays (where the long time between muons, $\sim 20$~s in each module, will allow clean samples), and neutron captures (where the high energy release, 6.1~MeV, makes the high-energy edge of the spectrum observable).  These will help calibrate the energy scale, energy resolution, and absolute size of the fiducial volume.  On the analysis side, control of systematics will be aided by the use of ratios (day-night, as well as the comparison of $\nu_e + \, ^{40}$Ar and $\nu_{e,\mu,\tau} + e^-$ spectra).


\subsection{Impact of Cross-Section Uncertainty}
\label{sec:impact_CC_uncertainty}

\begin{figure*}[t]
\begin{center}
\includegraphics[width=\columnwidth]{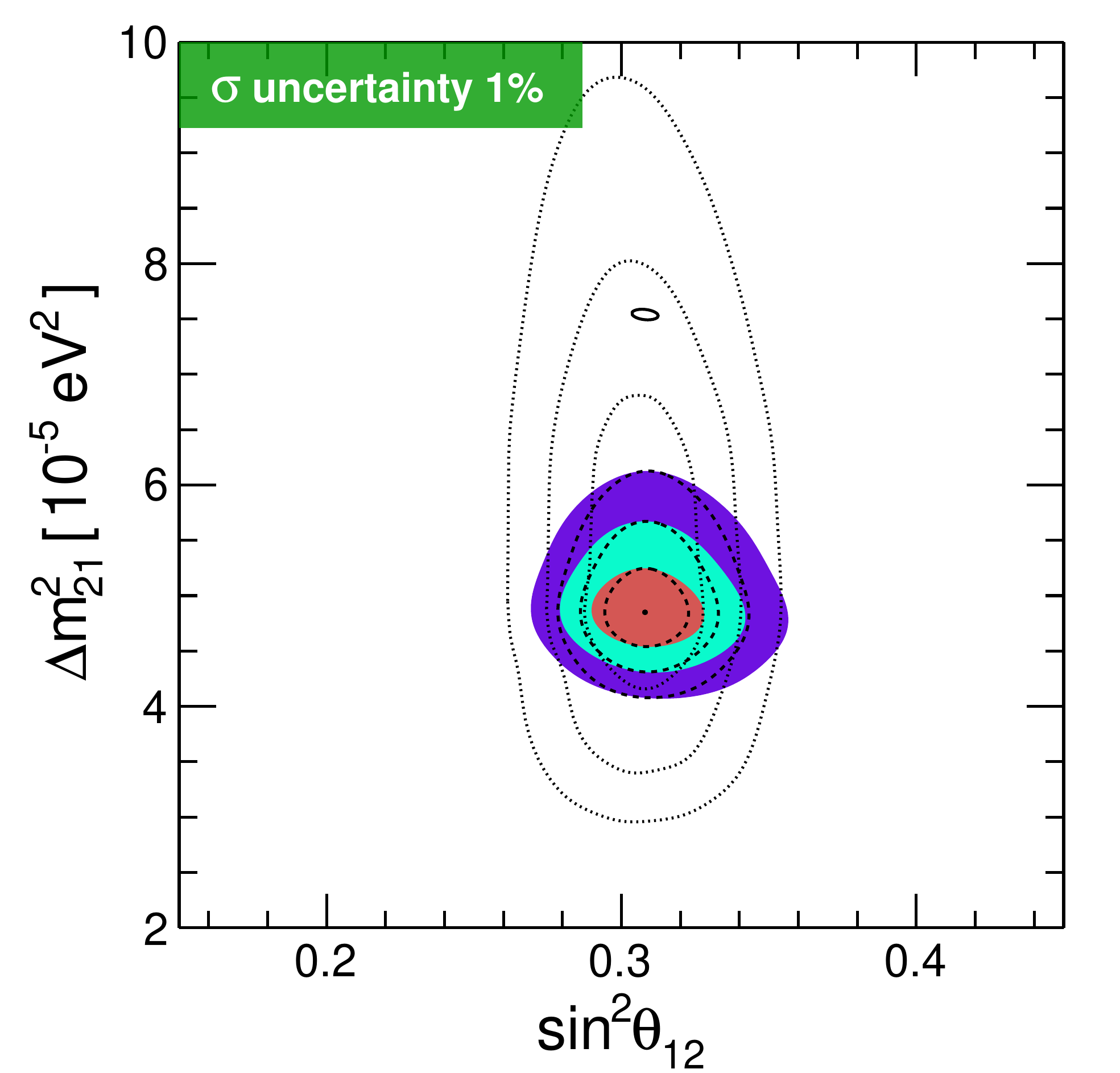}
\hspace{0.25cm}
\includegraphics[width=\columnwidth]{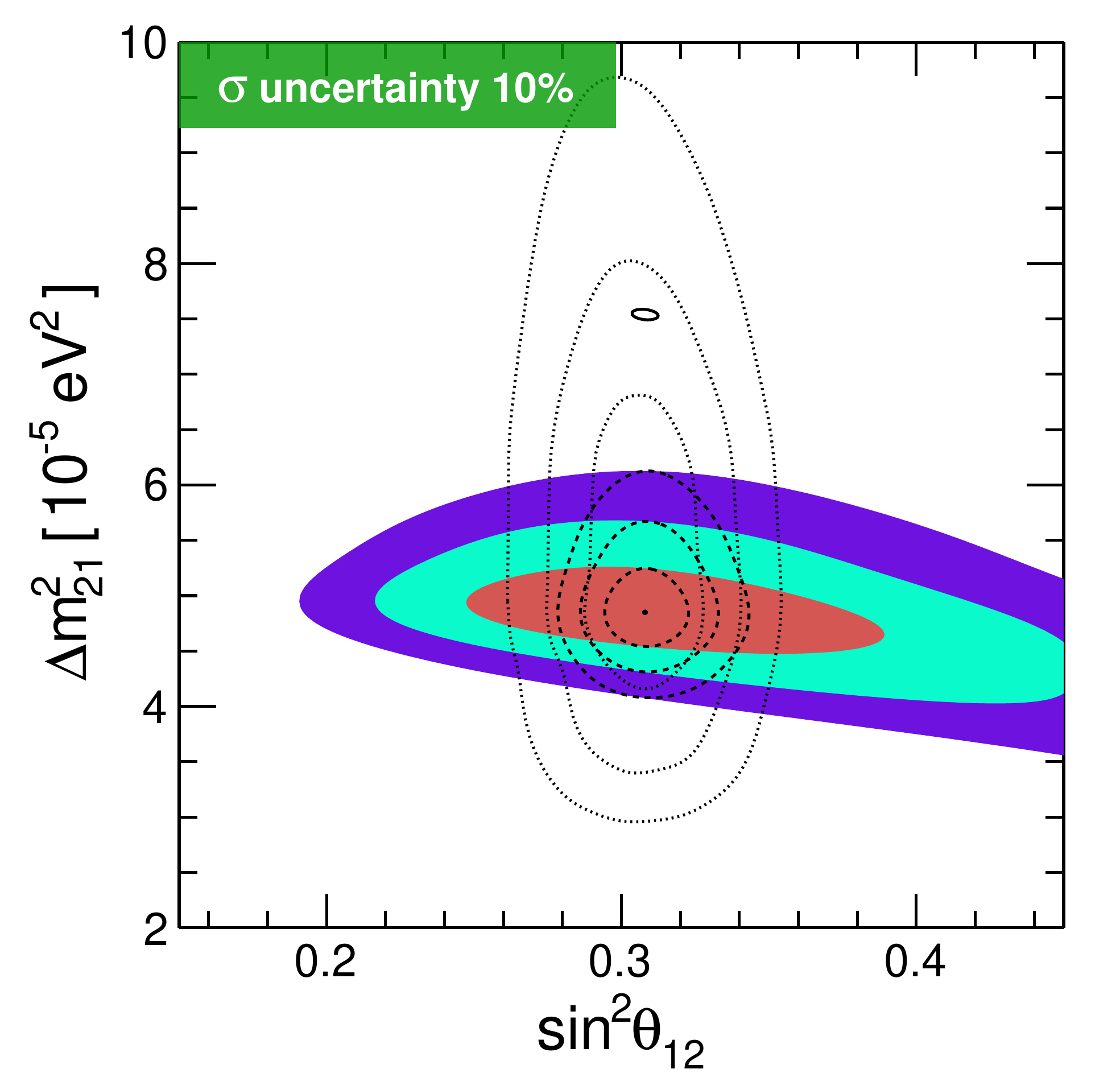}
\caption{Impact of the uncertainty on the charged-current cross section (1\% or 10\%) on the precision of the mixing parameters.  The dashed lines are for our nominal case, where this uncertainty is neglected, and for which results are shown in Fig.~2.}
\label{fig:CC_uncertainty}
\end{center}
\end{figure*}

\begin{figure*}
\begin{center}
\includegraphics[width=\columnwidth]{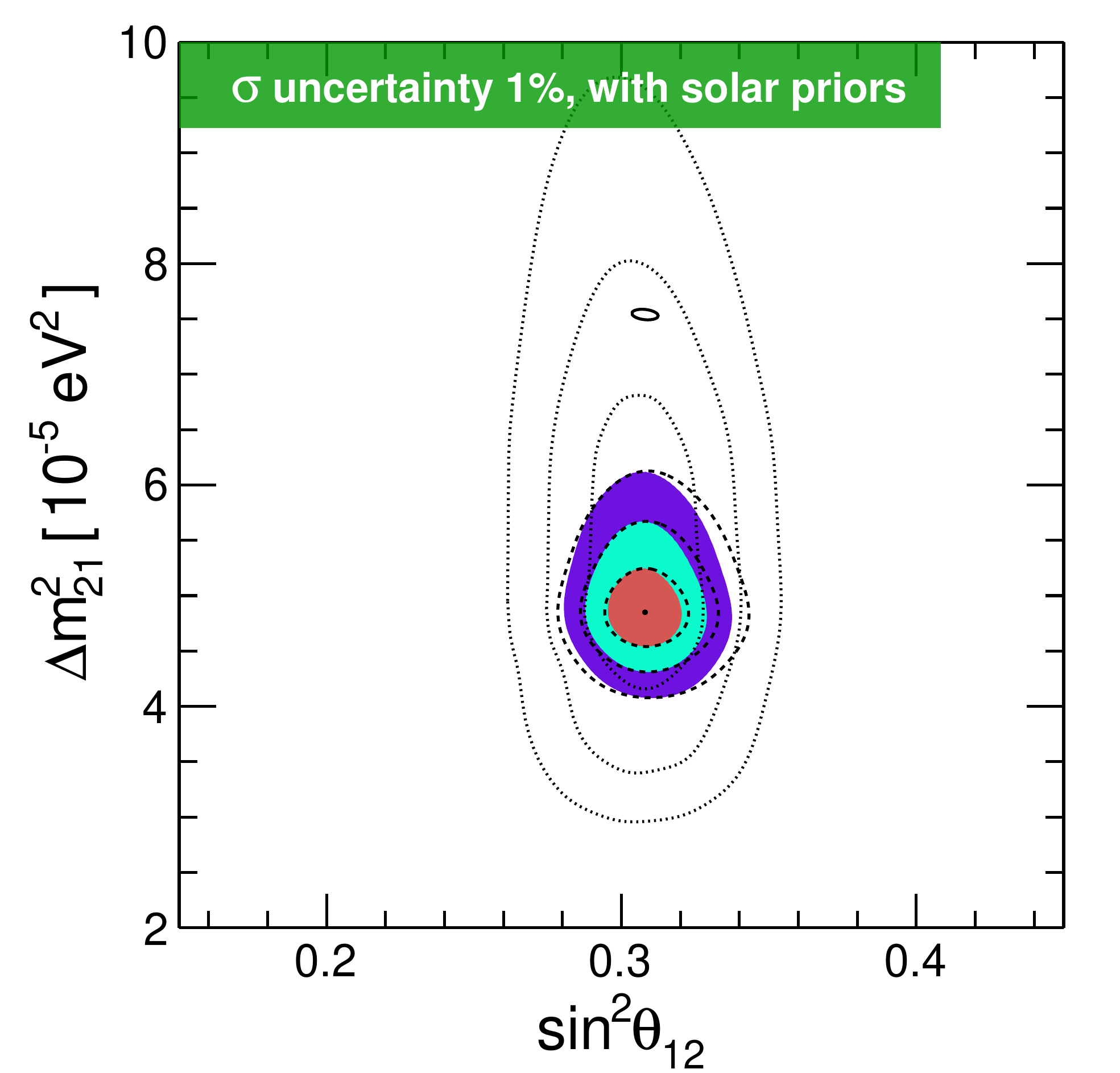}
\hspace{0.25cm}
\includegraphics[width=\columnwidth]{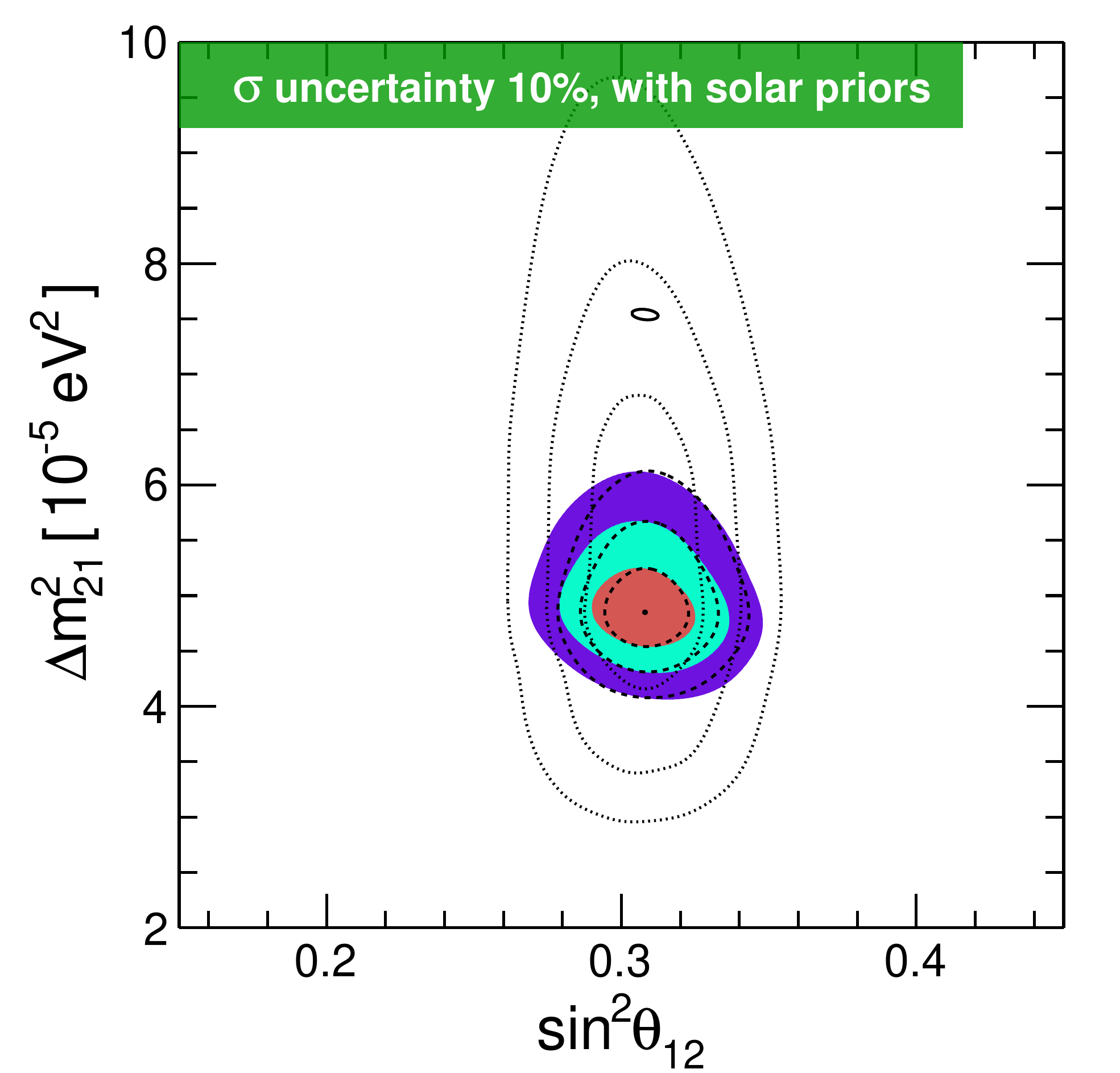}
\caption{Same as Fig.~\ref{fig:CC_uncertainty}, but now including uncorrelated priors of 4\% each on $\phi(^8$B) and $\sin^2\theta_{12}$, based on solar data.}
\label{fig:CC_uncertainty2}
\end{center}
\end{figure*}

In our main analysis, we neglected the uncertainty on the charged-current cross section.  Here we estimate the effects of a nonzero uncertainty by assuming a normalization uncertainty of 1\% or 10\%.  In a future work, we will consider the effects of uncertainties on individual transition strengths.  Figure~\ref{fig:CC_uncertainty} shows the allowed ranges of the mixing parameters in these cases.  For a cross-section uncertainty of 1\%, the effects are moderate, indicating that reaching this scale would be adequate.  For a cross-section uncertainty of 10\%, the uncertainty on $\sin^2\theta_{12}$ is worsened significantly, while that of $\Delta m^2_{21}$ is unchanged.  The normalization of the charged-current event spectrum depends on the product of $\phi(^8$B), the cross section, and $\sin^2\theta_{12}$, so they are degenerate.  The elastic-scattering event spectrum depends on $\phi(^8$B), a known cross section, and a different weighting of $\sin^2\theta_{12}$, so it can only be used to solve for two of the three parameters.  In the day-night asymmetry, all three factors cancel.

If the results of other detectors are included, then the effects of cross-section uncertainties are much less.  (All results above are based on results from DUNE alone, with no priors on the parameters.)  The reason is that these data help break the degeneracy in the normalization of the spectrum.  Figure~\ref{fig:CC_uncertainty2} repeats Fig.~\ref{fig:CC_uncertainty} but takes into account 4\% Gaussian priors on $\phi(^8$B) and $\sin^2\theta_{12}$ from solar data, and not taking advantage of their correlations.  A joint fit with other data would be even more powerful.

\begin{figure*}[t]
\begin{center}
\includegraphics[width=\columnwidth]{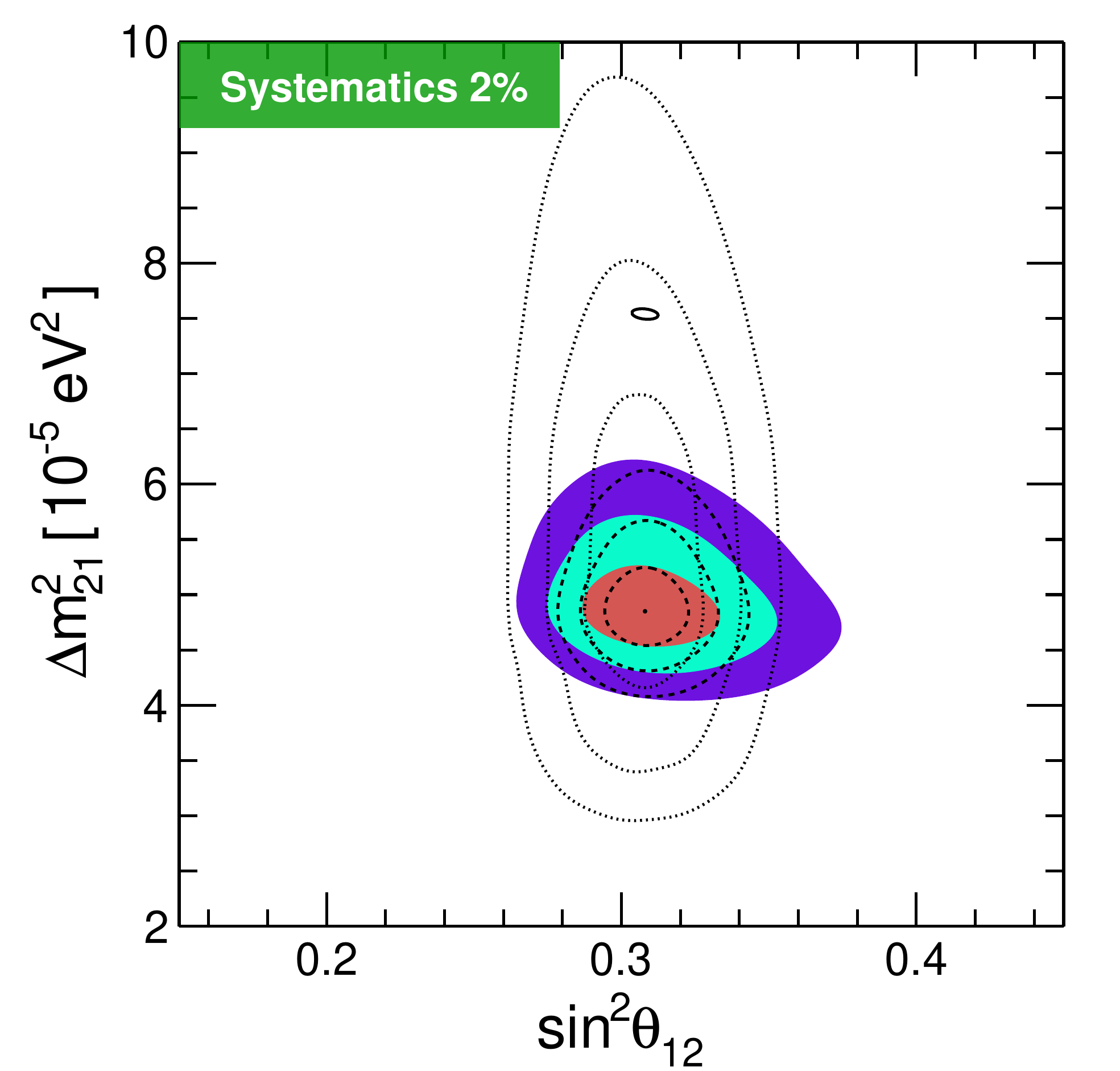}
\hspace{0.25cm}
\includegraphics[width=\columnwidth]{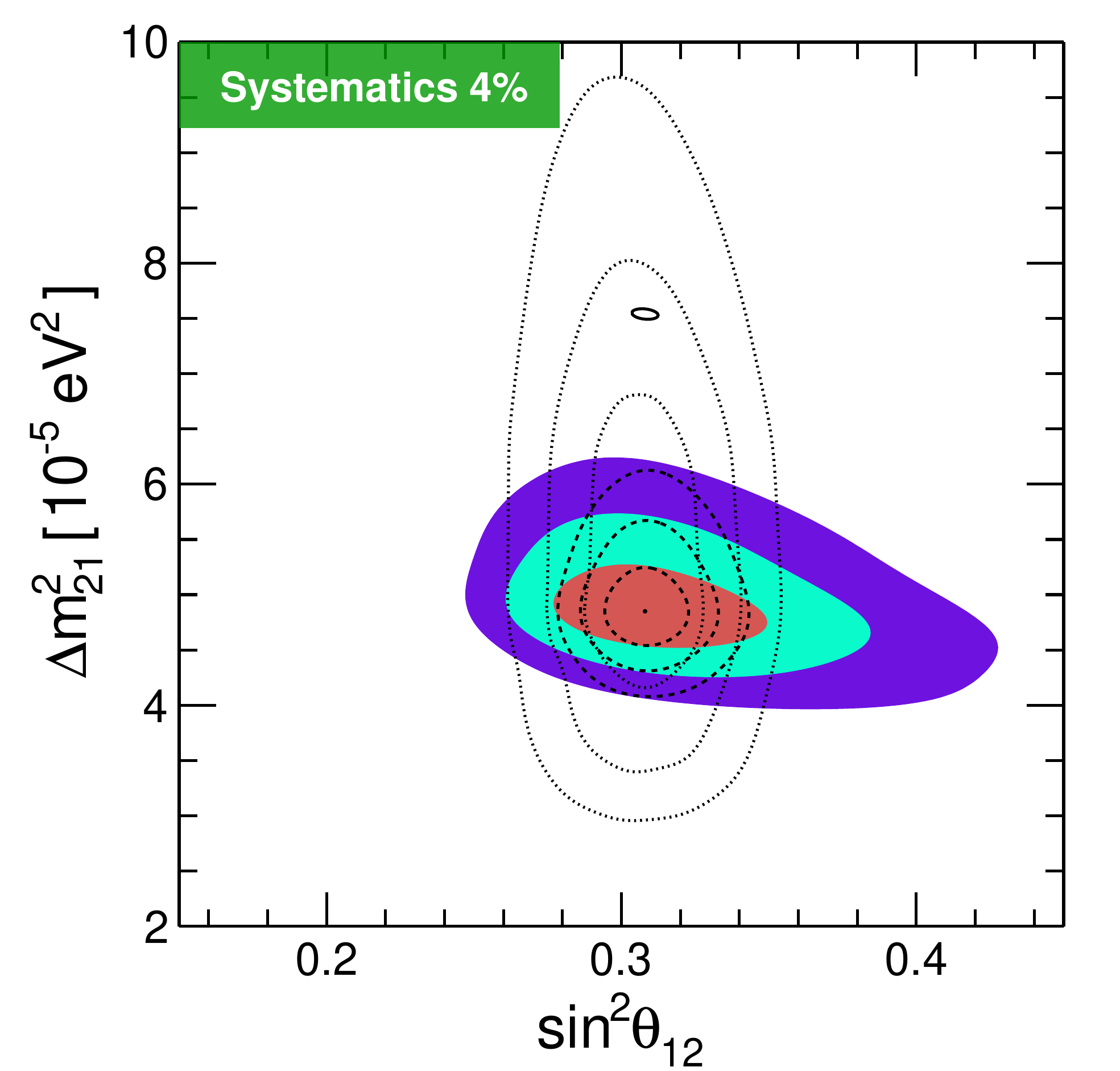}
\caption{Similar to Fig.~\ref{fig:CC_uncertainty}, but for the cases of 2\% and 4\% systematics.  With solar priors, these regions would shrink.}
\label{fig:syst_uncertainty}
\end{center}
\end{figure*}


\subsection{Impact of Other Systematic Uncertainties}
\label{sec:impact_systematics}

In our main analysis, we neglected systematic uncertainties.  Here we make estimates of their effects on our results, neglecting cross-section uncertainties, as those on the charged-current cross section are considered above and those on the neutrino-electron scattering cross section are tiny.  All other inputs are as in our main analysis.  To introduce a scale for expectations, the systematic uncertainties in Super-K and SNO are at or below a few percent~\cite{Aharmim:2005gt, Aharmim:2011vm, Abe:2016nxk}.

The most important energy correlated systematic uncertainties are typically those on the energy scale (e.g., nonlinearities) and energy resolution parameters.  Additionally, one needs to consider the uncertainty on the shape of the $^8$B energy spectrum.  The most important energy uncorrelated systematic uncertainties, i.e., those acting independently on each energy bin, are typically uncertainties on efficiencies, cuts, and background spectrum shape.  The main systematic uncertainties for DUNE are unknown, so we simply set a scale and repeat our analysis.  When we set a scale of $X \%$, we mean this for {\it each} of the energy scale, the energy resolution, and the energy uncorrelated systematic uncertainties.  This is conservative compared to what has been achieved in Super-K and SNO.  We take the systematic uncertainty on the $^8$B spectrum shape from Ref.~\cite{Winter:2004kf}.

Figure~\ref{fig:syst_uncertainty} shows how the allowed regions change with systematic uncertainties of 2\% and 4\%.  The effects are comparable to those due to an uncertainty on the charged-current cross section.  The precision on $\sin^2\theta_{12}$ is weakened, because it depends on the spectrum normalization.  On the other hand, the precision on $\Delta m^2_{21}$ is much less affected, because it mostly derives from the day-night effect, for which the systematics cancel almost completely in the ratio.

As above for the cross-section uncertainties, the effects of the systematic uncertainties are significantly mitigated if we include solar priors (not shown).  Because of this, and because we conservatively took all types of systematics at the same scale, the true results would be better.


\end{appendix}

\end{document}